\documentclass[twocolumn]{aastex61}
\usepackage{amssymb}
\usepackage{subfigure}
\submitjournal{ApJ}
\shorttitle{CO and Its Isotopologues toward Cas A}
\shortauthors{Ma et al.}

\begin{document}
\title{A Large-field $J=1-0$ Survey of CO and Its Isotopologues Toward the Cassiopeia A Supernova Remnant}
\correspondingauthor{Yuehui Ma}
\email{mayh@pmo.ac.cn}

\author[0000-0002-8051-5228]{Yuehui Ma}
\affil{Purple Mountain Observatory and Key Laboratory of Radio Astronomy, Chinese Academy of Sciences, 2 West Beijing Road, Nanjing 210008, China}
\affil{University of Chinese Academy of Sciences, 19A Yuquan Road, Shijingshan District, Beijing 100049, China}

\author[0000-0003-0746-7968]{Hongchi Wang}
\affil{Purple Mountain Observatory and Key Laboratory of Radio Astronomy, Chinese Academy of Sciences, 2 West Beijing Road, Nanjing 210008, China}

\author{Miaomiao Zhang}
\affil{Purple Mountain Observatory and Key Laboratory of Radio Astronomy, Chinese Academy of Sciences, 2 West Beijing Road, Nanjing 210008, China}

\author{Chong Li}
\affil{Purple Mountain Observatory and Key Laboratory of Radio Astronomy, Chinese Academy of Sciences, 2 West Beijing Road, Nanjing 210008, China}
\affil{University of Chinese Academy of Sciences, 19A Yuquan Road, Shijingshan District, Beijing 100049, China}

\author[0000-0001-7768-7320]{Ji Yang}
\affil{Purple Mountain Observatory and Key Laboratory of Radio Astronomy, Chinese Academy of Sciences, 2 West Beijing Road, Nanjing 210008, China}

\begin{abstract}
We have conducted a large-field simultaneous survey of $^{12}$CO, $^{13}$CO, and C$^{18}$O $J=1-0$ emission toward the Cassiopeia A (Cas A) supernova remnant (SNR), which covers a sky area of $3.5\arcdeg\times3.1\arcdeg$. The Cas giant molecular cloud (GMC) mainly consists of three individual clouds with masses on the order of 10$^4$$-$10$^5\ M_{\sun}$. The total mass derived from the $\rm{^{13}CO}$ emission of the GMC is 2.1$\times10^{5}\ M_{\sun}$ and is 9.5$\times10^5\ M_{\sun}$ from the $\rm{^{12}CO}$ emission. Two regions with broadened (6$-$7 km s$^{-1}$) or asymmetric $^{12}$CO line profiles are found in the vicinity (within a 10$\arcmin\times$10$\arcmin$ region) of the Cas A SNR, indicating possible interactions between the SNR and the GMC. Using the GAUSSCLUMPS algorithm, 547 $^{13}$CO clumps are identified in the GMC, 54$\%$ of which are supercritical (i.e. $\alpha_{\rm{vir}}<2$). The mass spectrum of the molecular clumps follows a power-law distribution with an exponent of $-2.20$. The pixel-by-pixel column density of the GMC can be fitted with a log-normal probability distribution function (N-PDF). The median column density of molecular hydrogen in the GMC is $1.6\times10^{21}$ cm$^{-2}$ and half the mass of the GMC is contained in regions with H$_2$ column density lower than $3\times10^{21}$ cm$^{-2}$, which is well below the threshold of star formation. The distribution of the YSO candidates in the region shows no agglomeration.

\end{abstract}
\keywords{ISM: clouds --- ISM: individual objects(Cas A) --- ISM: supernova remnants --- stars: formation --- surveys}

\section{Introduction} \label{sec:Sec1}
Molecular clouds are the densest part of the interstellar medium and the birthplace of stars. Although molecular hydrogen (H$_2$) is the primary component of molecular clouds, they are difficult to observe via rotational transition lines as H$_2$ possesses no permanent dipole moment. The most commonly used tracers of molecular clouds are the $^{12}$CO molecule and its isotopologues. In the past four decades, there have been plenty of large-scale CO surveys of the Milky Way (for a review, see \cite{Heyer+etal+2015}), such as the most extensive CfA-Chile survey \citep{Dame+etal+2001} and the FCRAO Outer Galaxy Survey (OGS) \citep{Heyer+etal+2001}. These surveys revealed the global distribution and statistical properties of molecular gas in the Milky Way. The majority of molecular clouds are accumulated into cloud complexes, which are called giant molecular clouds (GMCs). The sizes and masses of GMCs lie approximately in the range from 10 to 150 pc and 10$^3M_{\sun}$ to 10$^7M_{\sun}$, respectively. GMCs are composed of hierarchical structures like clouds, clumps, and cores \citep{Scalo+etal+1985, Falgarone+etal+1987, Blitz+etal+1999}. The mass function of molecular clouds follows a power law, with distinct power-law exponents for the inner and the outer Galaxy, $-$1.6 and $-$2.2, respectively \citep{Rice+etal+2016}. There exists a scaling relation between the line-widths and the sizes of molecular clouds over a wide range of cloud size \citep{Larson+etal+1981, Solomon+etal+1987}. Turbulence is considered to be one of the origins of this line-width-size relation and the hierarchical structure of molecular clouds \citep{Vazquez-Semadeni+etal+1994}. The physical properties and dynamical state of molecular clumps are of great importance as molecular clumps are observed to be the birthplace of massive clusters \citep{Williams+etal+1995}. Different statistical properties are expected for clumps under different supporting and confining mechanisms \citep{Bertoldi+etal+1992, Heyer+etal+2001, Kauffmann+etal+2013}. In addition, the temperature and number density of molecular clumps also vary with different interstellar environments \citep{Blitz+etal+1993, Williams+etal+1995}.

The structure and dynamics of molecular clouds are strongly influenced by dynamical processes like cloud$-$cloud collision, stellar winds from massive stars, and supernova explosion \citep{Truelove+etal+1999, Matzner+etal+2002, Bally+etal+2016}, among which the supernova explosion is the most intense. The blast wave of the explosion drives strong shocks into the surrounding interstellar medium. When the shock hits a molecular cloud, the shocked gas will be compressed, accelerated, and heated, leading to a variety of observable effects such as OH 1720 MHz maser emission, velocity broadening of the molecular-line emission, extended wing emission of $^{12}$CO low-$J$ transition lines, or high CO $J=2-1/J=1-0$ line intensity ratio \citep{Seta+etal+1998, Jiang+etal+2010}. However, among the 295 supernova remnants (SNRs) detected in the Milky Way \citep{Green+etal+2017}, only $\sim23\%$ of them are confirmed or suggested to be associated with molecular clouds \citep{Jiang+etal+2010, Chen+etal+2014}.

The Cassiopeia A SNR is the youngest known supernova remnant in our Galaxy and is the brightest discrete radio source at 100$-$1000 MHz in the sky, except for our Sun. It is the outcome of a Type IIb supernova explosion at around 1681 A.D. \citep{Fesen+etal+2006, Krause+etal+2008}. The morphological structures, physical properties, and chemical compositions of the Cas A SNR have been studied by many multiband observations. The SNR is mainly composed of a broken shell of size $\sim$4$\arcmin$ as observed in the X-ray, optical, and IR wavelengths. The main shell consists of the reverse shocked debris that is rich of O, Si, S, Ar, and Ca elements \citep{Thorstensen+etal+2001}. Furthermore, a lot of ``fast$-$moving knots (FMKs)" are distributed asymmetrically along the northeast and southwest directions \citep{Reed+etal+1995, Thorstensen+etal+2001, Ennis+etal+2006, Patnaude+etal+2014}. The outermost FMKs are located beyond the forward shock ($\sim$5$\arcmin$ in diameter) and have expansion velocities greater than $\sim$10,000 km s$^{-1}$ \citep{Fesen+etal+1996, Gotthelf+etal+2001, Fesen+etal+2006, Hammell+etal+2008}. Optical spectroscopic observations of the FMKs revealed that the Cas A SNR is located at the distance of 3.4 $^{+0.3}_{-0.1}$ kpc \citep{Reed+etal+1995}.

There have been many studies on the relationship between the Cas A SNR and the surrounding medium. X-ray observations reveal that Cas A is expanding in the wind of its progenitor star \citep{Lee+etal+2014}. The flickering and trailing emissions of the optical knots in the remnant's periphery indicate the initial stage of the interactions between the SN ejecta and the surrounding ISM \citep{Fesen+etal+2011}. Absorption lines of the molecules OH, H$_2$CO, and HCO$^+$ \citep{Goss+etal+1984, Bieging+etal+1986, Reynoso+etal+2002, Zhou+etal+2018}, and emission lines of CO and its isotopologues have been detected toward the SNR \citep{Wilson+etal+1993, Kilpatrick+etal+2014, Zhou+etal+2018}. These observations have shown that some molecular gas of the GMC coincides with the SNR along the line of sight. However, whether there is any interaction between the SNR and the molecular gas is still unclear. A supporting evidence for the interaction is the slightly broadened H$_2$CO lines found by \cite{Reynoso+etal+2002} and the broadened $^{12}$CO $J=2-1$ lines found by \cite{Kilpatrick+etal+2014}. Recently, \cite{Zhou+etal+2018} suggested that there is no distinctive physical evidence for an interaction between the Cas A SNR and molecular gas. Observations by both \cite{Kilpatrick+etal+2014} and \cite{Zhou+etal+2018} are limited to a sky coverage of only or less than $10\arcmin \times 10\arcmin$ centered on the SNR. Wide-field surveys of molecular gas around the Cas A SNR include the CfA-Chile survey \citep{Dame+etal+2001} and the FCRAO OGS survey \citep{Heyer+etal+1998}. \cite{Ungerechts+etal+2000} have provided an $\sim8\arcdeg\times8\arcdeg$ map of the GMCs in the region of Cas A and NGC 7538 as part of the CfA-Chile survey, but the mapping has quite coarse resolution and low sensitivity. The FCRAO OGS survey has only partly covered the MCs around Cas A (see Figure 1 in \citealt{Heyer+etal+2001}). Therefore, a detailed study of the structures and physical properties of the molecular gas toward the Cas A SNR is still lacking.

In this work, we present a large-scale (3.5$\arcdeg\times$3.1$\arcdeg$) survey of molecular gas toward the Cas A SNR, as part of the MWISP project \footnote{\url{http://www.radioast.nsdc.cn/mwisp.php}} \citep{Su+etal+2019}. The observation is described in Section \ref{sec:Sec2} and the results are presented in Section \ref{sec:Sec3}. We discuss the column density distribution and star formation activity in this region in Section \ref{sec:Sec4} and give our summary in Section \ref{sec:Sec5}.

\section{Observations} \label{sec:Sec2}
The observations were taken using the PMO-13.7 m millimeter-wavelength telescope located at Delingha in China during two observational seasons from 2015 October to 2016 April and 2016 October to 2017 April. The $^{12}$CO $J=1-0$, $^{13}$CO $J=1-0$, and C$^{18}$O $J=1-0$ line emission data were obtained simultaneously through the nine-beam Superconducting Spectroscopic Array Receiver (SSAR) mounted on the telescope \citep{Shan+etal+2012}. The front end of the receiver is a two-sideband Superconductor-Insulator-Superconductor (SIS) mixer. The $^{12}$CO $J=1-0$ line emission is contained in the upper sideband while $^{13}$CO $J=1-0$ and C$^{18}$O $J=1-0$ line emission in the lower sideband. The backend of the receiver is a Fast Fourier Transform Spectrometer (FFTS) with a total bandwidth of 1 GHz and 16,384 frequency channels, providing a spectral resolution of 61 kHz per channel. The resulting velocity resolution at 110 GHz is 0.17 km s$^{-1}$. During all the observational epochs, the half-power beam width (HPBW) of the telescope is about 52$\arcsec$ and 50$\arcsec$ at 110 GHz and 115 GHz, respectively. The pointing of the telescope has an accuracy of about 5$\arcsec$. The antenna temperature is calibrated according to $\rm{T_{mb}=T^{*}_{A}/\eta_{mb}}$ during the data collection process, where the main beam efficiency is 0.55 at 110 GHz and 0.51 at 115 GHz according to the status report \footnote{\url{http://english.dlh.pmo.cas.cn/fs/}} of the PMO-13.7 m millimeter telescope.

In the MWISP project, the sky area is divided into 30$\arcmin\times$30$\arcmin$ cells. For each cell, the observations were made in the position-switch on-the-fly (OTF) mode along the directions of the galactic longitude and galactic latitude, with a scanning rate of 50$\arcsec$ per second and a dump time of 0.3 s. The MWISP pipeline uses the GILDAS/CLASS\footnote{\url{http://www.iram.fr/IRAMFR/GILDAS}} software package to reduce the data. This process includes the subtraction of a first-order baseline from each spectrum and the combination of the spectra for the same sky position obtained at different times. For each cell, the data are then regridded into 30$\arcsec\times$30$\arcsec$ pixels in the directions of the galactic longitude and latitude. The final RMS noise level is $\sim$0.5 K per channel at the $^{12}$CO $J=1-0$ line wavelength and $\sim$0.3 K at the $^{13}$CO $J=1-0$ and C$^{18}$O $J=1-0$ line wavelengths.

\section{Results} \label{sec:Sec3}
\subsection{Spatial Distribution and Physical Properties of Cas GMC} \label{sec:Sec3.1}
\subsubsection{Spatial Distribution} \label{sec:Sec3.1.1}

\begin{figure*}[htb!]
\centering
\includegraphics[trim=0cm 0cm 0cm 1cm, width=0.6\linewidth , clip]{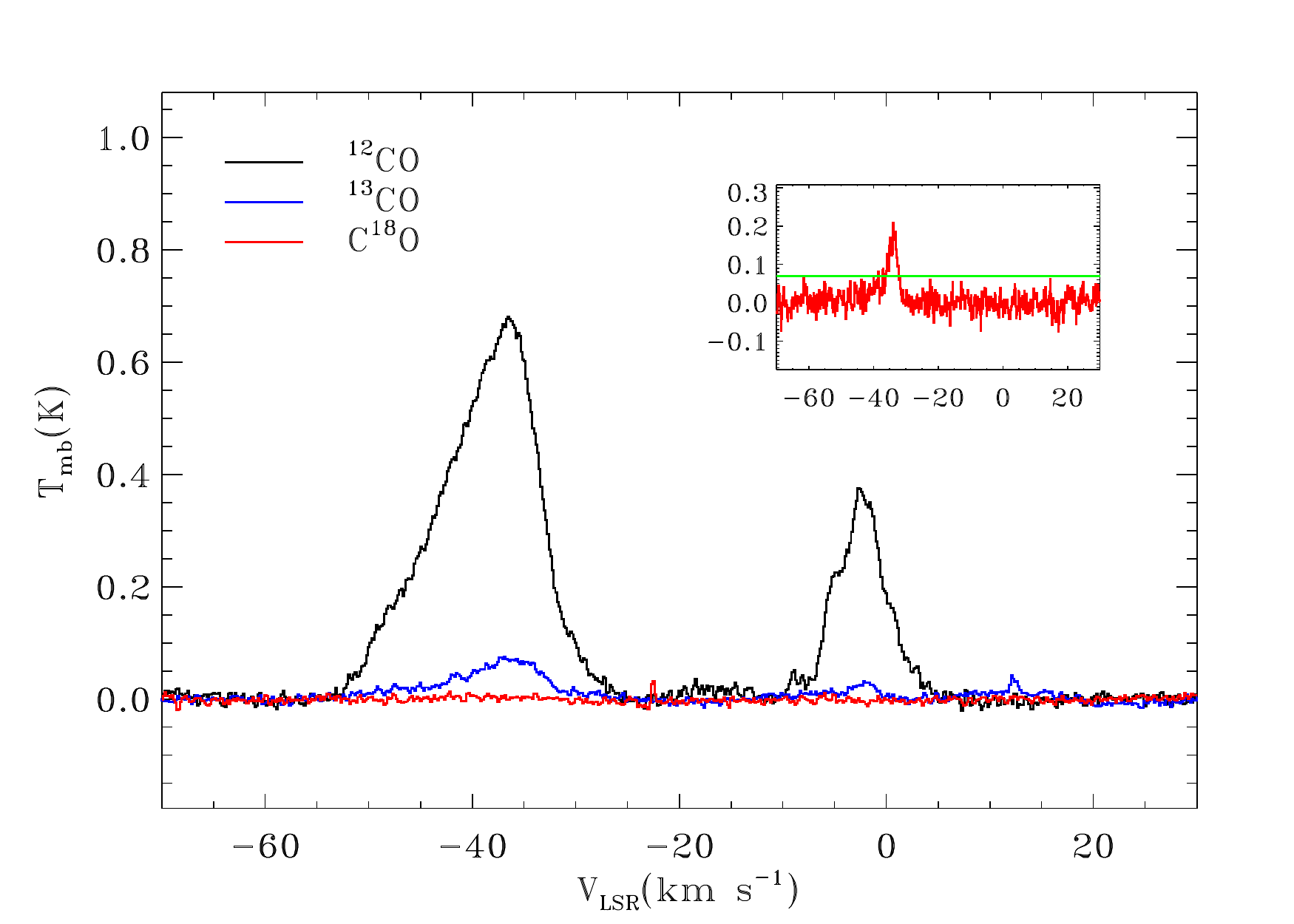}
\caption{Average spectrum of the observed 3.5$\arcdeg$$\times$3.1$\arcdeg$ region. Black, blue, and red lines indicate the $^{12}$CO, $^{13}$CO, and C$^{18}$O $J=1-0$ emission, respectively. The zoomed-in panel shows the average of the C$^{18}$O spectra that have at least four contiguous channels with C$^{18}$O emission above 1.5$\sigma$. The green line shows the 3$\sigma$ noise level of the average spectrum. The emission of $^{13}$CO at 13 km s$^{-1}$ and the emission of C$^{18}$O at $-$22.5 km s$^{-1}$ are due to bad channels.}
\label{fig:Fig1}
\end{figure*}

The average spectrum of the observed $3.5\arcdeg\times3.1\arcdeg$ region is shown in Figure \ref{fig:Fig1}. There are mainly two velocity components in this region, one from $-$60 to $-$20 km s$^{-1}$ and the other from -10 to 10 km s$^{-1}$. The first component is much stronger than the second one. Obviously, the second component is located in the solar neighborhood. In this work, we concentrate on the first component and the study on the second component is deferred to a future paper.

\begin{figure*}[htb!]
\centering
\includegraphics[trim=2cm 4cm 4cm 4cm, width=0.6\linewidth , clip]{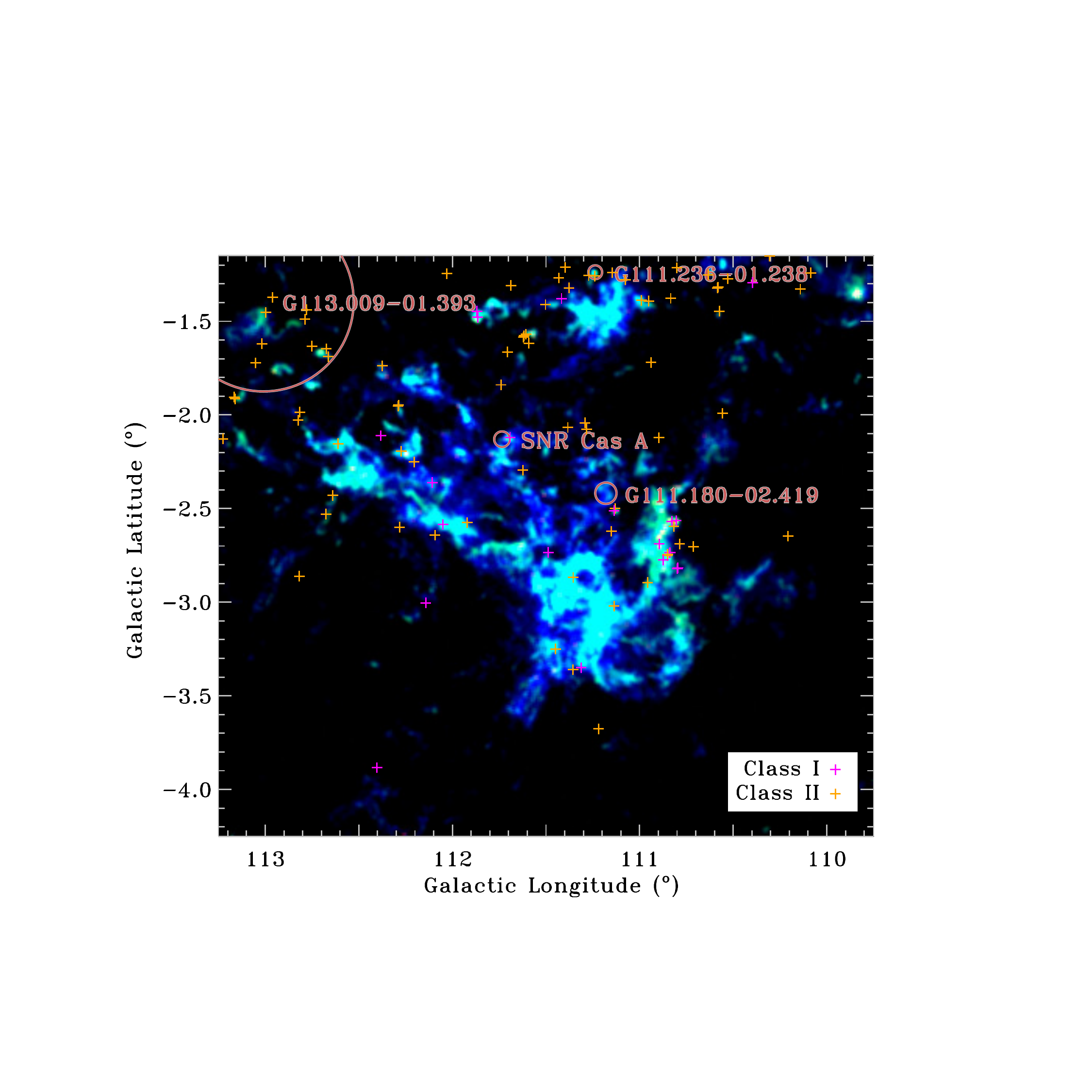}
\caption{Composite map of the $^{12}$CO, $^{13}$CO, and C$^{18}$O $J=1-0$ integrated intensity of the Cas GMC. The integrated velocity range is [$-$60, $-$24.5] km s$^{-1}$ for the $^{12}$CO and $^{13}$CO emission and [$-$40, $-$30] km s$^{-1}$ for C$^{18}$O. The blue, green, and red colors correspond to $^{12}$CO, $^{13}$CO, and C$^{18}$O emission, respectively. The Cas A SNR and the three \ion{H}{2} regions from the WISE catalog \citep{Anderson+etal+2014} in the region are indicated with red circles. The sizes of these circles correspond to their angular sizes. The Class II YSOs and Class I protostars identified in this work are marked with orange and magenta pluses, respectively.}
\label{fig:Fig2}
\end{figure*}

Figure \ref{fig:Fig2} illustrates the spatial distribution of the integrated intensity of the $^{12}$CO, $^{13}$CO, and C$^{18}$O $J=1-0$ emission. The blue, green, and red colors correspond to the $^{12}$CO, $^{13}$CO, and C$^{18}$O $J=1-0$ integrated intensity, respectively. The integrated velocity range for both $^{12}$CO and $^{13}$CO is from $-$60 to $-$24.5 km s$^{-1}$. Only spectra having at least five contiguous channels with brightness temperature above 1.5 $\rm\sigma_{rms}$ are used for the intensity integration. For C$^{18}$O, the integrated velocity range is from $-$40 to $-$30 km s$^{-1}$ with the same criterion as above. The known SNRs and \ion{H}{2} regions in the region \citep{Anderson+etal+2014, Green+etal+2014, Massaro+etal+2015} are indicated with red circles. As can be seen from Figure \ref{fig:Fig2}, the $^{12}$CO and $^{13}$CO emission in this region appears to be clumpy, and throughout the whole GMC, the C$^{18}$O emission is very weak. In the following analysis and calculation of this work, we only use the $^{12}$CO and $^{13}$CO data. Morphologically, there are two main separated substructures in this area, the northern one with a meniscus shape below the \ion{H}{2} region G111.236$-$01.238 and the southern one that consists of two filaments extending from the northeast to the southwest with two shell-like structures at their southwestern ends. The Cas A SNR is located at the center of the north filament, while the \ion{H}{2} region G113.009$-$01.393 is located near the northern end of this filament and the \ion{H}{2} region G111.180$-$02.419 is associated with the shell-like structure.

\begin{figure*}[ht!]
\centering
\includegraphics[trim=0cm 0.5cm 0cm 1cm, width=\linewidth, clip]{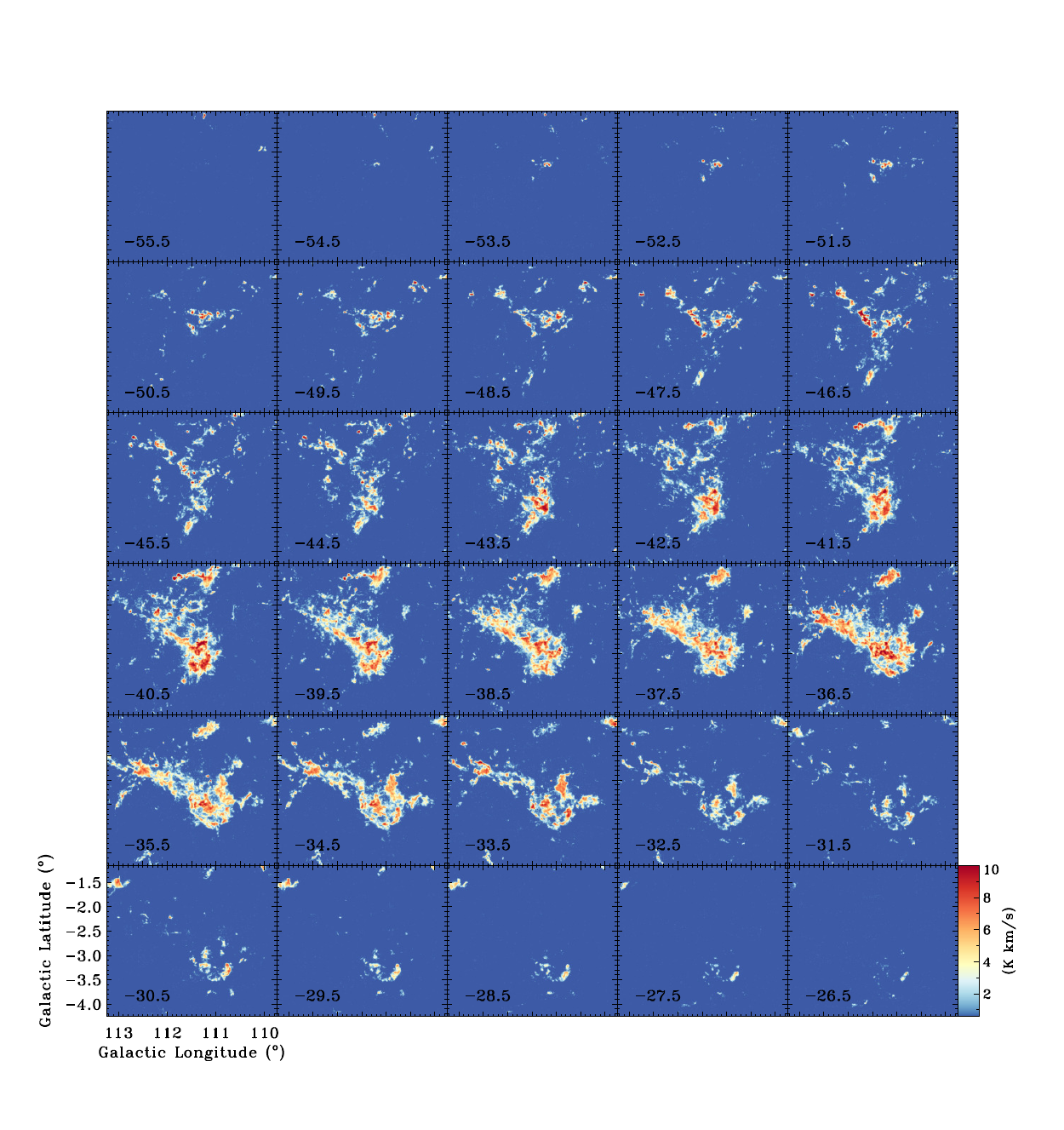}
\caption{$^{12}$CO $J=1-0$ velocity channel maps of the Cas GMC from $-$56 km s$^{-1}$ to $-$26 km s$^{-1}$ with a step of 1 km s$^{-1}$. The central velocity of each channel is shown at the bottom left corner of each pannel.}\label{fig:Fig3}
\end{figure*}

Figure \ref{fig:Fig3} displays the channel maps of the Cas GMC. We can see that the northern meniscus-shape cloud has velocity in the range from $-$45.5 to $-$33.5 km s$^{-1}$ while the southern part of the Cas GMC exhibits two velocity components. The first component ranges from $-$54.5 to $-$43.5 km s$^{-1}$ and is most discernible at the velocity channel $-$46.5 km s$^{-1}$. The second velocity component starts from $-$44.5 km s$^{-1}$ and ends at $-$26.5 km s$^{-1}$. Therefore, from the spatial distribution and the velocity structure, we can see that the Cas GMC can be divided into three major clouds: Cloud 1 near the \ion{H}{2} region G111.236$-$01.238 with a meniscus shape, Cloud 2 in the southern structure in the velocity range from $-$54.5 to $-$43.5 km s$^{-1}$, and Cloud 3 in the southern structure in the velocity range from $-$44.5 to $-$26.5 km s$^{-1}$. The channel maps of $^{13}$CO emission in the region are presented in Figure \ref{fig:Fig16} in the Appendix.

\begin{figure*}[htb!]
\centering
\subfigure[]{
\label{fig:Fig4a}
\includegraphics[trim=2cm 0cm 1cm 1cm, width=0.6\linewidth , clip]{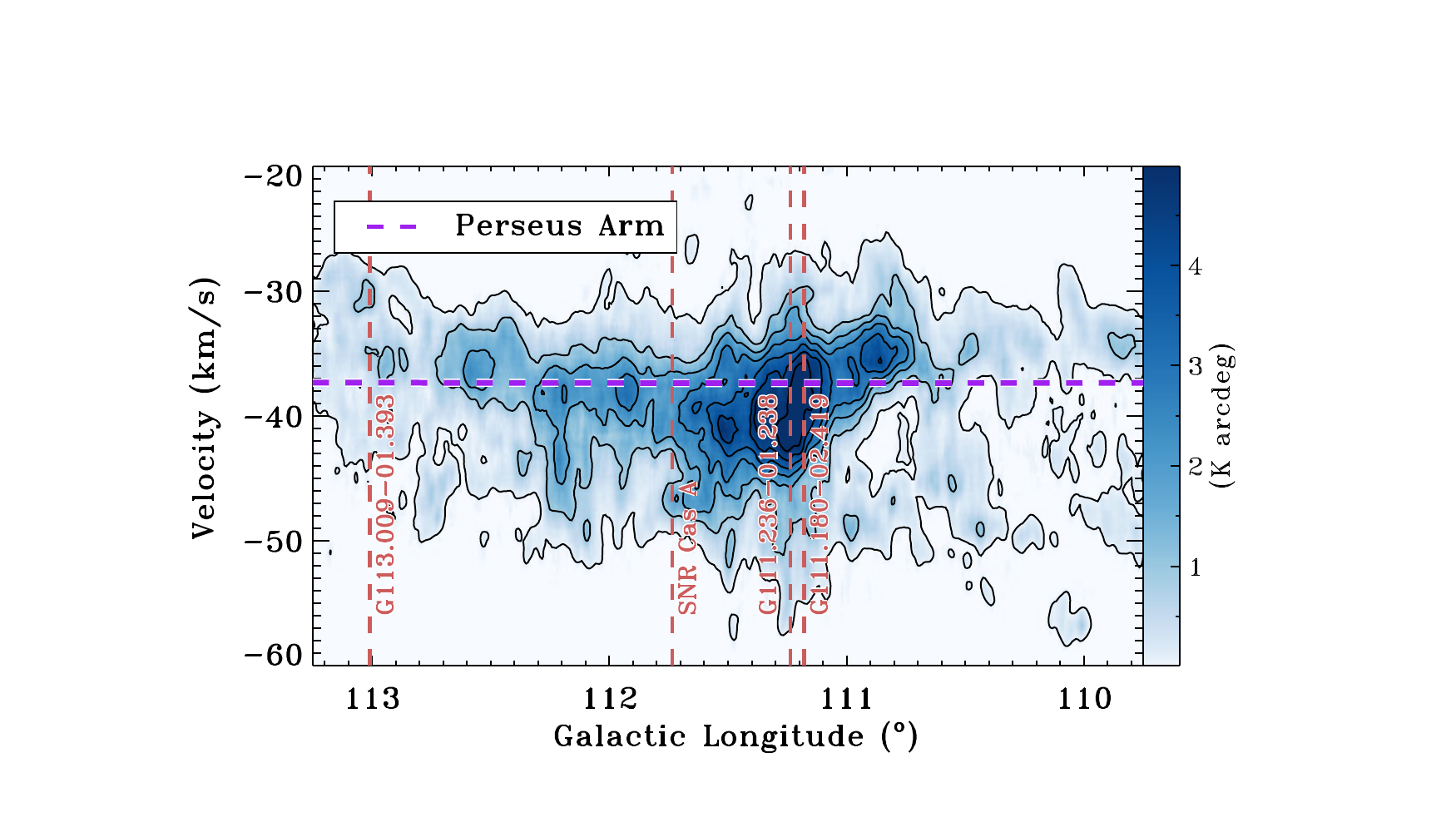}}
\subfigure[]{
\label{fig:Fig4b}
\includegraphics[trim=0.5cm 3cm 1cm 5cm, width=0.38\linewidth , clip]{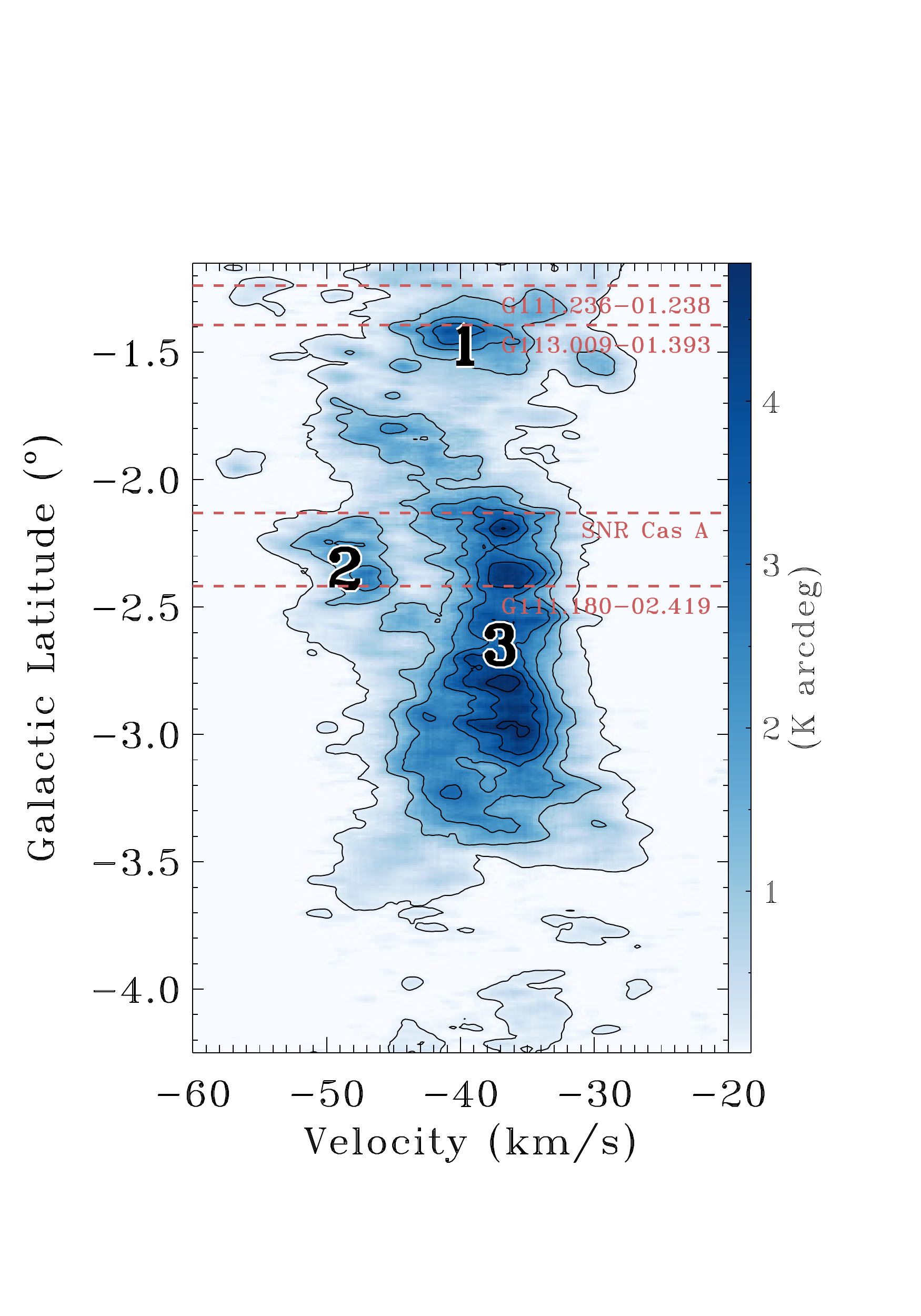}}
\caption{(a) $^{12}$CO l$-$v diagram. The purple dashed line is the l$-$v curve for the Perseus Arm derived from model A5 from \cite{Reid+etal+2014}. (b) Same as panel (a) but for the b$-$v diagram. The numbers indicate the three PPV components of Cas GMC (see the text for detailed information). Red dashed lines in the two panels mark the positions of the \ion{H}{2} regions and Cas A SNR. The contours start at 1.5$\sigma$ and then increase to 0.8 times the emission peak in six steps that have the same intervals.}\label{fig:Fig4}
\end{figure*}

The position-velocity diagrams of the $^{12}$CO emission of the Cas GMC are present in Figure \ref{fig:Fig4}. Figure \ref{fig:Fig4a} displays the l-v diagram, and Figure \ref{fig:Fig4b} displays the b-v diagram. The purple dashed line in Figure \ref{fig:Fig4a} shows the velocity of the Perseus Arm of model A5 from \cite{Reid+etal+2014}. We can see that the majority of the Cas GMC has velocities consistent with the Perseus Arm, which suggests that the Cas GMC is located in the Perseus Arm. No distinct structures can be identified in the l-v diagram. It appears that there are velocity broadenings near the galactic longitudes of the \ion{H}{2} regions G111.236$-$01.238 and G111.180$-$02.419. In fact, we find these velocity broadenings are caused by the velocity difference between Cloud 2 and Cloud 3. No velocity broadening is identified near the Cas A SNR or the \ion{H}{2} region G113.009$-$01.393. In contrast to Figure \ref{fig:Fig4a}, three distinct structures that correspond to Clouds 1-3 can be identified in the b-v diagram. As in the l-v diagram, no velocity broadening is found near the SNR or the \ion{H}{2} regions. For comparison, the p-v diagrams of the $^{13}$CO emission are presented in Figure \ref{fig:Fig17} in the Appendix. From the b-v diagram of the $^{13}$CO emission, we can see a ring structure in the southern end of Cloud 3 at b = $-$3.1$\arcdeg$ with velocity from $-$45 to $-$33 km s$^{-1}$. This ring coincides with the shell-like structure located at [l, b] = [111.3, $-$3.0]$\arcdeg$, as seen clearly in the $-$46.5 to $-$41.5 km s$^{-1}$ panels in Figure \ref{fig:Fig3}. A ring structure in a p-v diagram is often considered to be an expanding sphere \citep{Arce+etal+2011}. In this case, however, we have not found any source in the region that is responsible for the expansion.

\begin{figure*}[htb!]
\centering
\subfigure[]{
\label{fig:Fig5a}
\includegraphics[trim= 1cm 3.5cm 2cm 3cm, width=0.48\linewidth , clip]{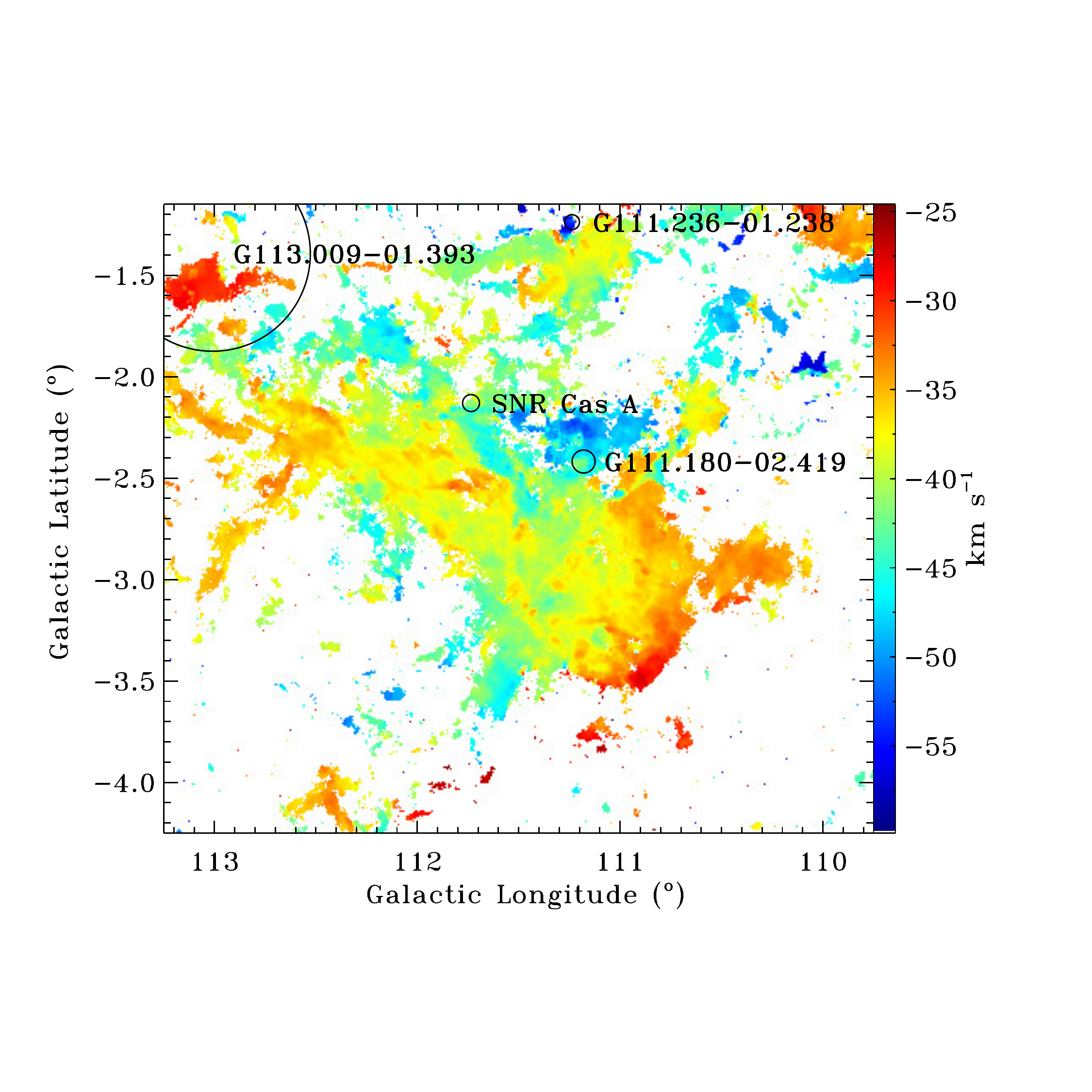}}
\subfigure[]{
\label{fig:Fig5b}
\includegraphics[trim= 1cm 3.5cm 2cm 3cm, width=0.48\linewidth , clip]{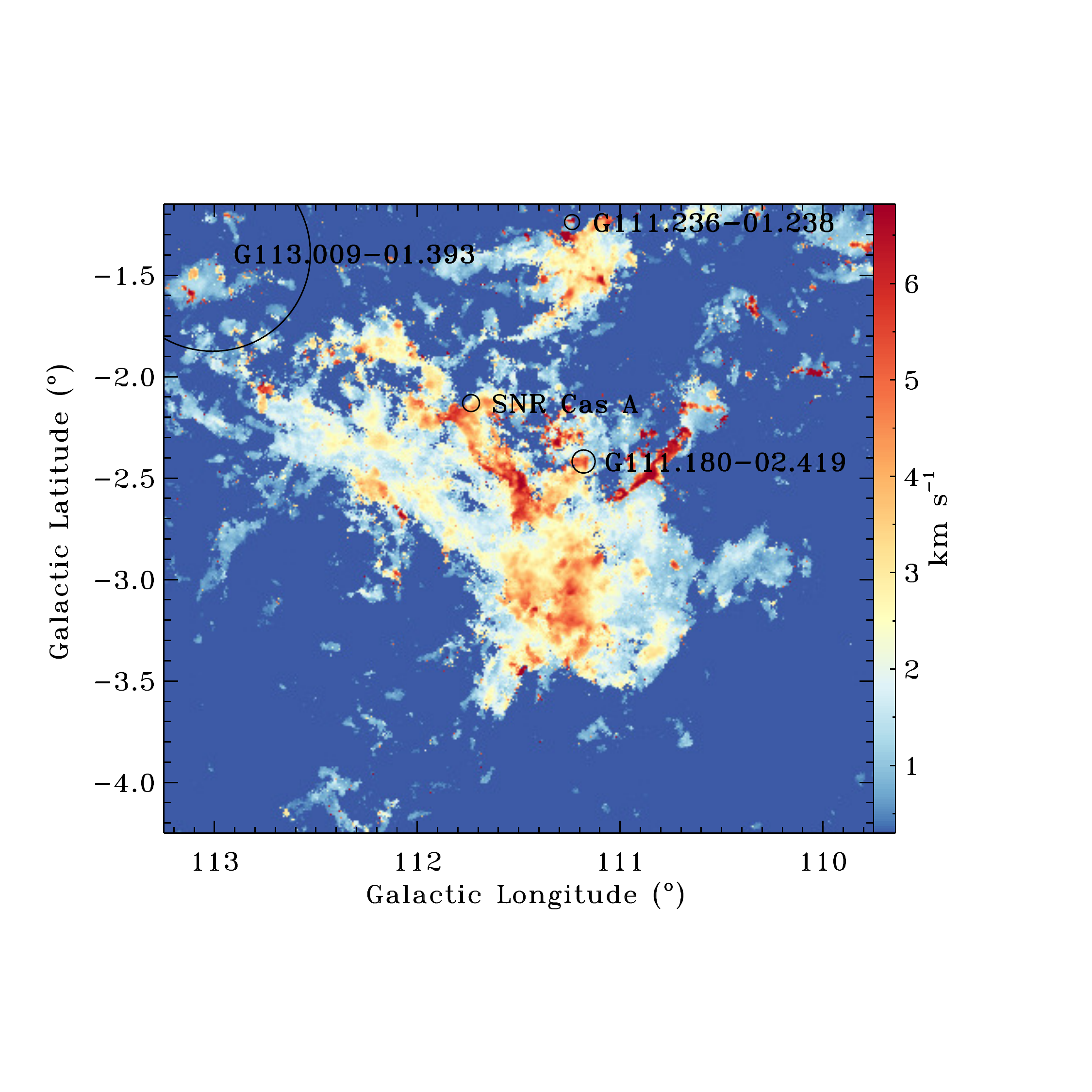}}
\caption{(a) Intensity-weighted centroid velocity map and (b) velocity dispersion map of the $^{12}$CO emission over the velocity range from $-$60 to $-$24.5 km s$^{-1}$.}
\label{fig:Fig5}
\end{figure*}

\begin{figure*}[htb!]
\centering
\subfigure[]{
\label{fig:Fig6a}
\includegraphics[trim= 1cm 3.5cm 2cm 3cm, width=0.48\linewidth , clip]{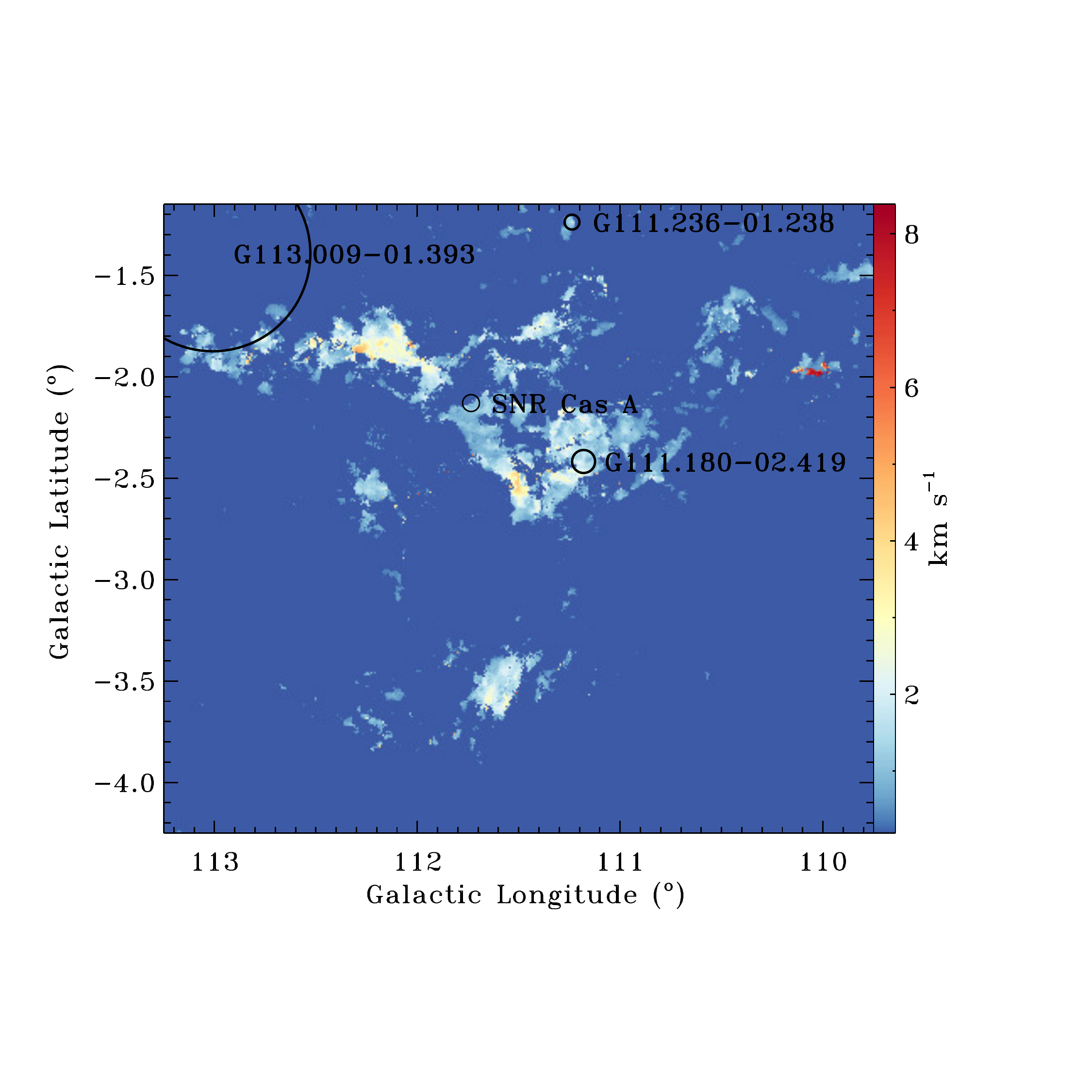}}
\subfigure[]{
\label{fig:Fig6b}
\includegraphics[trim= 1cm 3.5cm 2cm 3cm, width=0.48\linewidth , clip]{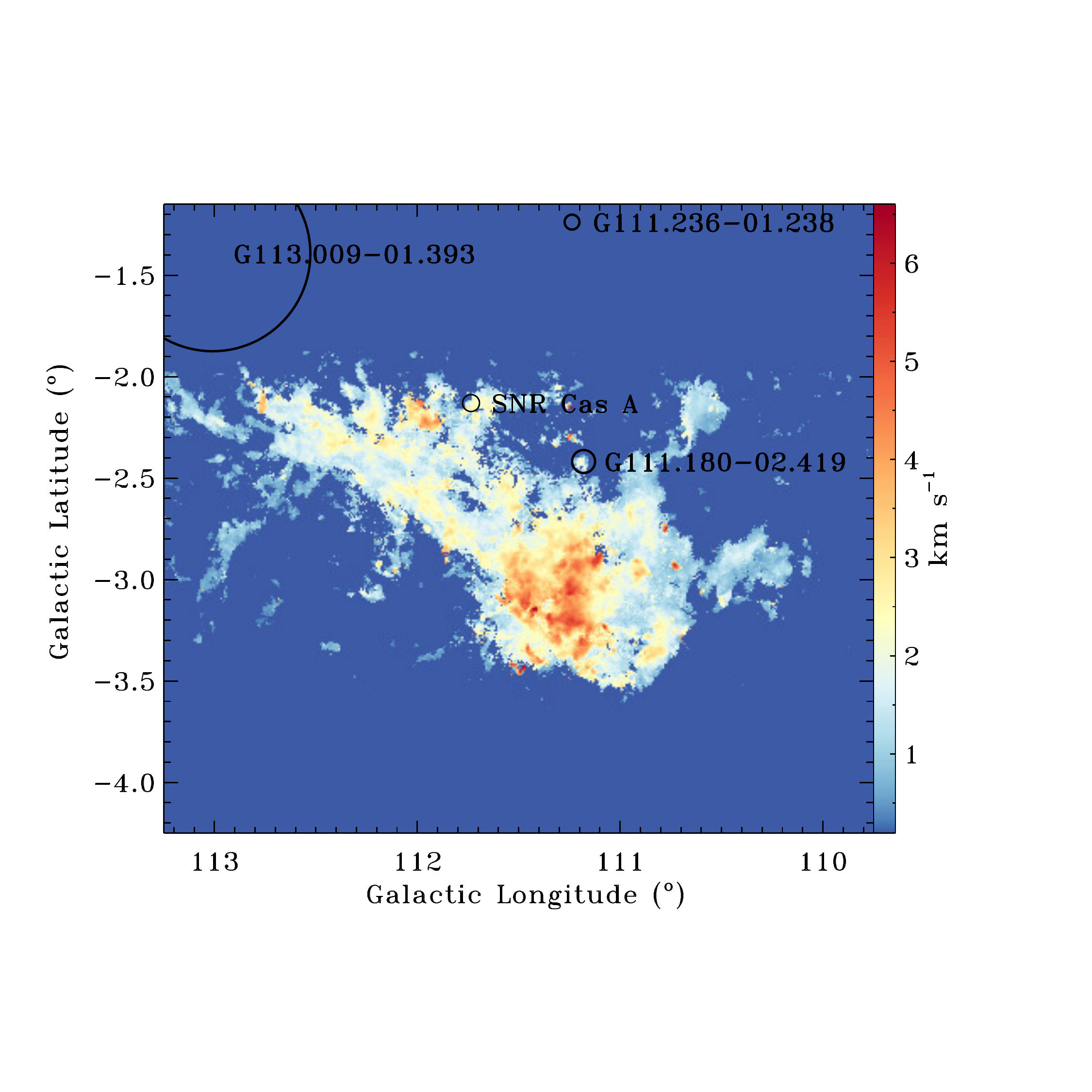}}
\caption{Velocity dispersion map of $^{12}$CO emission of (a) Cloud 2 and (b) Cloud 3 in the velocity range from $-$60 to $-$24.5 km s$^{-1}$.}
\label{fig:Fig6}
\end{figure*}

The centroid velocity and the velocity dispersion maps of the $^{12}$CO emission are presented in Figure \ref{fig:Fig5}. As shown in Figure \ref{fig:Fig5a}, although Clouds 2 and 3 somewhat coincide in spatial distribution, they exhibit distinct velocities with Cloud 2 shown in blue and Cloud 3 in yellow and orange. The southwestern and northeastern ends of Cloud 3 show red-shift velocities. In Figure \ref{fig:Fig5b}, we can see that large velocity dispersions ($>$3.5 km s$^{-1}$) are mostly due to the spatial coincidence of Clouds 2 and 3. In order to examine the velocity dispersion of individual clouds, we separate Cloud 2 from Cloud 3, and present their velocity dispersion maps in Figure \ref{fig:Fig6}. The median velocity dispersion of the $^{12}$CO emission is about 0.6 km s$^{-1}$ for Cloud 2 and 1.3 km s$^{-1}$ for Cloud 3. No apparent line broadening exists near the three \ion{H}{2} regions. A clump with large velocity dispersions (up to 4 km s$^{-1}$) can be found to the east of the Cas A SNR in Cloud 3.

\subsubsection{Physical Properties} \label{sec:Sec3.1.2}

\begin{deluxetable*}{CCLCCRCCCC}[htb!]
\tablecaption{Derived cloud parameters \label{tab:Tab1}}.
\tablecolumns{8}
\tablenum{1}
\tablewidth{0pt}
\tablehead{
\colhead{Cloud} & \colhead{$l_c$} & \colhead{$b_c$} & \colhead{$v_c$} & \colhead{$\sigma_v$} & \colhead{$d_{kin}$} & \colhead{$T_{ex}$} & \colhead{$M_{LTE}$} & \colhead{$M_{X_{CO}}$} \\
\colhead{} & \colhead{$(\arcdeg)$} & \colhead{$(\arcdeg)$} & \colhead{(km s$^{-1}$)} & \colhead{(km s$^{-1}$)} & \colhead{(kpc)} & \colhead{(K)} & \colhead{(M$_{\sun}$)} & \colhead{(M$_{\sun}$)}}
\startdata
$\rm{Cloud\ 1}$ & 111.301 & $-$1.454 & $-$39.38 & 2.62 & 3.17 & 8.1 & 2.4$\times10^{4}$ & 8.8$\times10^{4}$ \\
$\rm{Cloud\ 2}$ & 111.578 & $-$2.271 & $-$46.13 & 2.46 & 3.73 & 7.4 & 2.4$\times10^{4}$ & 1.6$\times10^{5}$ \\
$\rm{Cloud\ 3}$ & 111.519 & $-$2.770 & $-$37.49 & 3.29 & 3.00 & 7.8 & 1.5$\times10^{5}$ & 6.3$\times10^{5}$ \\
\enddata
\tablecomments{Columns 2$-$4 give the centroid position and column 5 the velocity dispersion of the clouds. The velocity dispersion $\sigma_v$ is derived from $^{13}$CO data. The kinematic distance, the mean excitation temperature, and the mass  of each cloud are given in Columns 6$-$9. Masses of the clouds are derived using a uniform distance of d = 3.4 kpc (see details in the text).}
\end{deluxetable*}

\begin{figure}[htb!]
\centering
\subfigure[]{
\label{fig:Fig7a}
\includegraphics[trim= 1cm 3.5cm 2cm 3cm, width=0.8\linewidth , clip]{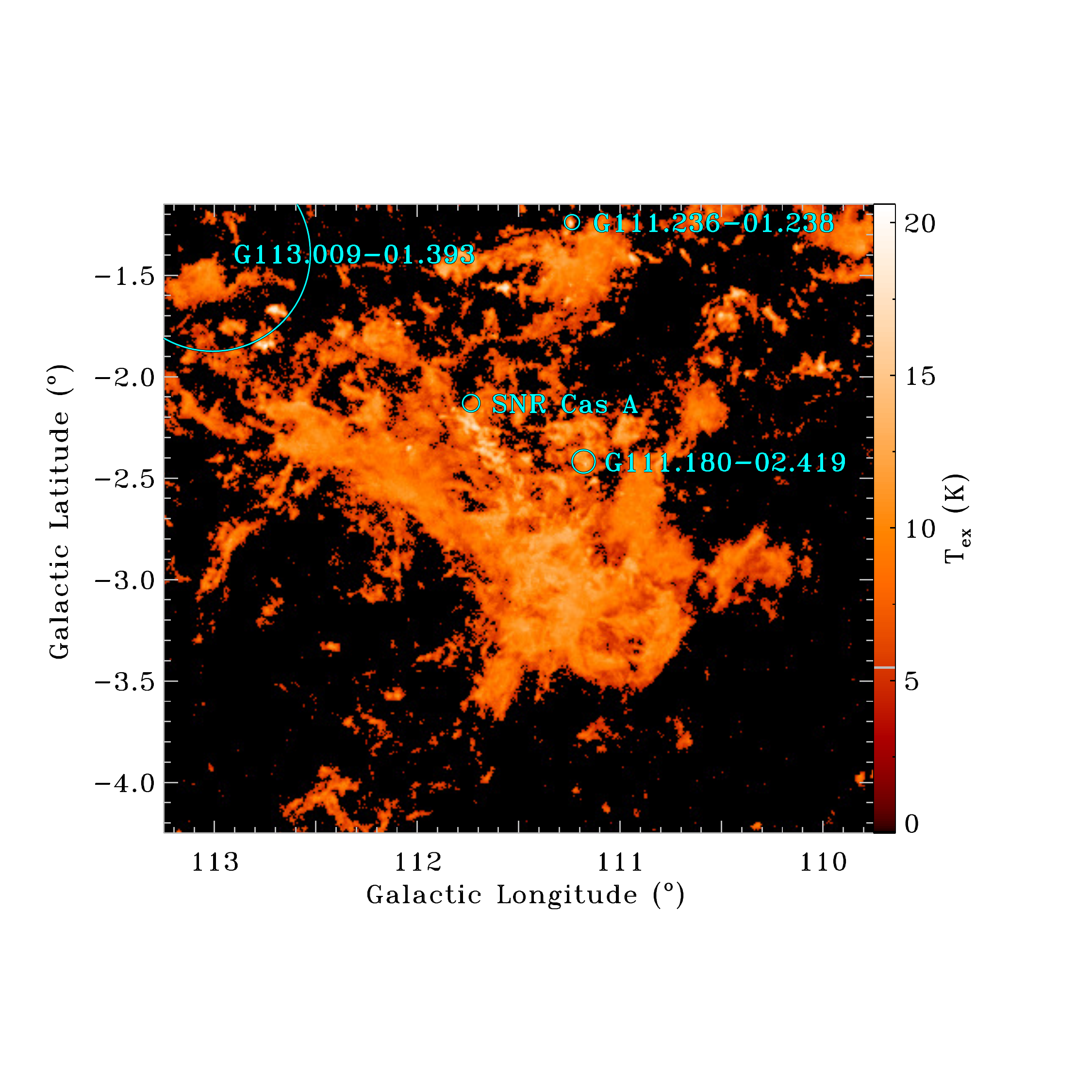}}
\subfigure[]{
\label{fig:Fig7b}
\includegraphics[trim= 0cm 0cm 0cm 1cm, width=0.8\linewidth , clip]{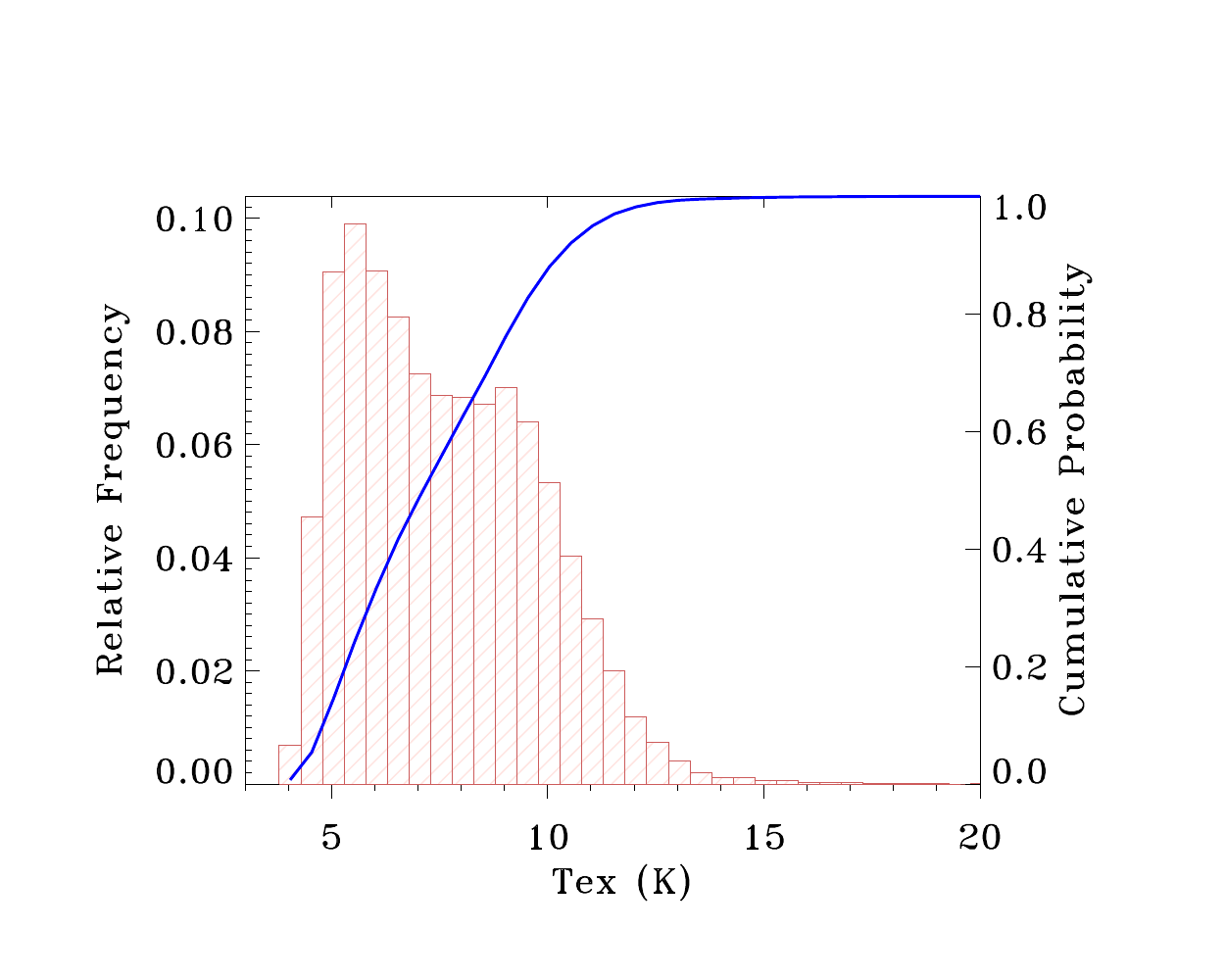}}
\caption{(a) Spatial distribution and (b) histogram of the excitation temperatures of the Cas GMC.}
\label{fig:Fig7}
\end{figure}

Following the method of \cite{Heyer+etal+2001}, the centroid position, [$l_c, b_c, v_c$], of a cloud that occupies N voxels in PPV space can be calculated through $\emph{\textbf{x}}_c = \Sigma_{i=1}^{N} T_i\emph{\textbf{x}}_i/\Sigma_{i=1}^{N} T_i$, where \emph{\textbf{x}} refers to $l$, $b$, or $v$ and $T_i$ is the brightness temperature of voxel $i$. The equivalent line width of a cloud can be obtained from the composite spectrum by dividing its integrated intensity by its emission peak. For the calculation, we only use the spectra that have at least five contiguous channels with brightness temperatures above 1.5$\sigma$. Based on the derived centroid position, the kinematic distance, $\rm{d_k}$, of a cloud can be obtained under the assumption of a flat rotation curve of the Galaxy (we use model A5 of \citealt{Reid+etal+2014} in this work).

The measured centroid positions and kinematic distances of Clouds 1$-$3 are listed in Table \ref{tab:Tab1}. The kinematic distances of Clouds 1$-$3 are 3.17, 3.73, and 3.00 kpc, respectively. We can see that the velocity and distance of Cloud 2 are not consistent with those of Clouds 1 and 3. We checked the trigonometric parallax distances of the two galactic masers, G111.23$-$01.23 and G111.25$-$00.76, which are located in or adjacent to the observed region. The parallax distances of G111.23$-$01.23 and G111.25$-$00.76, 3.472 and 3.401 kpc respectively \citep{Reid+etal+2014}, are consistent with each other, although their radial velocity difference amounts to 10 km s$^{-1}$ ($-$53 km s$^{-1}$ for maser G111.23$-$01.23 and $-$43 km s$^{-1}$ for maser G111.25$-$00.76).  Considering the non-circular motion of the Milky Way and the intrinsic velocity dispersions of molecular clouds, the method with the Galactic rotation model will induce significant uncertainties in distance estimation. As the centroid velocities of Clouds 1-3 are consistent with the maser G111.25$-$00.76, we adopt a uniform distance of 3.4 kpc for the three main clouds 1$-$3 and the Cas GMC as a whole in the following calculations and discussions.

The column density of the clouds is calculated in two different ways. In the first way, we assume that the clouds are under the local thermodynamic equilibrium (LTE) conditions and the $^{12}$CO $J=1-0$ emission is optically thick. The excitation temperature is calculated with the peak intensity of the $^{12}$CO $J=1-0$ emission, and the column density of $^{13}$CO is calculated with the integrated intensity of the $^{13}$CO $J=1-0$ emission. The detailed calculation processes are the same as in \cite{Li+etal+2018}. The derived column density of $^{13}$CO is then converted into the column density of molecular hydrogen with the abundance ratios $\rm{[^{12}C/^{13}C]=}$ 6.21$d_{GC}+18.71$ and $\rm{H_2/^{12}CO} = 1.1\times10^4$ \citep{Frerking+etal+1982, Milam+etal+2005, Gong+etal+2016}, where $d_{GC}$ is the cloud distance from the Galactic center. In our case, the distance $d_{GC}$ of the clouds is 10.1 kpc and the $\rm{[^{12}C/^{13}C]}$ ratio is 81. In the second way, we use a constant conversion factor $X\rm_{CO} = 2\times 10^{20}$ cm$^{-2}$ (K km s$^{-1}$)$^{-1}$\citep{Bolatto+etal+2013} to convert the integrated intensity of $^{12}$CO emission into the H$_2$ column density. The mass of the cloud is obtained through the integration of the H$_2$ column density over the effective emission area of the cloud.

The total mass for the whole Cas GMC derived using the LTE method from $^{13}$CO is $2.1\times10^5$ $M_{\sun}$, and that using the $X$ factor method is $9.5\times10^5$ $M_{\sun}$. The masses calculated with the above two methods and the average excitation temperatures for Clouds 1-3 are listed in Table \ref{tab:Tab1}. The masses of Clouds 1-3 lie between $10^4$ and $10^5$ M$_{\sun}$. The $X\rm_{CO}$ masses are 3.7 to 6.7 times the LTE masses, which is common for Galactic GMCs.

As introduced in Section \ref{sec:Sec1}, increasing temperature is one of the indicators of SNR-MC interaction and stellar feedback. We present the distribution and the statistics of the excitation temperature of the Cas GMC in Figure \ref{fig:Fig7}. The Cas GMC is generally cold, with a median temperature of about 7.5 K and a maximum of 20 K. No significant increase in temperature is found around the \ion{H}{2} regions or the Cas A SNR except for two clumps at the southwestern edge of the \ion{H}{2} region G113.009-01.393 and the ridge located below the SNR in the orientation of northeast-southwest. This ridge is part of Cloud 2. This temperature increase should be an intrinsic property of Cloud 2, rather than due to an interaction with the Cas A SNR. Therefore, the Cas GMC is not affected by the Cas A SNR or \ion{H}{2} regions at large scales in terms of increased temperature.

\subsection{Interactions between the Cas A SNR and the GMC} \label{sec:Sec3.2}

\begin{figure*}[htb!]
\centering
\includegraphics[trim=1cm 2cm 2cm 2cm, width= 12cm, clip]{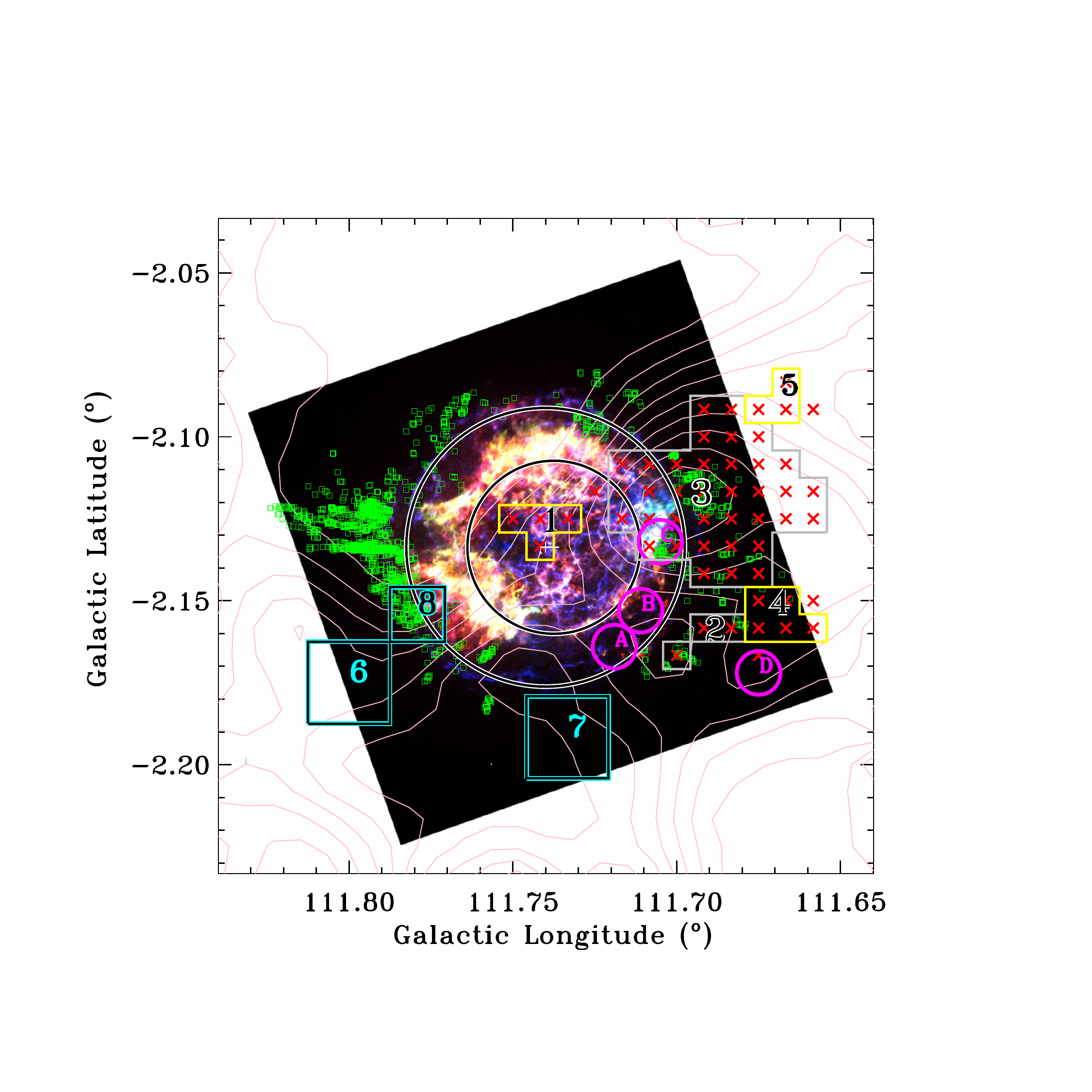}
\caption{X-ray image of the Cas A SNR. The red, green, and blue colors correspond to the energy bands of 0.5$-$1.5, 1.5$-$3.5, and 4$-$6 KeV, respectively. The data are from the \emph{Chandra X-ray Observatory} (\url{http://chandra.harvard.edu/photo/openFITS/xray_data.html}). The forward and the reverse shocks identified by \cite{Gotthelf+etal+2001} are represented with black circles. The pink contours represent $^{12}$CO emission. The green squares designate the positions of the outer optical knots \citep{Hammell+etal+2008}. The red crosses indicate the positions with broadened or asymmetric $^{12}$CO $J=1-0$ spectra that we found in this work. The gray and the yellow polygons mark the CO patches that show moderately broadened or asymmetric $^{12}$CO $J=1-0$ spectra. The blue-black rectangles 6$-$8 mark the control regions. The positions of the broadened $^{12}$CO $J=2-1$ spectra found by \cite{Kilpatrick+etal+2014} are shown with magenta circles.}
\label{fig:Fig8}
\end{figure*}

\begin{figure*}[htb!]
\centering
\includegraphics[trim=1cm 1.5cm 2cm 2cm, width= 18cm, clip]{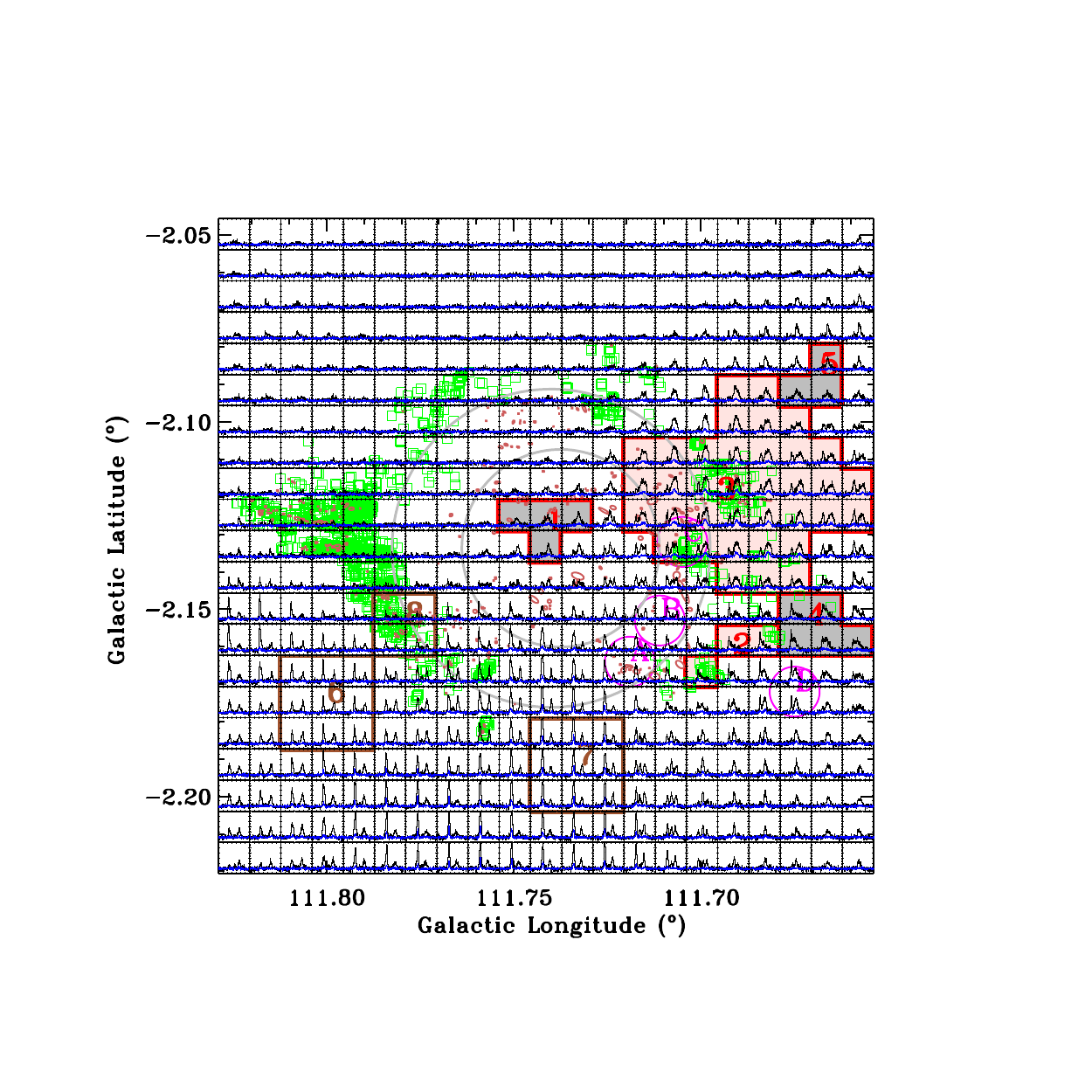}
\caption{Line grid map of the 10$\arcmin\times$10$\arcmin$ region that we examined. The black and blue lines in each grid are the $^{12}$CO $J=1-0$ and the $^{13}$CO $J=1-0$ spectra, respectively. The circles, squares, and polygons are the same as in Figure \ref{fig:Fig8}. The adjacent polygons are filled with different colors for easy in distinguishing. The dark red ellipses are the NIR knots identified by \cite{Koo+etal+2018} in their deep [\ion{Fe}{2}]+[\ion{Si}{1}] map toward the Cas A SNR.} 
\label{fig:Fig9}
\end{figure*}

\subsubsection{Spectral Features}\label{sec:Sec3.2.1}

\begin{figure*}[htb!]
	\centering
	\subfigure[]{\label{Fig:knot_1}
		\includegraphics[trim=1.5cm 1cm 3.4cm 3cm, width = 0.3\linewidth , clip]{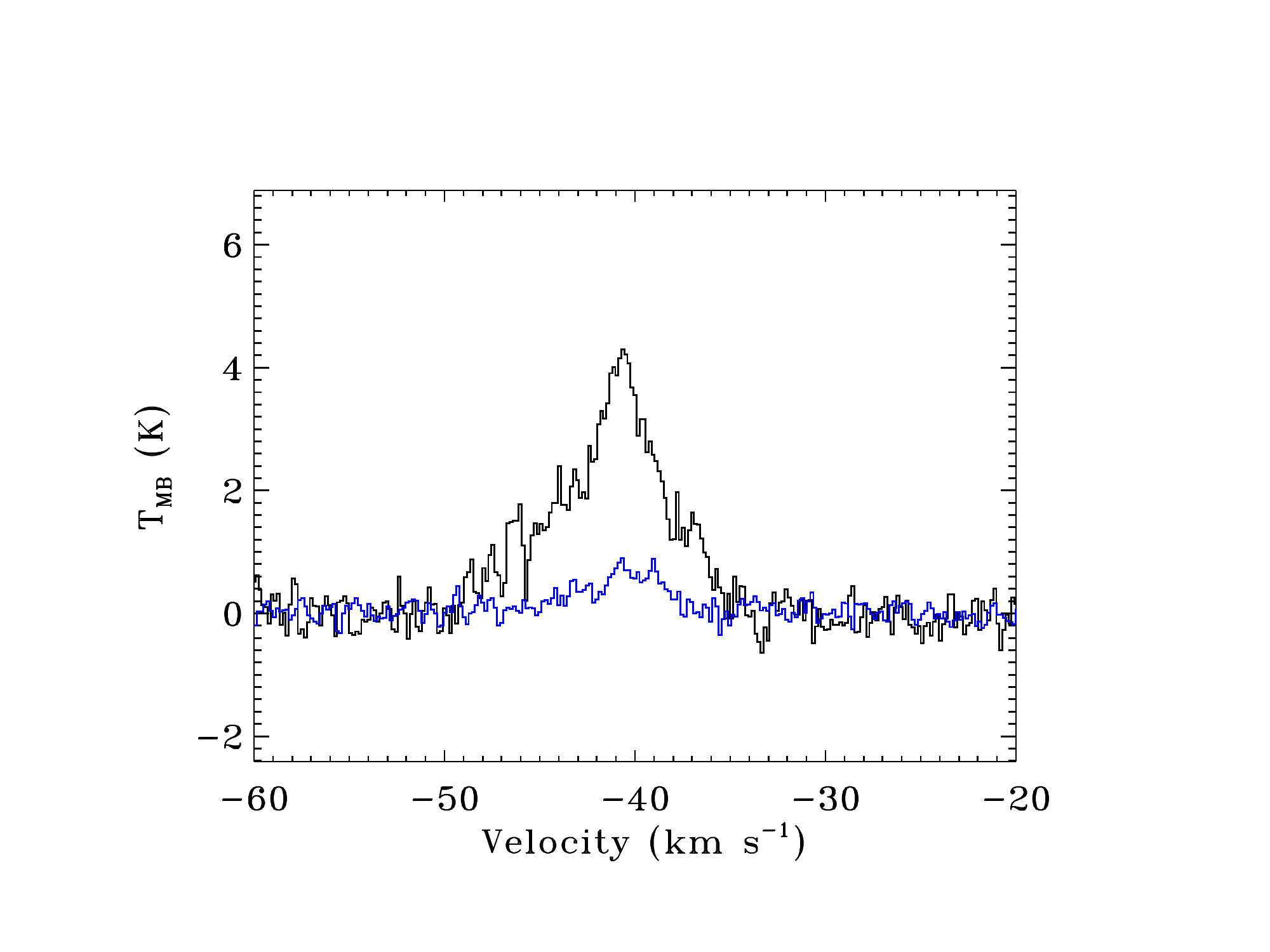}}
	\subfigure[]{\label{Fig:knot_2}
		\includegraphics[trim=1.5cm 1cm 3.4cm 3cm, width = 0.3\linewidth , clip]{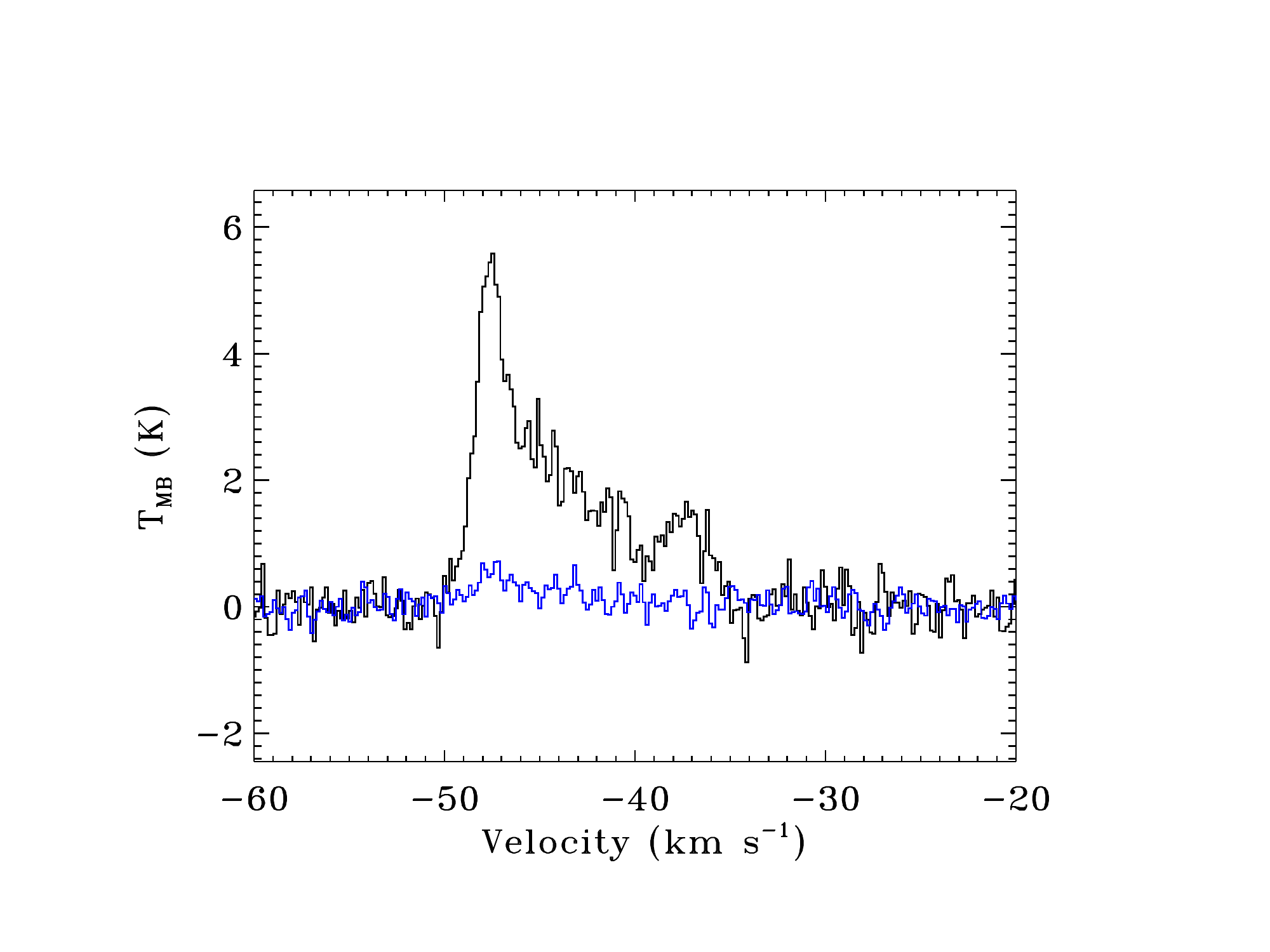}}
	\subfigure[]{\label{Fig:knot_3}
		\includegraphics[trim=1.5cm 1cm 3.4cm 3cm, width = 0.3\linewidth , clip]{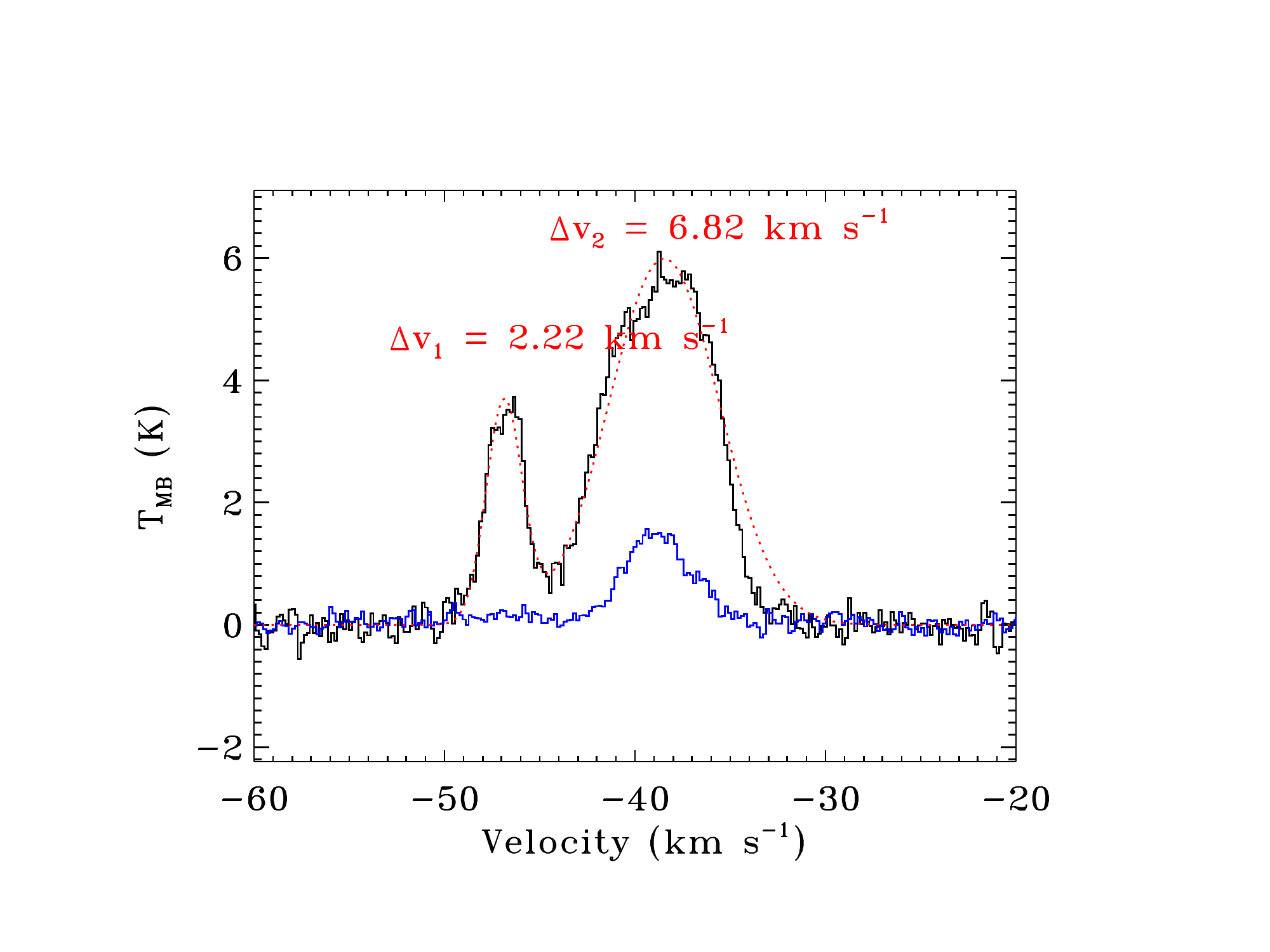}}
	\subfigure[]{\label{Fig:knot_4}
		\includegraphics[trim=1.5cm 1cm 3.4cm 3cm, width = 0.3\linewidth , clip]{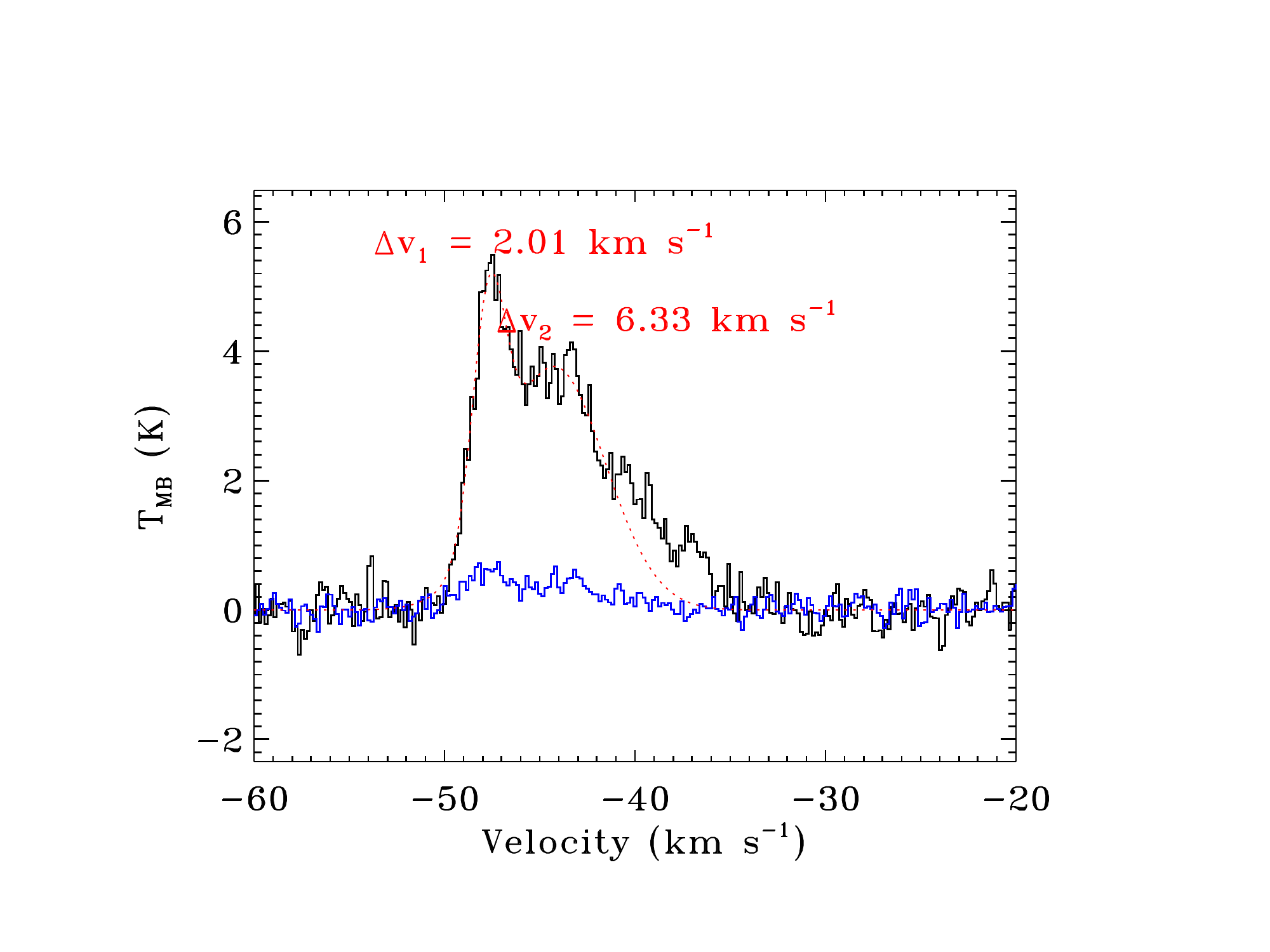}}
	\subfigure[]{\label{Fig:knot_5}
		\includegraphics[trim=1.5cm 1cm 3.4cm 3cm, width = 0.3\linewidth , clip]{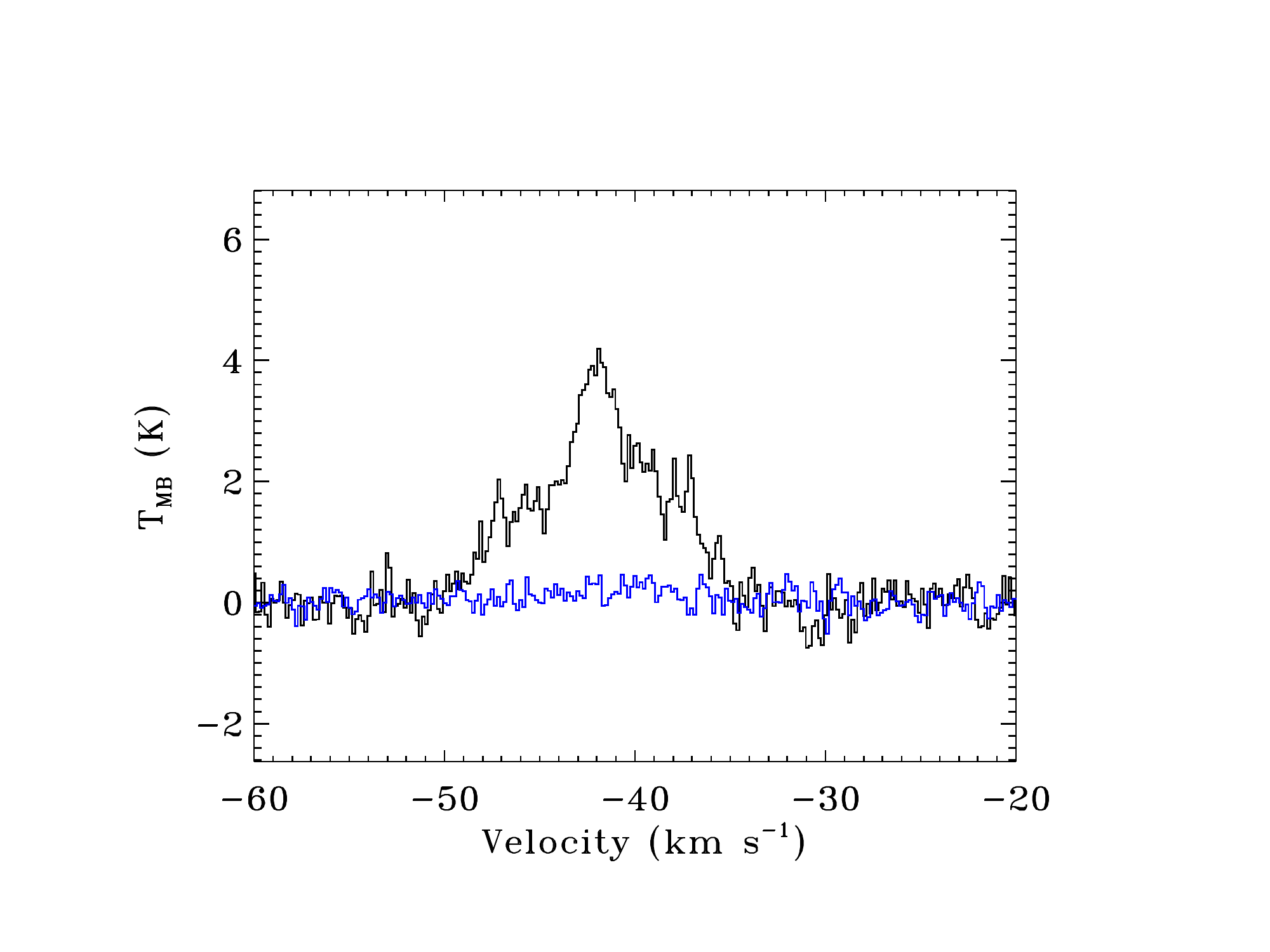}}
	\\
	\subfigure[]{\label{Fig:knot_6}
		\includegraphics[trim=1.5cm 1cm 3.4cm 3cm, width = 0.3\linewidth , clip]{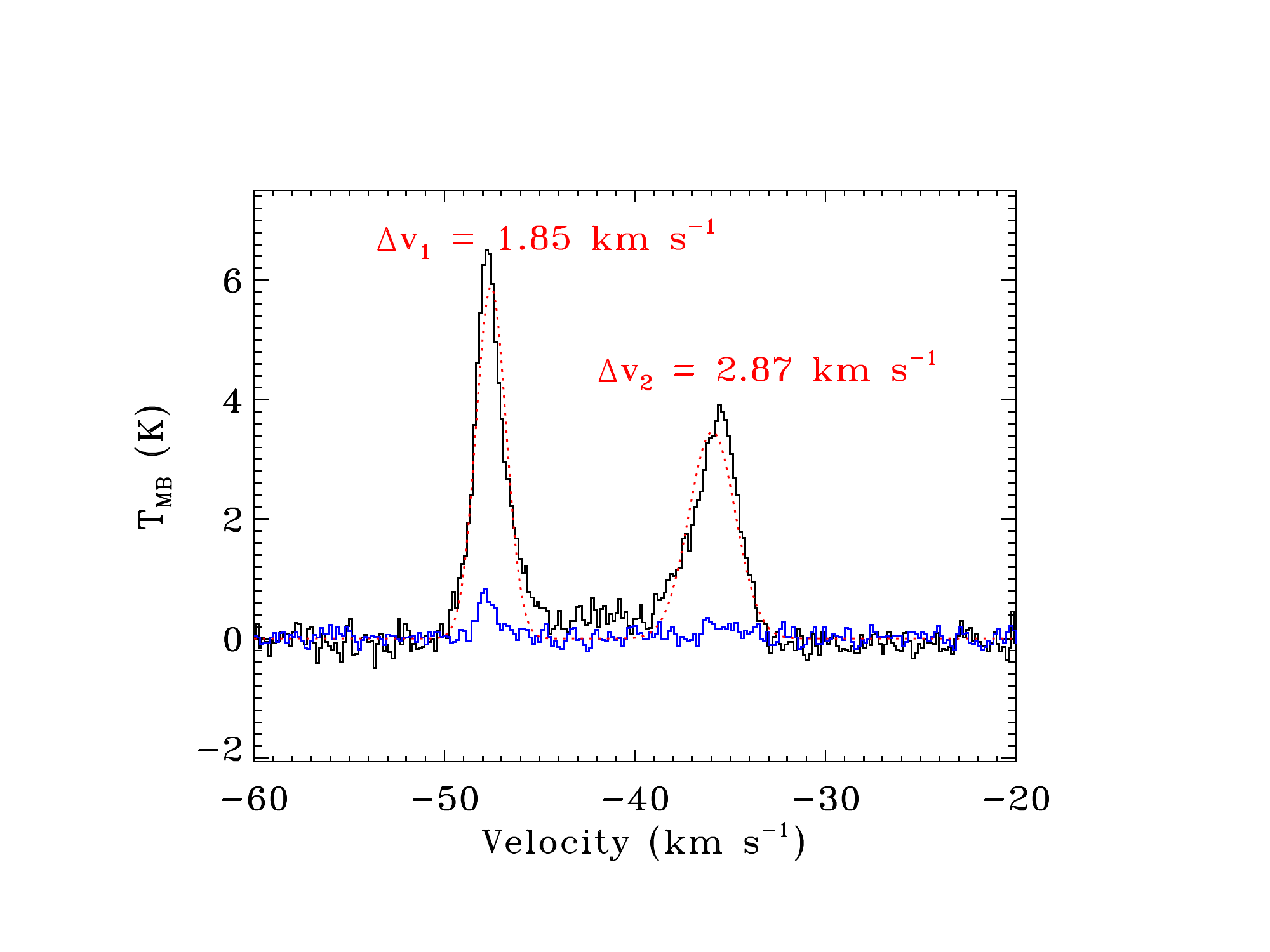}}
	\subfigure[]{\label{Fig:knot_7}
		\includegraphics[trim=1.5cm 1cm 3.4cm 3cm, width = 0.3\linewidth , clip]{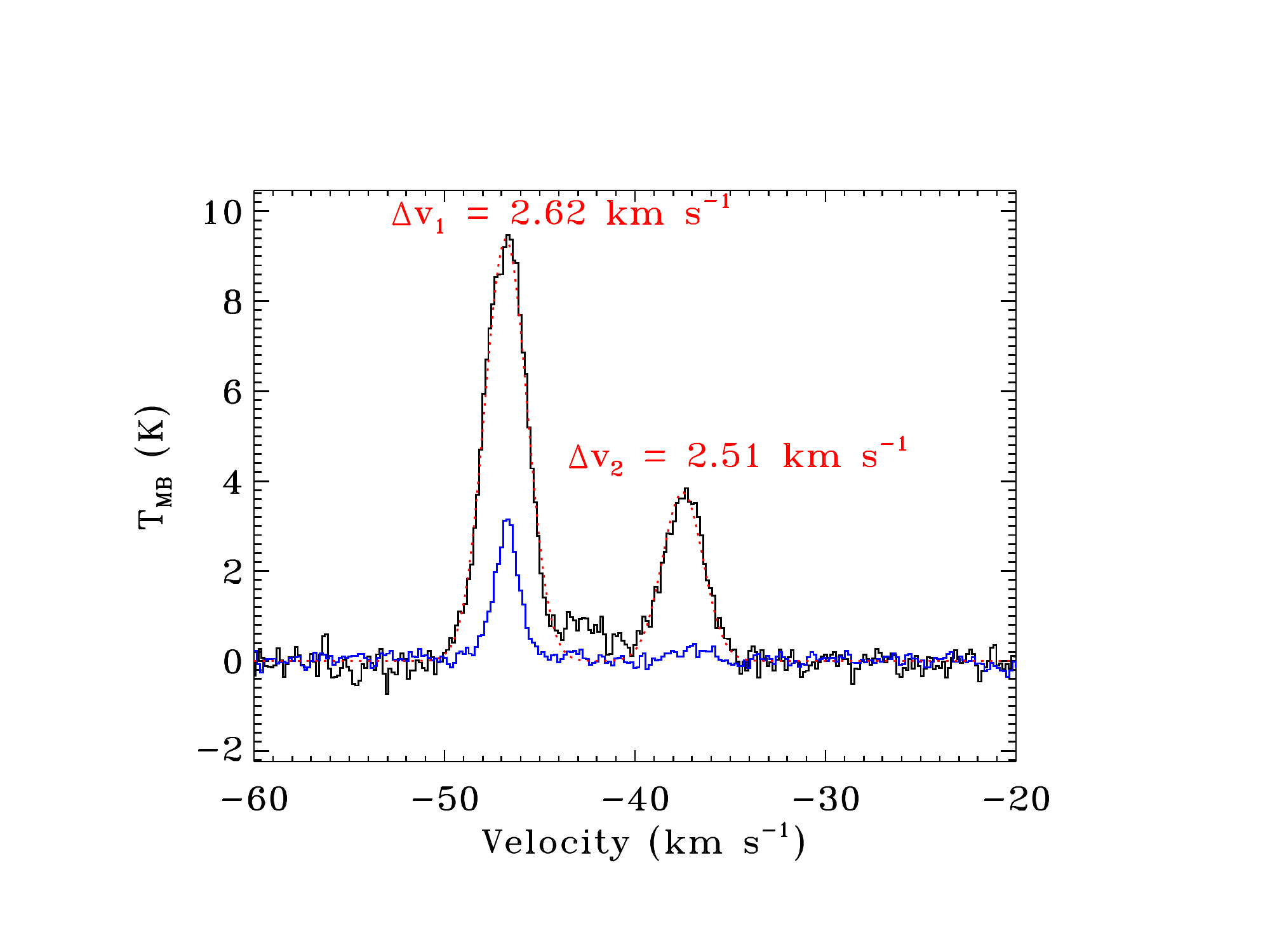}}
	\subfigure[]{\label{Fig:knot_8}
		\includegraphics[trim=1.5cm 1cm 3.4cm 3cm, width = 0.3\linewidth , clip]{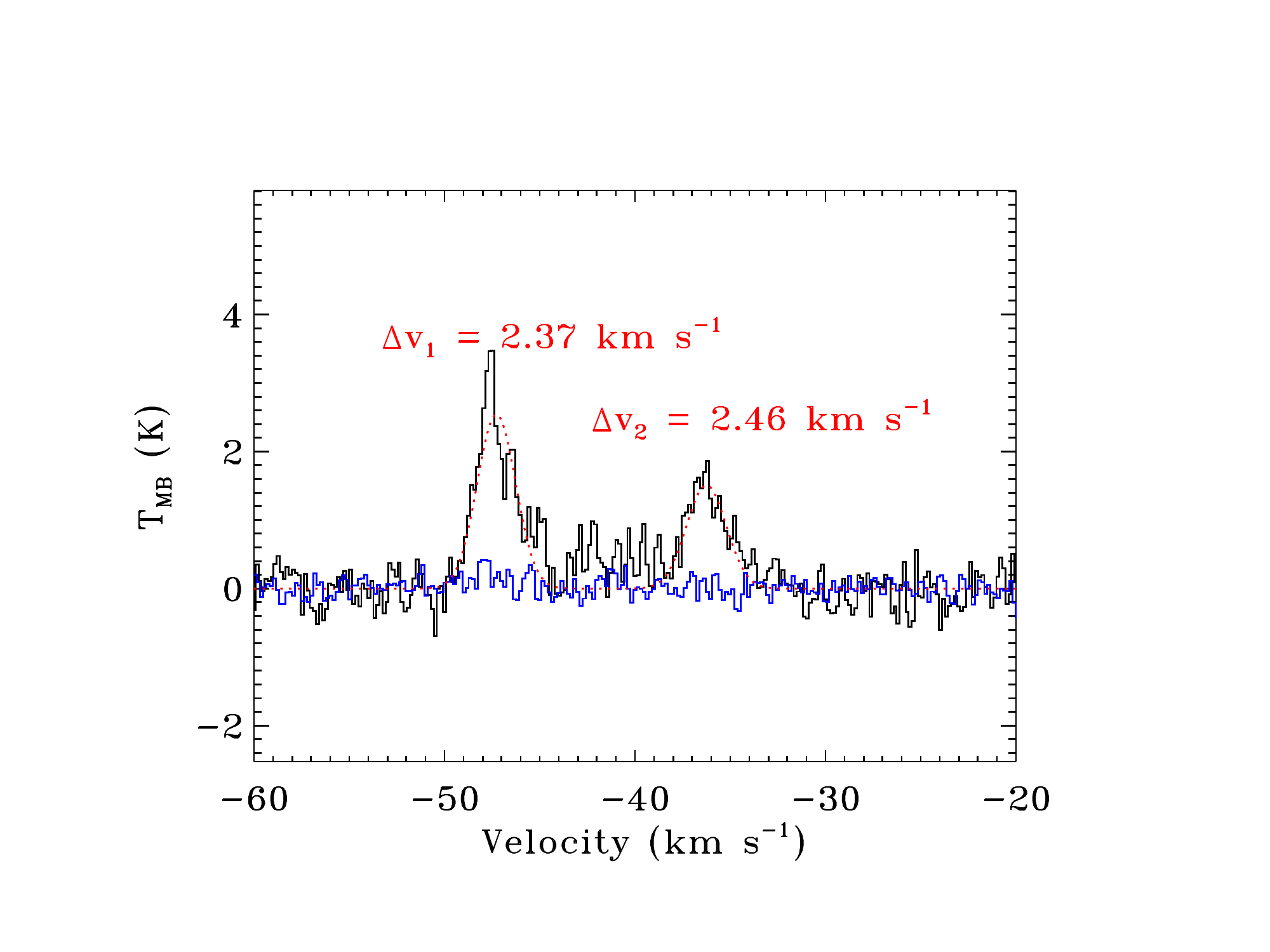}}
	\caption{Average spectra of the disturbed regions and the comparison regions that are indicated in Figure \ref{fig:Fig9}, with panels (a) $-$ (h) corresponding to regions 1$-$8. The black and blue spectra in each panel are the $^{12}$CO and $^{13}$CO $J=1-0$ lines, respectively. For those regions with Guassian$-$like line profiles, we fitted the $^{12}$CO $J=1-0$ line with gauss functions. The fitted profiles are plotted in red dotted lines, and the FWHM of each velocity component is given in the corresponding panel.}
	\label{fig:Fig10}
\end{figure*}

\begin{figure*}[htb!]
	\centering
	\includegraphics[trim=5cm 0.5cm 2cm 0cm, width= 21cm, clip]{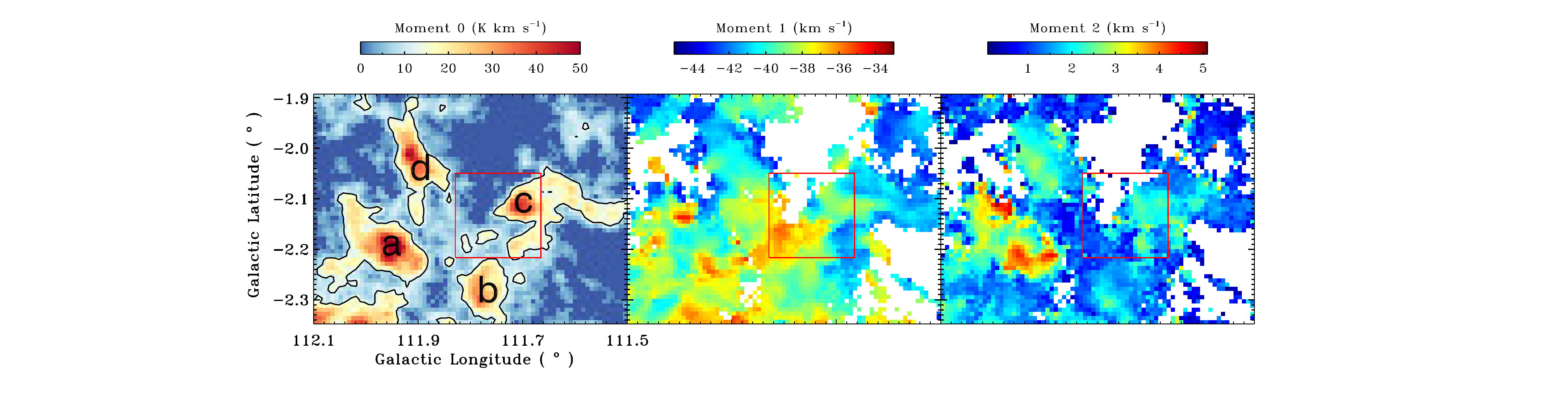}
	\caption{From left to right, the moment 0, 1, and 2 maps of a 36$\arcmin\times$30$\arcmin$ field centered on the SNR, showing the four clumps, a$-$d, at the vicinity of the Cas A SNR. In the first image, the black contour is 12 K km s$^{-1}$ (15 $\sigma$). The central 10$\arcmin\times$10$\arcmin$ region for which we have done the line fitting is indicated with the red rectangle.}
	\label{fig:Fig11}
\end{figure*}

Observationally, there are several multiwavelength indicators for the diagnosis of SNR-MC interactions, such as the H$_2$ rovibrational lines and the [\ion{Fe}{2}] emission line at the near infrared wavelength and the broadenings or the asymmetric profiles of the optically thin $^{12}$CO rotational lines at the radio wavelength \citep{Denoyer+etal+1979, Dubner+etal+2004, Reach+etal+2005, Su+etal+2009, Zhou+etal+2009, Jiang+etal+2010, Chen+etal+2014, Kilpatrick+etal+2014, Su+etal+2017, Koo+etal+2018}. In this work, we manually examined all the $^{12}$CO $J=1-0$ spectra for extra velocity broadenings or asymmetric wing emissions within a square region of $15\arcmin\times15\arcmin$ centered at $\alpha$(J2000) = 23$^h$23$^m$27.$^s$77, $\delta$(J2000) = 58$\arcdeg$48$\arcmin$49$\arcsec$.4, i.e., the center of the Cas A SNR \citep{Thorstensen+etal+2001}. For the search of line broadenings, we use Gaussian fitting to get the FWHM velocities when the spectra have Gaussian profiles. We use the criterion $\Delta v > 6$ km s$^{-1}$ \citep{Kilpatrick+etal+2014} to screen out the broadened spectra.

We found some moderately broadened or asymmetric $^{12}$CO lines in our examination. They are all limited to a 10$\arcmin\times$10$\arcmin$ region around the Cas A SNR, as indicated in Figure \ref{fig:Fig8}. Figure \ref{fig:Fig9} gives the detailed line grid map of this region. As shown in Figure \ref{fig:Fig9}, lines at the the northwestern part of the region have more complex profiles than the southeastern lines. Two narrow velocity components occur in the southeastern lines, one with a centroid velocity of -47 km s$^{-1}$ and the other with $-$37 km $^{-1}$. Some positions near the center and to the southwest of the SNR have three narrow velocity components at $-$47 km s$^{-1}$, $-$42 km s$^{-1}$, and $-$37 km s$^{-1}$. The counterparts of the $-$37 km s$^{-1}$ component in the northwestern lines are much broader than those in the southeastern and the southwestern lines. The lines that we found to be broadened or asymmetric can be further divided into five groups according to their locations in the region. For comparison, we also selected three control areas (6$-$8 in Figure \ref{fig:Fig9}). The spectra in these areas are characterized by narrow $^{12}$CO lines. For convenience, we name the five areas with distinctive spectra and the three control areas as CO patches 1$-$8 hereafter. The polygons and the squares in Figures \ref{fig:Fig8}-\ref{fig:Fig9} serve to mark the locations of the CO patches 1-8.

The average spectra of CO patches 1$-$8 are present in Figure \ref{fig:Fig10}. In Figure \ref{fig:Fig10}, the $^{12}$CO emission of the control areas, CO patches 6$-$8, consists of two velocity components, $-$47 km $^{-1}$ and $-$37 km $^{-1}$, with the $-$47 km $^{-1}$ component being much stronger than the $-$37 km $^{-1}$ component. As shown in Figure \ref{fig:Fig10}, the line profiles of CO patches 1$-$5 are different from those of the CO patches 6$-$8. The spectra of CO patches 1$-$5 can be divided into three categories. The first category is represented by CO patches 1 and 5, which have one velocity component centered at $-$41 km s$^{-1}$ and has wing emission extended toward both the red-shift and the blue-shift for about 7 km s$^{-1}$. However, the redshift wing of CO patch 1 is contaminated by another velocity component at $-$37 km s$^{-1}$. The spectra of CO patches 2 and 4 represent the second category, characterized by an asymmetric line profile with extended wing emission toward the red-shift. Nevertheless, the redshift wing of CO patch 2 is indistinguishable from the surrounding velocity components. The asymmetric profile of patch 4 is likely composed of two velocity components at $-$47 and $-$43 km s$^{-1}$. We have superimposed the Gaussian fittings of the two velocity components on the average spectrum of CO patch 4. The line width of the $-$43 km s$^{-1}$ component is clearly larger than the counterpart of its spectra to the east and south of CO patch 4. Further, the $-$43 km s$^{-1}$ velocity component also shows a Gaussian excess at the red-shift wing. Patch 3 represents the third category. We can see that the line widths of the two velocity components in CO patches 6$-$8 are all less than 3 km s$^{-1}$, but the $-$37 km s$^{-1}$ component in CO patch 3 is broadened to 6.82 km s$^{-1}$, a factor of 2 larger than the the $-$37 km s$^{-1}$ component in CO patches 6$-$8. The velocity of the $^{12}$CO peak is red shifted relative to the $^{13}$CO peak by about 2 km s$^{-1}$. 

\cite{Kilpatrick+etal+2014} also found four regions with $^{12}$CO $J=2-1$ line broadenings around the Cas A SNR, which are marked in Figure \ref{fig:Fig8} and \ref{fig:Fig9} with magenta circles ``A" to ``D." Except for region C, the positions of the broadened $^{12}$CO $J=2-1$ spectra do not coincide with those of the broadened $^{12}$CO $J=1-0$ spectra found in this work. For regions A and B, the -47 km s$^{-1}$ velocity component of the $^{12}$CO $J=2-1$ emission is mixed with the $-$37 km s$^{-1}$ component. But in Figure \ref{fig:Fig9} we can see that the two velocity components in the $^{12}$CO $J=1-0$ emission at the positions of A-D are separated, and no significant velocity broadenings are present. \cite{Zhou+etal+2018} also mapped the Cas A SNR with $^{12}$CO/$^{13}$CO $J=1-0$ and $J=2-1$ emission lines and searched for line broadening. They checked the spectra near the shock front, in the regions used in \cite{Kilpatrick+etal+2014}, and at the positions of the CO brightness peaks. They did not find any prominent line broadening at these positions and proposed that the line widths of the spectra found by \cite{Kilpatrick+etal+2014} are not broader than those of the environment. They took $\Delta v > 10$ km s$^{-1}$ as the criterion for significant line broadening, and they did not show the spectra at the positions of CO patches 1, 4, and 5. Moreover, they searched for line broadenings at CO patch 3 by comparing the spectra in the pre- and postshocked regions. We note that in the CO patch 3 region there are many optical knots. These optical knots have very high velocities ($>$10000 km s$^{-1}$) and carry a considerable amount of kinetic energy \citep{Hammell+etal+2008}. As was summarized by \cite{Chevalier+etal+1977}, the actual interaction between an SNR and its surrounding medium may show a combination of a smooth shell and some clumpy ejecta (Figs 1 and 2 in their paper). When the clumpy ejecta interact with the surrounding medium, the bow shocks produced in front of them will change the dynamics of the surrounding gas. Therefore, it is possible that CO patches 3 and 4 have been disturbed by the Cas A SNR. CO patch 5 and the northwestern part of CO patch 3 are relatively far from the Cas A SNR, and they might be unreachable by the FMKs or the forward shock. The possible disturbing mechanisms for these regions are discussed in Section \ref{sec:Sec3.2.2}.

The SNR-MC interactions can also be observed in the near-infrared band. The shock-excited [\ion{Fe}{2}] 1.64 $\mu m$ emission is usually observed together with the H$_2$ 2.12 $\mu m$ rovibrational lines in SNRs. It is found that the [\ion{Fe}{2}] emission is often distributed close to but inside the H$_2$ NIR emission in SNRs \citep{Reach+etal+2002, Lee+etal+2009}. We have superimposed the [\ion{Fe}{2}]+[\ion{Si}{1}] knots identified by \cite{Koo+etal+2018} in Figure \ref{fig:Fig9} for comparison. Some [\ion{Fe}{2}]+[\ion{Si}{1}] knots spatially coincide with CO patches 1 and 3, but these knots cannot be taken as evidence for the SNR-MC interactions in this region, as \cite{Koo+etal+2018} found that the [\ion{Fe}{2}]+[\ion{Si}{1}] emission in this region is from the shocked dense atomic gas, rather than molecular gas. Due to the lack of other observational evidence such as increased temperature and the raised intensity ratio between CO J = 2-1 and J=1-0 emission\citep{Zhou+etal+2018}, further proof is still needed for interactions in the region.

\subsubsection{Large-Scale Kinematics of Molecular Gas around Cas A SNR}\label{sec:Sec3.2.2}
One advantage of our observation toward Cas A SNR compared with previous CO studies, such as those by \cite{Kilpatrick+etal+2014} and by \cite{Zhou+etal+2018}, is the much larger sky coverage. The kinematics of the surrounding molecular gas can be investigated on a large scale. Outside the 10$\arcmin\times$10$\arcmin$ region (hereafter the central zone), there are four main clumpy structures in the velocity range of v = [$-$44, $-$30] km s$^{-1}$. We have intercepted the 36$\arcmin\times$30$\arcmin$ area that covers the four main clumps and have created three moment maps for this subregion in the velocity range of v= [$-$44, $-$30] km s$^{-1}$. The results are displayed in Figure \ref{fig:Fig11}, where the four main clumps are marked clumps ``a" to ``d". In the moment 0 map, it can be seen that the four clumps are located 2.5$-$13 pc away from the SNR center and that they encircle a less CO luminous region, which includes the southeast corner of the central zone. The moment 1 map reveals that all these clumps are blue-shifted by $\sim$5 km s$^{-1}$ relative to the center of the less luminous region, which forms a void-like structure in the position-velocity space. At the distance of the SNR, this velocity difference corresponds to a velocity gradient of about 0.6 km s$^{-1}$ pc$^{-1}$. The velocity dispersions (moment 2) of these clumps are larger than the average velocity of the nearby clouds (see Figure \ref{fig:Fig6b}), and the values are as large as 5 km s$^{-1}$ at the northeast and the southwest ends of clump a.

Although the clumps a, b, and d are far outside the forward shock front of the Cas A SNR, and thus they cannot be impacted by the ejecta of the Cas A SNR, it is interesting to explore the possible interaction between the wind from the progenitor of the Cas A SNR and these clumps. Although binary scenarios have been discussed for the progenitor of the Cas A SNR \citep{Young+etal+2006}, most authors argued that the progenitor is a single massive star with a ZAMS mass between 20 and 35 $M_{\sun}$ \citep{Garcia+etal+1996, Perez-Rendon+etal+2009, Veelen+etal+2009}. The progenitor may have experienced the main-sequence (MS), red supergiant (RSG), and Wolf-Rayet (W-R) evolution stages prior to the final explosion \citep{Garcia+etal+1996, Perez-Rendon+etal+2009, Veelen+etal+2009}. The winds from the progenitor at the MS, RSG, and W$-$R stages can produce bubbles and inject energy to the surrounding medium. Following \cite{Whitworth+etal+1994}, we use the analytic solution to calculate the radii of the bubbles produced by the winds (see equation 7 in their work). For the wind parameters, we use values given by stellar evolution models \citep{Garcia+etal+1996}. We adopt a lifetime of $5\times10^6\ yr$, a mass-loss rate of $5\times10^{-7}\ M_{\sun}$ $ yr^{-1}$, and a wind terminal velocity of 2000 $\rm{km\ s^{-1}}$ for the MS stage. When a number density of 10 cm$^{-3}$ is used for the circumstellar material, the radius of the MS bubble is derived to be 38 pc. Even if a much higher density of the circumstellar matter, n = 100 cm$^{-3}$, is used, the radius of the MS bubble is still up to 24 pc, which means that all the clumps a$-$d, including CO patch 5 and the northwestern part of CO patch 3, are in the reach of the MS wind of the progenitor, if the clumps and the SNR are at the same distance. When we use the same number density n = 10 cm$^{-3}$ for the circumstellar matter and the typical values of $\tau\sim10^{5}$ yr, $dM/dt\sim$ 10$^{-4}\ M_{\sun}\ yr^{-1}$, and $v$ = $\rm{50}$ km s$^{-1}$ for the RSG stage and $\tau\sim10^{4}\ yr$, $dM/dt\sim$ 10$^{-5}\ M_{\sun}\ yr^{-1}$, $v=\rm{1000}$ km s$^{-1}$ for the W-R stage \citep{Garcia+etal+1996}, the radii of the RSG and W-R bubbles are 2.3 pc and 1.25 pc, respectively, which means that the clumps a, b, and d are well beyond the influence range of the RSG and W-R winds. Based on the aforementioned values for the wind parameters, the total kinetic energy that the MS wind has injected into the interstellar medium is $5\times10^{6}\ M_{\sun}\rm{\ km^2\ s^{-2}}$ (corresponding to $1\times10^{50}\ \rm erg$). The masses of the CO clumps a$-$d are in the range of 2000$-$3200 M$_{\sun}$; thus only 2$-$3 \% of the injected energy from the MS wind is needed to cause a line broadening of 10 km s$^{-1}$ for these clumps. For the eastern part of the CO patch 3 in clump c, which is close to the forward shock, its line broadening is likely caused by the influence of the SNR ejecta, because only 1\% of the SN explosion energy ($\sim$10$^{51}$ erg) or 10\% of the FMKs' kinetic energy ($\sim$10$^{50}$ erg, \citealp{Laming+etal+2006}) is capable of a 10 km s$^{-1}$ broadening for a cloud with mass as large as 10$^4\ M_{\sun}$.

\begin{figure*}[htb!]
\centering
\includegraphics[trim=0cm 0.5cm 0cm 1cm, width= 11cm, clip]{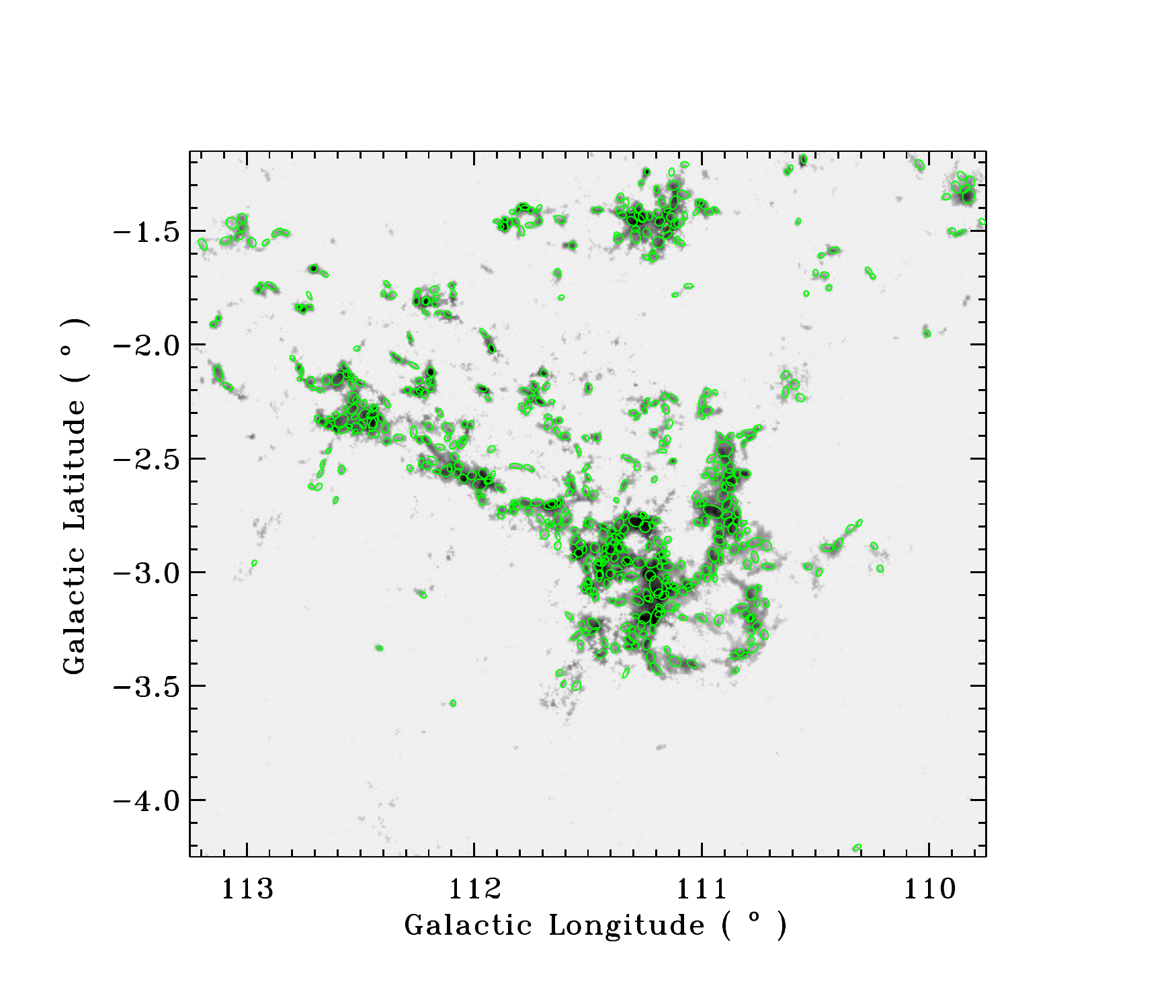}
\caption{Spatial distribution of the $^{13}$CO clumps identified in the Cas GMC using the GAUSSCLUMPS algorithm. The background is the column density distribution of the Cas GMC, and the green ellipses represent the fitted clumps.}
\label{fig:Fig12}
\end{figure*}

\subsection{Properties of $^{13}$CO Clumps} \label{sec:Sec3.3}
In this section, we decompose the Cas GMC into a set of clumps and discuss their statistical and dynamical properties.

\subsubsection{Method and Calculation}\label{sec:Sec3.3.1}
\cite{Kramer+etal+1998} present a brief review of the different methods for decomposing molecular clouds into discrete clumps, from by-eye inspection to automated clump-finding algorithms, such as the widely used GAUSSCLUMPS \citep{Stutzki+etal+1990} and CLUMPFIND \citep{Williams+etal+1994}. \cite{Williams+etal+1994} suggested that both GAUSSCLUMPS and CLUMPFIND perform well in identifying intermediate- to high-mass clumps, whereas GAUSSCLUMPS can avoid velocity blending. To avoid velocity blending, we therefore chose to use the GAUSSCLUMPS task in GILDAS software to identify $^{13}$CO clumps in this work. We used the following settings in the calculation, T = 1.61 K (7$\sigma$) for the termination temperature, ($s_0, s_a, s_c$) = (1, 1, 1) for the stiffness parameters, and 3 (in units of velocity or spatial resolution) for the FWHMs of the weighting function. The stiffness parameters confine the peak temperature and the peak position of a fitted clump. The FWHMs are used to make sure that the Gaussian fitting in each iteration actually uses the data points that are physically related to the observed peak position. The method for the column density calculation has been discussed in Section \ref{sec:Sec3.1.2}. Here we only give a brief description of the calculation of the other parameters of the detected clumps. The mass of a clump is obtained through its total $\rm{^{13}CO}$ flux, $T_{peak}\Delta v\theta_a \theta_b [\pi/(4ln2)]^{3/2}$, where $\Delta v,\theta_a$, and $\theta_b$ are the FWHM along each axis of a triaxial gaussian clump. The effective radius of a clump is obtained through the deconvolution of the fitted clump size, $R_{eff}= 1/2d(\theta_a\theta_b - \theta_{beam}^2)^{1/2}$ \citep{Stutzki+etal+1990}, where $\theta_{beam}$ is the HPBW of the PMO-13.7 m telescope and $d=3.4$ kpc is the distance of the Cas GMC. The surface density and volume number density of the clump are then derived as $\Sigma = M/(\pi R_{eff}^2)$ and $n(H_2) = 3M/(4\pi R_{eff}^3\mu m_H)$, respectively.

\begin{figure*}[htb!]
\centering
\subfigure[]{\label{fig:Fig13a}
\includegraphics[trim=1.3cm 0.7cm 3cm 2.5cm, width = 0.3\linewidth , clip]{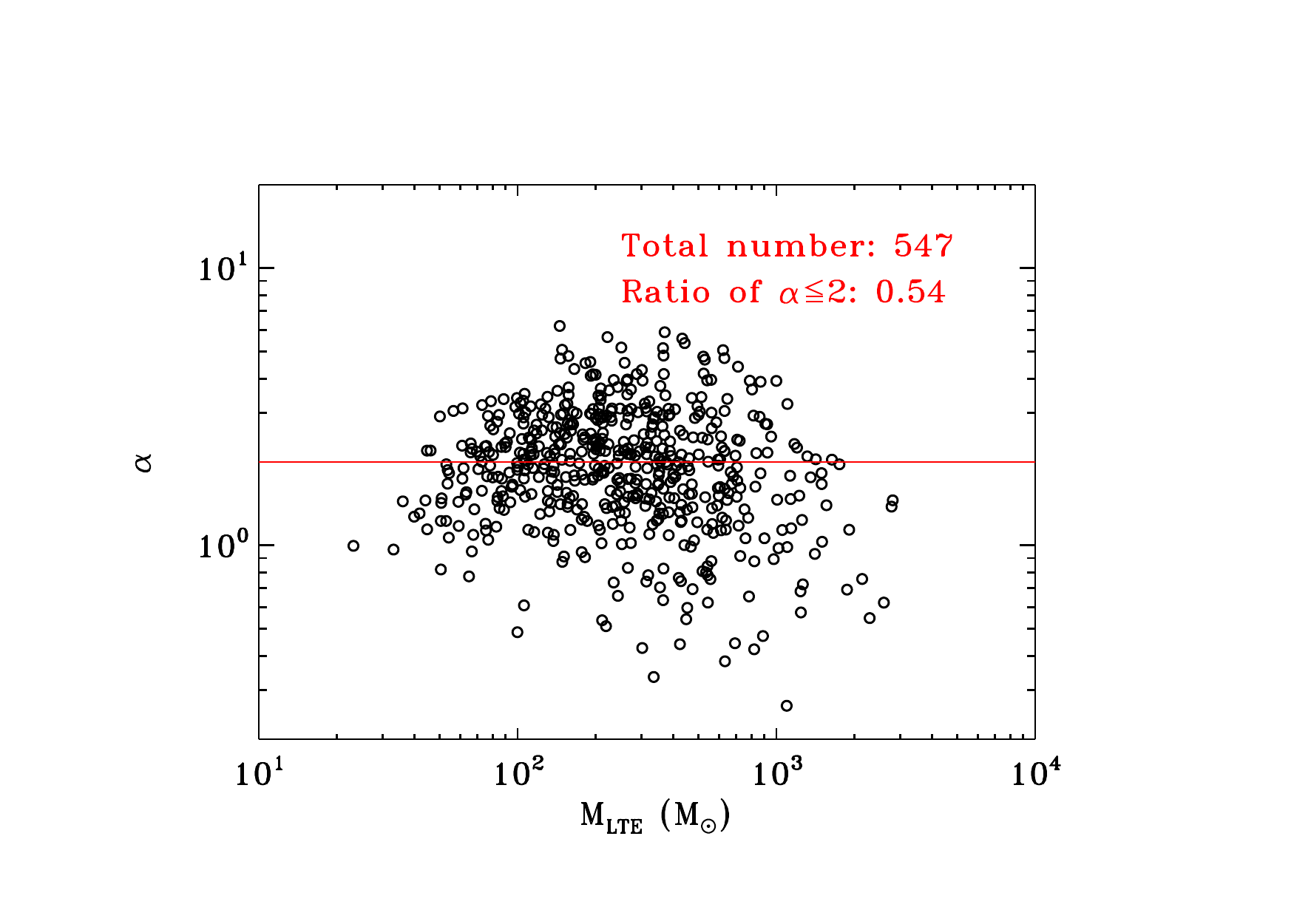}}
\subfigure[]{\label{fig:Fig13b}
\includegraphics[trim=1.3cm 0.7cm 3cm 2.5cm, width = 0.3\linewidth , clip]{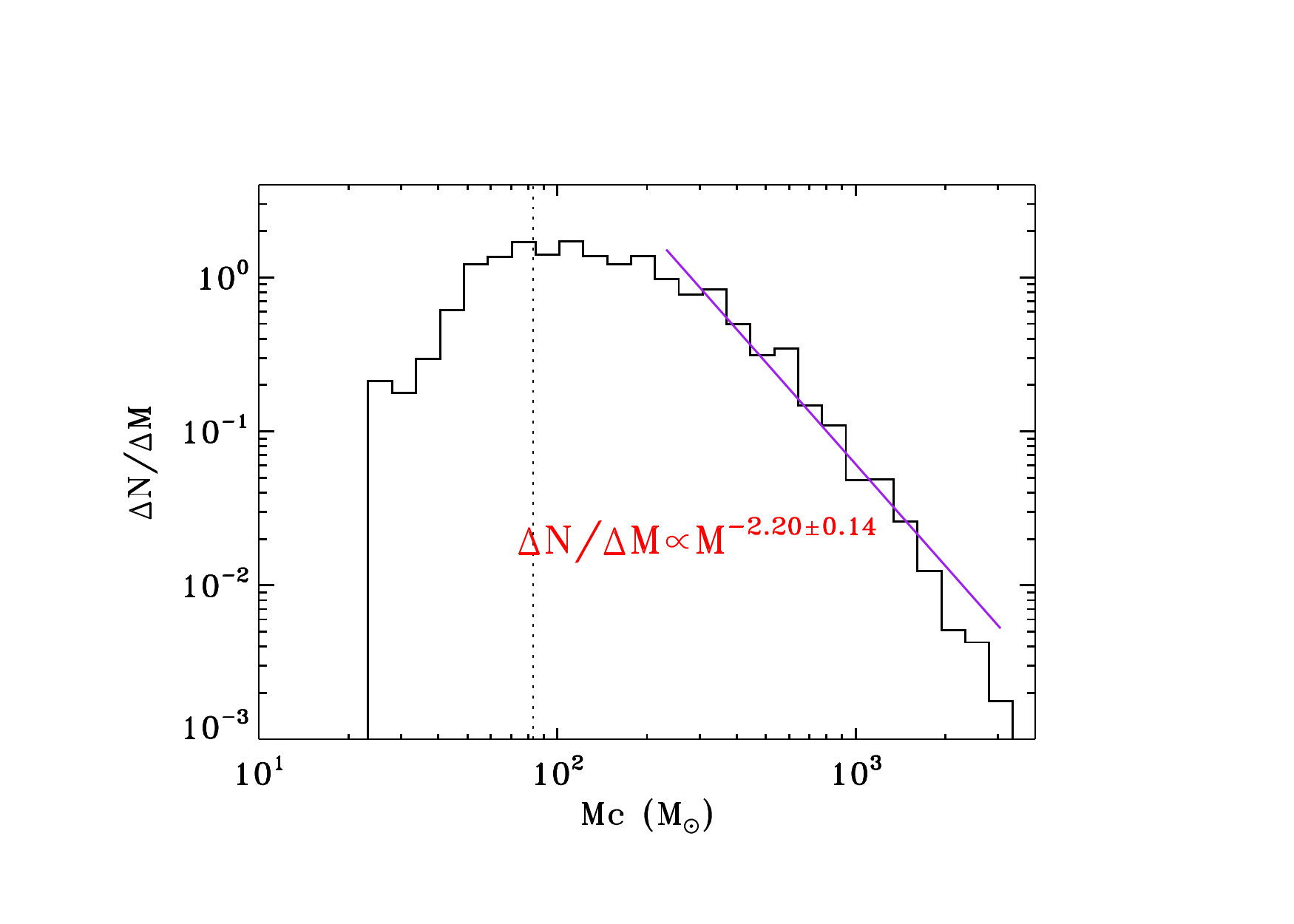}}
\subfigure[]{\label{fig:Fig13c}
\includegraphics[trim=1.3cm 0.7cm 3cm 2.5cm, width = 0.3\linewidth , clip]{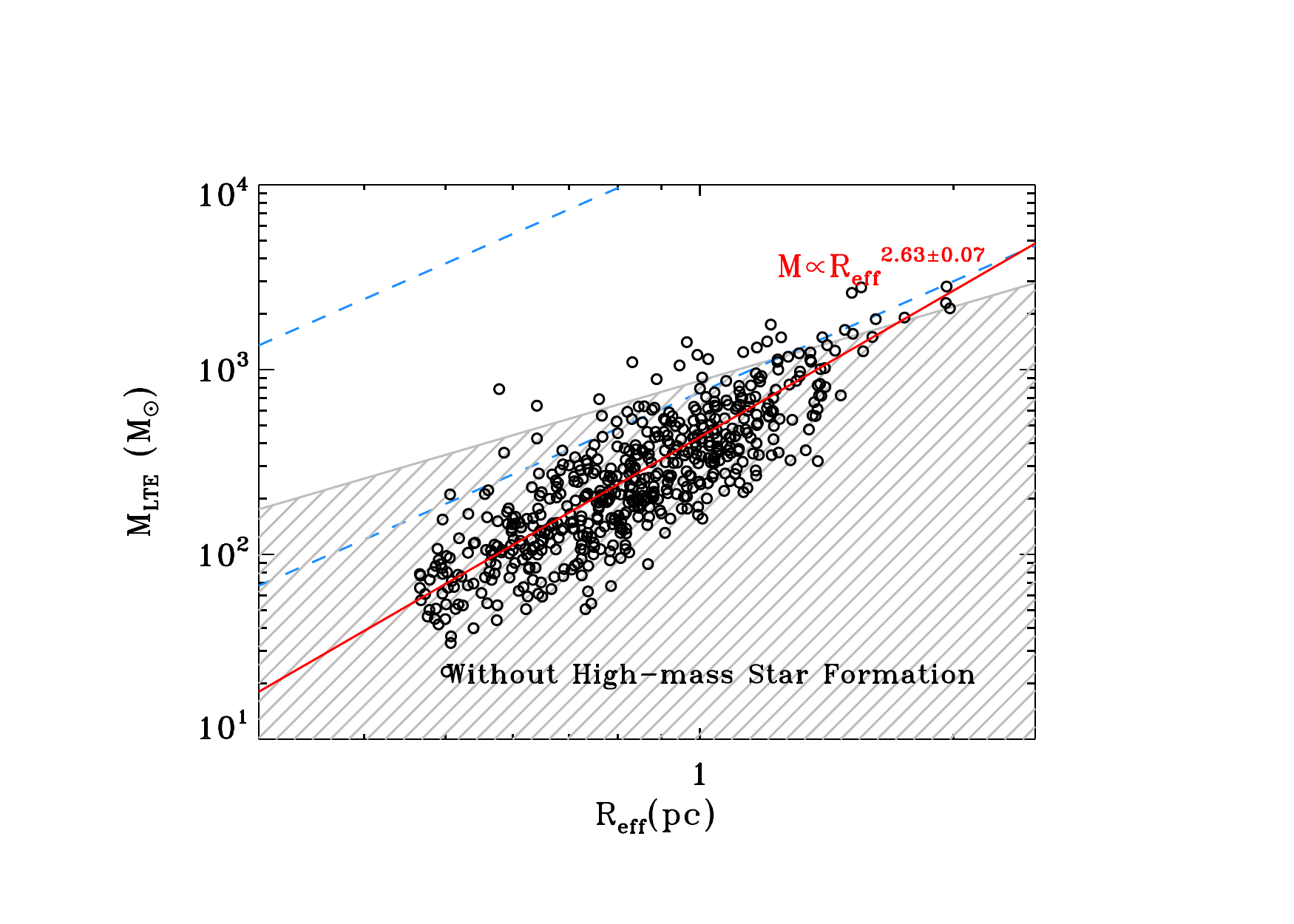}}
\caption{(a) Relationship between the virial parameters and the masses of the clumps. The red horizontal line indicates $\alpha_{vir} = 2$. The total number of the clumps and the fraction of the clumps with $\alpha_{vir}$ no larger than 2 are shown in the top$-$right corner in this panel. (b) Clump Mass Function (CMF) of the Cas GMC. The purple line indicates the power$-$law fitting of the CMF with an index of $-$2.20. The vertical dashed line marks the detection limit derived from the settings used in the GAUSSCLUMPS algorithm. (c) M$-$R relation of the clumps. The shaded area is the region where the clump masses follow the empirical relationship $M(R) = 870M_\sun(r/pc)^{1.33}$ found by \cite{Kauffmann+etal+2010a}. The blue dotted lines indicate the empirical upper and lower bounds of the clump surface density of 1 g cm$^{-2}$ and 0.05 g cm$^{-2}$ for massive star formation \citep{Urquhart+etal+2013}}\label{fig:Fig13}
\end{figure*}

\subsubsection{Statistical Properties of the Clumps}\label{sec:Sec3.3.2}

In this work, we ignore the clumps that have fitted angular sizes smaller than 1.5 $\theta_{beam}$. Ultimately, we identified 547 $^{13}$CO clumps in total. The spatial distribution of these clumps is displayed in Figure \ref{fig:Fig12}. Table \ref{tab:Tab2}, as an example, lists the physical parameters from the GAUSSCLUMPS algorithm and also the derived parameters for some clumps. The whole table for all identified clumps is available in machine-readable format.

The most widely used indicator of the equilibrium state of a molecular clump is the virial parameter \citep{Bertoldi+etal+1992},
\begin{equation}
\alpha = \frac{5\sigma_v^2 R}{GM} \approx 209(\frac{R}{pc}\frac{\Delta v^{2}}{km\ s^{-1}}\frac{M_{\sun}}{M_{LTE}})
\end{equation}
where $\sigma_v$ is the velocity dispersion of the fitted clump, which is related to $\Delta v$ via $\Delta v = \sqrt{8ln2} \sigma_v$. The virial parameter measures the ratio between the internal kinetic energy and the gravitational energy. A clump with virial parameter $\alpha > \alpha_{cr}$ is subcritical, which will expand or need external pressure to be confined, while a clump with virial parameter $\alpha < \alpha_{cr}$ is supercritical, which will collapse under its self-gravity. In the simplest equilibrium analysis of molecular clumps that only considers internal pressure and gravity, the result of $\alpha_{cr} = 1$ is obtained. When detailed models of stability and the response of models to perturbations are considered, the critical value for a nonmagnetized isothermal hydrostatic equilibrium sphere is $\alpha_{cr}=2$ \citep{Kauffmann+etal+2013}. Figure \ref{fig:Fig13a} shows that 54$\%$ of the clumps are supercritical and are unstable to collapse unless they are supported by other forces, such as magnetic fields. No correlation between the virial parameters and masses of the clumps is found in Figure \ref{fig:Fig13a}. By comparison, \cite{Bertoldi+etal+1992} and \cite{Kauffmann+etal+2013} found an anticorrelation, $\alpha\propto M^{h_{\alpha}}$, for the pressure-confined clumps (with an index $h_{\alpha}=-2/3$) and the supercritical clumps in high-mass star forming regions.

\begin{longrotatetable}
\begin{deluxetable*}{lllrrrrrrrrrrrrrr}
\tablecaption{Properties of $^{13}$CO clumps\label{tab:Tab2}}
\tablewidth{\linewidth}
\tablenum{2}
\tabletypesize{\scriptsize}
\tablehead{
\colhead{Name} & \colhead{l} &
\colhead{b} & \colhead{$\theta_a$} &
\colhead{$\theta_b$} & \colhead{PA} &
\colhead{$v_{lsr}$} & \colhead{$T_{peak}$} &
\colhead{$\Delta v$} & \colhead{$R_{eff}$} & \colhead{$T_{ex}$} &\colhead{$\tau$} &\colhead{$N_{H_2}$} &\colhead{$M_{LTE}$} &\colhead{$\Sigma$} &\colhead{n} &\colhead{$\alpha_{vir}$}  \\
\colhead{} & \colhead{$\arcdeg$} & \colhead{$\arcdeg$} & \colhead{$\arcsec$} &
\colhead{$\arcsec$} & \colhead{$\arcdeg$} & \colhead{(km s$^{-1}$)} &
\colhead{(K)} & \colhead{(km s$^{-1}$)} & \colhead{(pc)} & \colhead{(K)} & \colhead{} & \colhead{cm$^{-2}$} & \colhead{(M$_{\sun}$)} & \colhead{(M$_{\sun}$ pc$^{-2}$)} &\colhead{cm$^{-3}$}&\colhead{}
}
\startdata
MWISP G109.758$-$01.475$-$048.48 & 109.758 &  $-$1.475 &  82 &  82 &  $-$74 &  $-$48.48 &  1.8 &  0.67 &  0.5 & 10.1 & 0.3 & 8.6$\rm{\times 10^{20}}$ & 4.0$\rm{\times 10^{1}}$ & 4.4$\rm{\times 10^{1}}$ & 8.8$\rm{\times 10^{2}}$ &  1.27\\
MWISP G109.767$-$01.459$-$051.14 & 109.767 &  $-$1.459 &  76 & 109 &  139 &  $-$51.14 &  1.9 &  0.93 &  0.6 & 10.0 & 0.3 & 1.3$\rm{\times 10^{21}}$ & 7.3$\rm{\times 10^{1}}$ & 5.8$\rm{\times 10^{1}}$ & 9.9$\rm{\times 10^{2}}$ &  1.57\\
MWISP G109.818$-$01.283$-$034.86 & 109.818 &  $-$1.283 & 122 & 178 &  107 &  $-$34.86 &  3.0 &  1.15 &  1.1 & 10.1 & 0.6 & 2.8$\rm{\times 10^{21}}$ & 4.3$\rm{\times 10^{2}}$ & 1.0$\rm{\times 10^{2}}$ & 9.8$\rm{\times 10^{2}}$ &  0.74\\
MWISP G109.827$-$01.349$-$033.55 & 109.827 &  $-$1.349 & 188 & 217 &   80 &  $-$33.55 &  3.8 &  1.96 &  1.6 & 10.2 & 0.8 & 6.7$\rm{\times 10^{21}}$ & 1.9$\rm{\times 10^{3}}$ & 2.3$\rm{\times 10^{2}}$ & 1.5$\rm{\times 10^{3}}$ &  0.69\\
MWISP G109.843$-$01.316$-$033.38 & 109.843 &  $-$1.316 & 106 & 159 &   41 &  $-$33.38 &  2.0 &  1.05 &  1.0 &  9.9 & 0.4 & 1.5$\rm{\times 10^{21}}$ & 1.8$\rm{\times 10^{2}}$ & 5.8$\rm{\times 10^{1}}$ & 6.4$\rm{\times 10^{2}}$ &  1.27\\
MWISP G109.858$-$01.258$-$034.70 & 109.858 &  $-$1.258 & 109 & 158 &  138 &  $-$34.70 &  1.8 &  1.01 &  1.0 & 10.1 & 0.3 & 1.3$\rm{\times 10^{21}}$ & 1.6$\rm{\times 10^{2}}$ & 4.9$\rm{\times 10^{1}}$ & 5.3$\rm{\times 10^{2}}$ &  1.38\\
MWISP G109.867$-$01.509$-$046.99 & 109.867 &  $-$1.509 &  62 & 187 &   26 &  $-$46.99 &  1.9 &  1.93 &  0.8 &  9.7 & 0.4 & 2.6$\rm{\times 10^{21}}$ & 2.1$\rm{\times 10^{2}}$ & 1.1$\rm{\times 10^{2}}$ & 1.5$\rm{\times 10^{3}}$ &  2.90\\
MWISP G109.883$-$01.308$-$033.37 & 109.883 &  $-$1.308 & 130 & 192 &  111 &  $-$33.37 &  1.8 &  1.66 &  1.2 &  9.2 & 0.4 & 2.1$\rm{\times 10^{21}}$ & 3.6$\rm{\times 10^{2}}$ & 7.4$\rm{\times 10^{1}}$ & 6.5$\rm{\times 10^{2}}$ &  2.01\\
MWISP G109.900$-$01.500$-$048.98 & 109.900 &  $-$1.500 & 116 &  88 &   $-$5 &  $-$48.98 &  2.0 &  1.28 &  0.7 &  9.8 & 0.4 & 1.8$\rm{\times 10^{21}}$ & 1.3$\rm{\times 10^{2}}$ & 7.6$\rm{\times 10^{1}}$ & 1.1$\rm{\times 10^{3}}$ &  1.96\\
MWISP G109.925$-$01.350$-$031.88 & 109.925 &  $-$1.350 & 128 &  95 &   20 &  $-$31.88 &  1.7 &  1.08 &  0.8 &  9.4 & 0.3 & 1.3$\rm{\times 10^{21}}$ & 1.1$\rm{\times 10^{2}}$ & 5.1$\rm{\times 10^{1}}$ & 6.8$\rm{\times 10^{2}}$ &  1.86\\
MWISP G110.008$-$01.950$-$056.45 & 110.008 &  $-$1.950 &  83 & 115 &  110 &  $-$56.45 &  4.5 &  0.94 &  0.7 & 16.6 & 0.4 & 4.6$\rm{\times 10^{21}}$ & 3.0$\rm{\times 10^{2}}$ & 2.0$\rm{\times 10^{2}}$ & 3.1$\rm{\times 10^{3}}$ &  0.43\\
MWISP G110.042$-$01.209$-$030.22 & 110.042 &  $-$1.209 & 111 & 183 &  128 &  $-$30.22 &  2.1 &  1.39 &  1.1 &  8.6 & 0.5 & 2.1$\rm{\times 10^{21}}$ & 2.9$\rm{\times 10^{2}}$ & 7.5$\rm{\times 10^{1}}$ & 7.4$\rm{\times 10^{2}}$ &  1.54\\
MWISP G110.216$-$02.984$-$034.87 & 110.216 &  $-$2.984 &  90 & 112 &  121 &  $-$34.87 &  2.1 &  0.73 &  0.7 &  7.8 & 0.6 & 1.1$\rm{\times 10^{21}}$ & 7.7$\rm{\times 10^{1}}$ & 4.7$\rm{\times 10^{1}}$ & 7.0$\rm{\times 10^{2}}$ &  1.05\\
MWISP G110.242$-$02.884$-$034.04 & 110.242 &  $-$2.884 &  86 & 121 &  132 &  $-$34.04 &  1.7 &  1.45 &  0.7 & 10.1 & 0.3 & 1.7$\rm{\times 10^{21}}$ & 1.2$\rm{\times 10^{2}}$ & 7.3$\rm{\times 10^{1}}$ & 1.1$\rm{\times 10^{3}}$ &  2.60\\
MWISP G110.249$-$01.700$-$049.81 & 110.249 &  $-$1.700 &  96 &  68 &  $-$34 &  $-$49.81 &  2.2 &  0.91 &  0.5 & 10.9 & 0.3 & 1.5$\rm{\times 10^{21}}$ & 6.8$\rm{\times 10^{1}}$ & 7.7$\rm{\times 10^{1}}$ & 1.6$\rm{\times 10^{3}}$ &  1.35\\
MWISP G110.266$-$01.675$-$049.48 & 110.266 &  $-$1.675 &  72 & 134 &  130 &  $-$49.48 &  2.9 &  0.94 &  0.7 & 11.4 & 0.4 & 2.2$\rm{\times 10^{21}}$ & 1.5$\rm{\times 10^{2}}$ & 9.6$\rm{\times 10^{1}}$ & 1.5$\rm{\times 10^{3}}$ &  0.87\\
MWISP G110.308$-$02.784$-$036.53 & 110.308 &  $-$2.784 &  69 & 128 &   57 &  $-$36.53 &  1.7 &  1.70 &  0.7 & 10.6 & 0.3 & 2.1$\rm{\times 10^{21}}$ & 1.3$\rm{\times 10^{2}}$ & 9.3$\rm{\times 10^{1}}$ & 1.5$\rm{\times 10^{3}}$ &  3.12\\
MWISP G110.317$-$04.209$-$037.20 & 110.317 &  $-$4.209 &  77 & 131 &   35 &  $-$37.20 &  2.1 &  1.07 &  0.7 &  8.8 & 0.5 & 1.6$\rm{\times 10^{21}}$ & 1.1$\rm{\times 10^{2}}$ & 6.9$\rm{\times 10^{1}}$ & 1.0$\rm{\times 10^{3}}$ &  1.53\\
MWISP G110.349$-$02.809$-$036.03 & 110.349 &  $-$2.809 &  95 & 151 &   51 &  $-$36.03 &  1.8 &  1.23 &  0.9 & 10.8 & 0.3 & 1.7$\rm{\times 10^{21}}$ & 1.7$\rm{\times 10^{2}}$ & 6.5$\rm{\times 10^{1}}$ & 7.9$\rm{\times 10^{2}}$ &  1.71\\
MWISP G110.400$-$02.875$-$034.87 & 110.400 &  $-$2.875 & 103 & 200 &   58 &  $-$34.87 &  1.6 &  2.36 &  1.1 & 10.0 & 0.3 & 2.6$\rm{\times 10^{21}}$ & 3.7$\rm{\times 10^{2}}$ & 9.6$\rm{\times 10^{1}}$ & 9.4$\rm{\times 10^{2}}$ &  3.48\\
MWISP G110.417$-$01.584$-$049.14 & 110.417 &  $-$1.584 & 159 &  91 &    6 &  $-$49.14 &  3.5 &  1.55 &  0.9 & 15.9 & 0.3 & 5.4$\rm{\times 10^{21}}$ & 5.4$\rm{\times 10^{2}}$ & 2.1$\rm{\times 10^{2}}$ & 2.5$\rm{\times 10^{3}}$ &  0.84\\
MWISP G110.441$-$01.750$-$049.81 & 110.441 &  $-$1.750 &  86 &  93 &   63 &  $-$49.81 &  2.2 &  1.02 &  0.6 & 10.3 & 0.4 & 1.7$\rm{\times 10^{21}}$ & 9.4$\rm{\times 10^{1}}$ & 7.8$\rm{\times 10^{1}}$ & 1.4$\rm{\times 10^{3}}$ &  1.43\\
MWISP G110.450$-$02.892$-$034.20 & 110.450 &  $-$2.892 & 117 & 194 &  174 &  $-$34.20 &  1.9 &  1.66 &  1.2 &  8.9 & 0.4 & 2.2$\rm{\times 10^{21}}$ & 3.5$\rm{\times 10^{2}}$ & 8.0$\rm{\times 10^{1}}$ & 7.4$\rm{\times 10^{2}}$ &  1.96\\
MWISP G110.459$-$01.693$-$049.15 & 110.459 &  $-$1.693 &  90 & 126 &  169 &  $-$49.15 &  3.8 &  0.83 &  0.8 & 11.2 & 0.6 & 2.8$\rm{\times 10^{21}}$ & 2.2$\rm{\times 10^{2}}$ & 1.1$\rm{\times 10^{2}}$ & 1.6$\rm{\times 10^{3}}$ &  0.51\\
MWISP G110.475$-$01.608$-$048.98 & 110.475 &  $-$1.608 &  80 & 100 &   30 &  $-$48.98 &  1.8 &  1.53 &  0.6 & 11.5 & 0.3 & 2.2$\rm{\times 10^{21}}$ & 1.2$\rm{\times 10^{2}}$ & 1.0$\rm{\times 10^{2}}$ & 1.8$\rm{\times 10^{3}}$ &  2.53\\
MWISP G110.484$-$03.000$-$034.04 & 110.484 &  $-$3.000 &  88 & 136 &   64 &  $-$34.04 &  1.7 &  1.51 &  0.8 &  9.9 & 0.3 & 1.8$\rm{\times 10^{21}}$ & 1.5$\rm{\times 10^{2}}$ & 7.4$\rm{\times 10^{1}}$ & 1.0$\rm{\times 10^{3}}$ &  2.52\\
MWISP G110.500$-$01.684$-$049.48 & 110.500 &  $-$1.684 &  78 & 104 &   92 &  $-$49.48 &  2.8 &  0.61 &  0.6 & 17.2 & 0.2 & 1.8$\rm{\times 10^{21}}$ & 1.0$\rm{\times 10^{2}}$ & 8.2$\rm{\times 10^{1}}$ & 1.4$\rm{\times 10^{3}}$ &  0.49\\
MWISP G110.533$-$02.975$-$034.53 & 110.533 &  $-$2.975 & 122 & 101 &  $-$20 &  $-$34.53 &  2.0 &  1.63 &  0.8 & 10.3 & 0.3 & 2.4$\rm{\times 10^{21}}$ & 2.1$\rm{\times 10^{2}}$ & 9.8$\rm{\times 10^{1}}$ & 1.3$\rm{\times 10^{3}}$ &  2.20\\
MWISP G110.541$-$01.775$-$046.32 & 110.541 &  $-$1.775 &  71 &  83 &   88 &  $-$46.32 &  2.0 &  0.71 &  0.5 & 10.9 & 0.3 & 1.1$\rm{\times 10^{21}}$ & 4.5$\rm{\times 10^{1}}$ & 6.0$\rm{\times 10^{1}}$ & 1.4$\rm{\times 10^{3}}$ &  1.14\\
MWISP G110.549$-$01.916$-$045.66 & 110.549 &  $-$1.916 &  56 & 138 &  151 &  $-$45.66 &  2.0 &  1.00 &  0.6 & 12.0 & 0.3 & 1.6$\rm{\times 10^{21}}$ & 8.4$\rm{\times 10^{1}}$ & 7.4$\rm{\times 10^{1}}$ & 1.4$\rm{\times 10^{3}}$ &  1.49\\
MWISP G110.551$-$01.184$-$042.68 & 110.551 &  $-$1.184 &  83 & 141 &   94 &  $-$42.68 &  2.3 &  3.62 &  0.8 & 10.7 & 0.4 & 6.5$\rm{\times 10^{21}}$ & 5.2$\rm{\times 10^{2}}$ & 2.6$\rm{\times 10^{2}}$ & 3.6$\rm{\times 10^{3}}$ &  4.17\\
\enddata
\tablecomments{The source name is defined under the MWISP standard. Columns 2$-$6 give the centroid positions, the FWHMs of the two principal axes, and the position angles of the clumps. The centroid velocity, the peak intensity of $^{13}$CO emission, and the FWHM line width of the clumps are presented in columns 7$-$9. Columns 10$-$17 list the derived parameters of the clumps, i.e., effective radius, excitation temperature, optical depth, H$_2$ column density, total mass derived from the LTE method, surface density, number density, and the virial parameter, respectively. This table is available in its entirety in machine$-$readable form.}
\end{deluxetable*}
\end{longrotatetable}

The mass distribution, ranging from about 20 M$_{\sun}$ to 3$\times$ 10$^{3}$ M$_{\sun}$, of the identified clumps is displayed in Figure \ref{fig:Fig13b}. Previously, clumps identified using different methods and on different scales exhibit a universal power-law mass spectrum, $dN/dM \propto M^{-\alpha}$, with a power-law index $\alpha$ ranging from 1.4 to 2.2 \citep{Blitz+etal+1993, Rice+etal+2016}. The power-law index of the clump mass function (CMF) is very sensitive to the fit range. We fit the CMF from the mass derived with the median values of all the physical parameters to the mass end of the histogram in Figure \ref{fig:Fig13b}. The resulting power-law index is 2.20. This value is similar to the that of the CMF of the molecular clouds in the outer galaxy (2.2) \citep{Rice+etal+2016}, but is larger than the indexes (1.6$-$1.8) found by \cite{Kramer+etal+1998}. A CMF index greater than 2.0 indicates that the majority of the mass of clumps is contained in the low-mass clumps.

The relation between the masses and the sizes of clumps is a probe of the density distribution of cloud fragments \citep{Kauffmann+etal+2010a, Kauffmann+etal+2010b}. \cite{Kauffmann+etal+2010a} studied the mass-radius relation of clumps in several solar neighborhood clouds and found that clouds with masses under the empirical relationship $M(R)\leqslant 870M_\sun(R/pc)^{1.33}$ are not capable of hosting massive star formation (see shaded area in Figure \ref{fig:Fig13c}). Figure \ref{fig:Fig13c} also shows the empirical upper and lower bounds of the clump surface density, 1 g cm$^{-2}$ and 0.05 g cm$^{-2}$, required for massive star formation \citep{Urquhart+etal+2013}. Almost all clumps identified in this work fall below the Kauffmann's relation and the lower bound surface density of 0.05 g cm$^{-2}$ (corresponds to 1.07$\times$10$^{22}$cm$^{-2}$). This result is consistent with the lack of ongoing massive star formation in this region. The red line in Figure \ref{fig:Fig13c} shows the power-law fit of the mass-size relation $M\varpropto R_{eff}^{2.63}$ of the clumps. With a power-law index of 2.63, this result indicates the non-uniform surface densities of clumps on parsec scales.

\begin{figure*}[htb!]
	\centering
	\subfigure[]{\label{fig:Fig14a}
		\includegraphics[trim=1.3cm 0.7cm 3cm 2.5cm, width = 0.32\linewidth , clip]{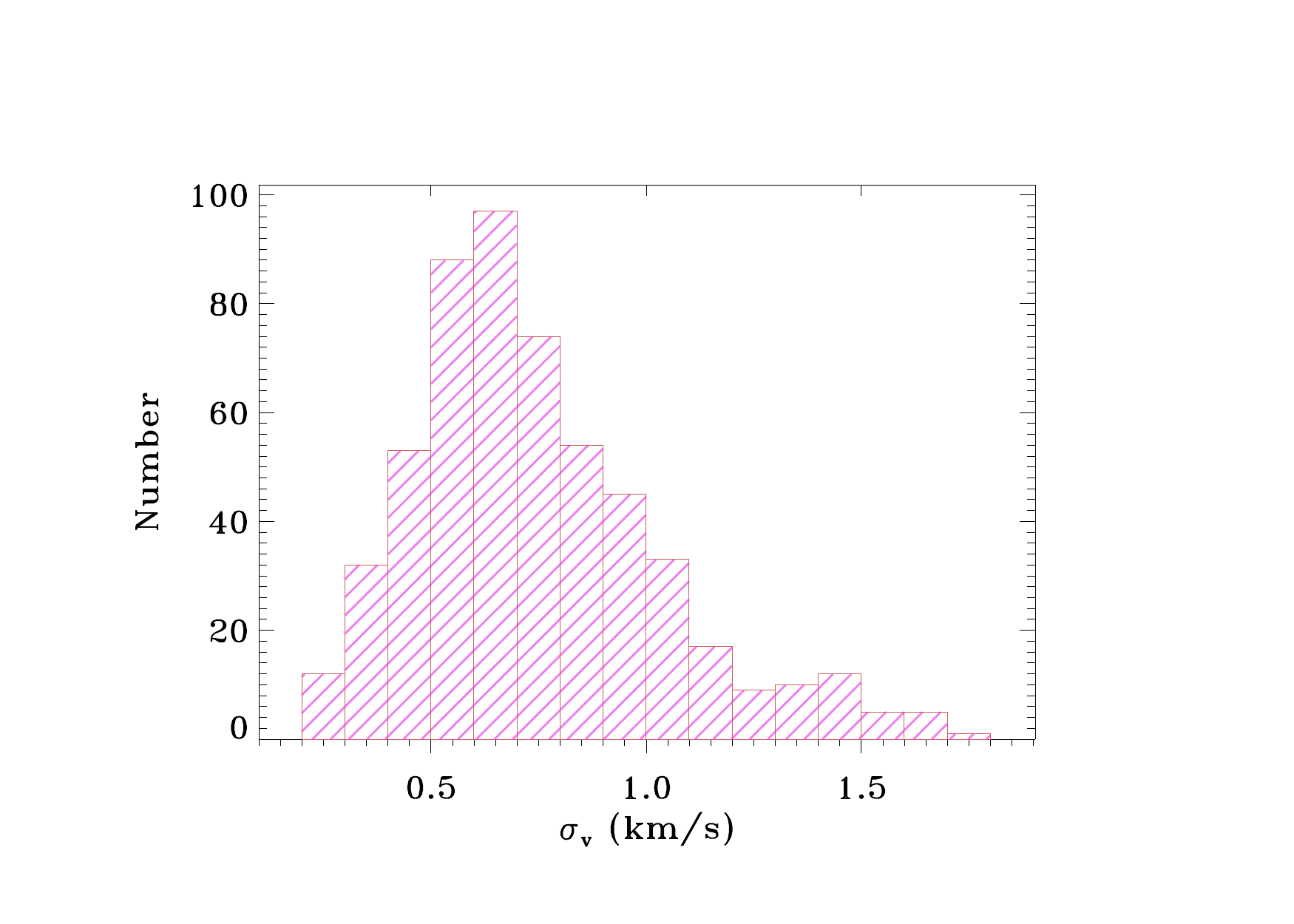}}
	\subfigure[]{\label{fig:Fig14b}
		\includegraphics[trim=1.3cm 0.7cm 3cm 2.5cm, width = 0.32\linewidth , clip]{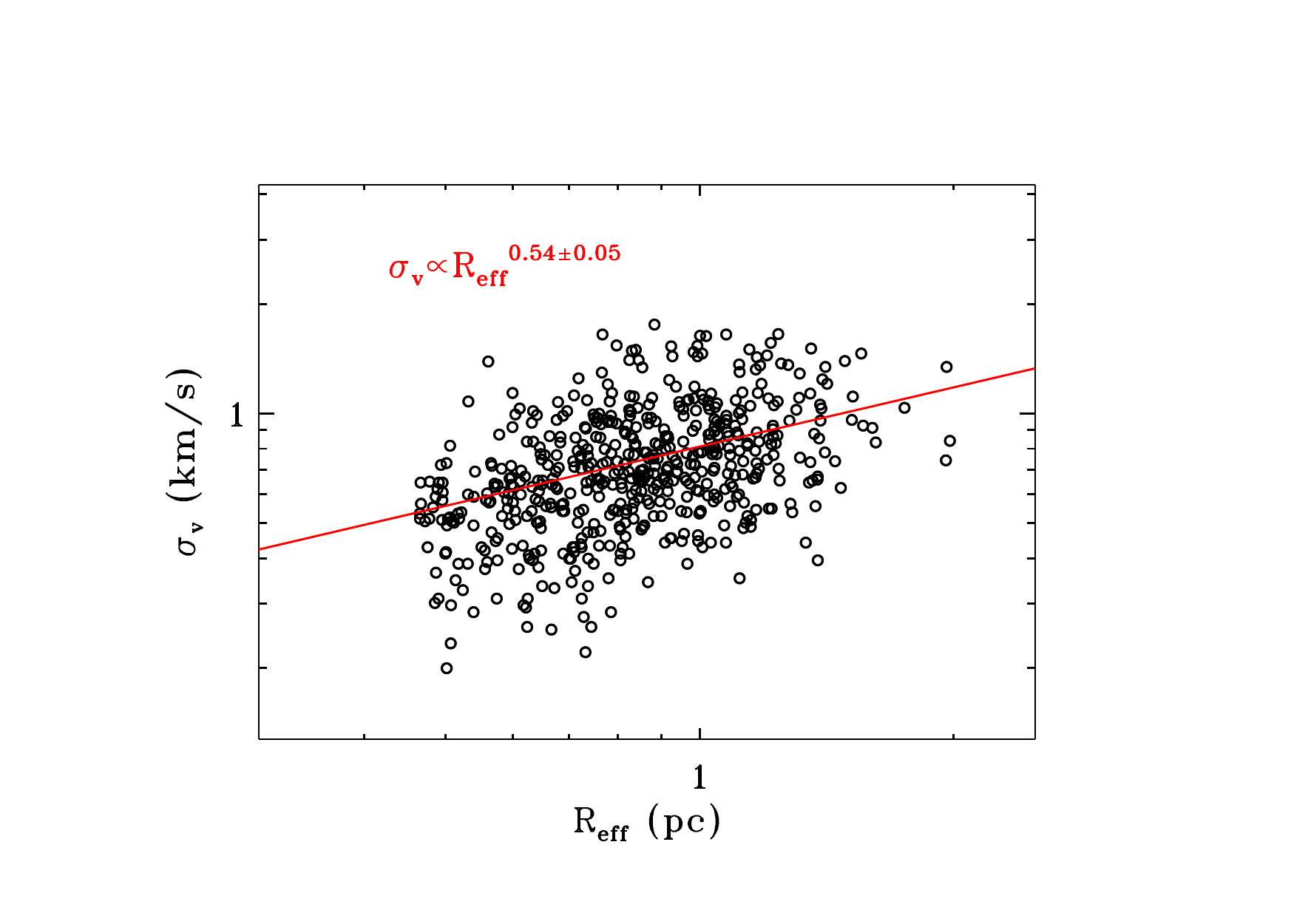}}
	\subfigure[]{\label{fig:Fig14c}
		\includegraphics[trim=1.1cm 0.7cm 3cm 2.3cm, width = 0.32\linewidth , clip]{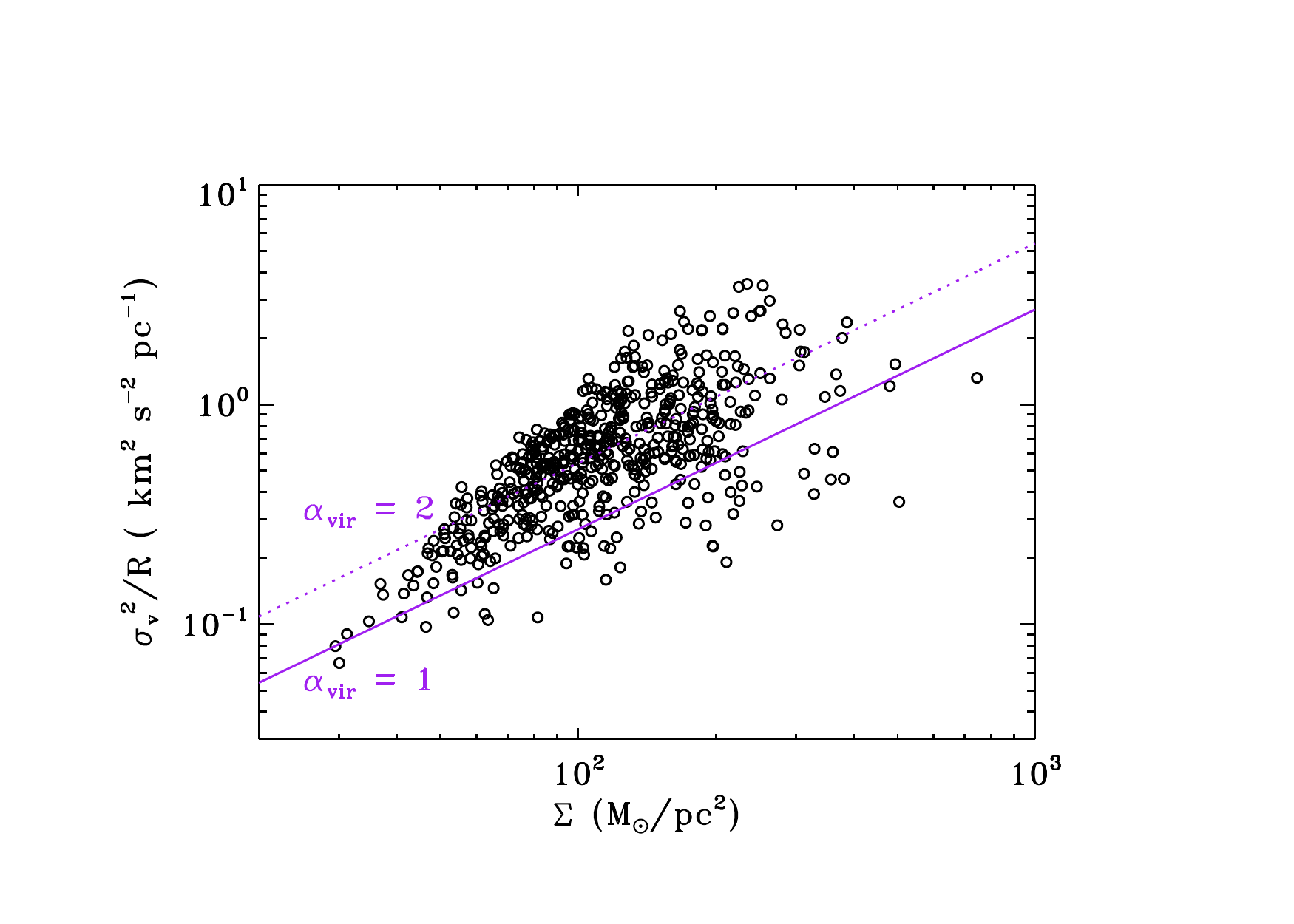}}
	\caption{(a) Distribution of velocity dispersion, (b) line-width$-$size relation, and (c) $\sigma_v^2/R$ vs. surface density for the fitted $\rm{^{13}CO}$ clumps in the Cas GMC.}
	\label{fig:Fig14}
\end{figure*}

Figure \ref{fig:Fig14a} shows the distribution of the velocity dispersions of the identified clumps. The observed velocity dispersions range from 0.2 km s$^{-1}$ to 1.7 km s$^{-1}$, with a median value around 0.6 km s$^{-1}$. \cite{Larson+etal+1981} found a line-width$-$size relation $\sigma_ v\propto R_{eff}^{0.38}$ for molecular clouds and suggested that the internal turbulence of the clouds is responsible for this relation. The fitted line-width$-$radius relation for the clumps identified in this work is shown with the red solid line in Figure \ref{fig:Fig14b}. The exponent is 0.54, larger than the Larson's exponent of 0.38, but the Pearson correlation coefficient of $log\ R_{eff}$ and $log\ \sigma_v$ is low (0.42). Figure \ref{fig:Fig14c} gives the relationship between $\sigma_v^{2}/R_{eff}$ and the surface density $\Sigma$ of the clumps. We can see that $\sigma_v^{2}/R_{eff}$ varies systematically with $\Sigma$, which is similar to the result found by \cite{Heyer+etal+2009}. This relationship conflicts with the Larson's third law that molecular clouds have constant surface densities. Our results indicate that the internal motions of molecular clouds depend both on their sizes and surface densities. The solid and dashed purple lines in Figure \ref{fig:Fig14c} show the relationships corresponding to simple virial equilibrium (SVE; $\alpha = 1$) and marginally bound clouds ($\alpha = 2$), respectively. The majority of the identified clumps are distributed above the SVE line but along the marginally bound line ($\alpha = 2$), which implies that gravity plays a dominant role in confining the identified clumps. This result is different from that obtained in the OGS survey, where \cite{Heyer+etal+2001} found that external pressure is needed to confine the clouds. 

\section{Discussion}\label{sec:Sec4}
\subsection{Probability Density Function (PDF) of the Column Density}\label{sec:Sec4.1}
\begin{figure*}[t!]
\centering
\subfigure[]{
\label{fig:Fig15a}
\includegraphics[trim=1cm 3.5cm 0.8cm 4cm, width = 0.46\linewidth , clip]{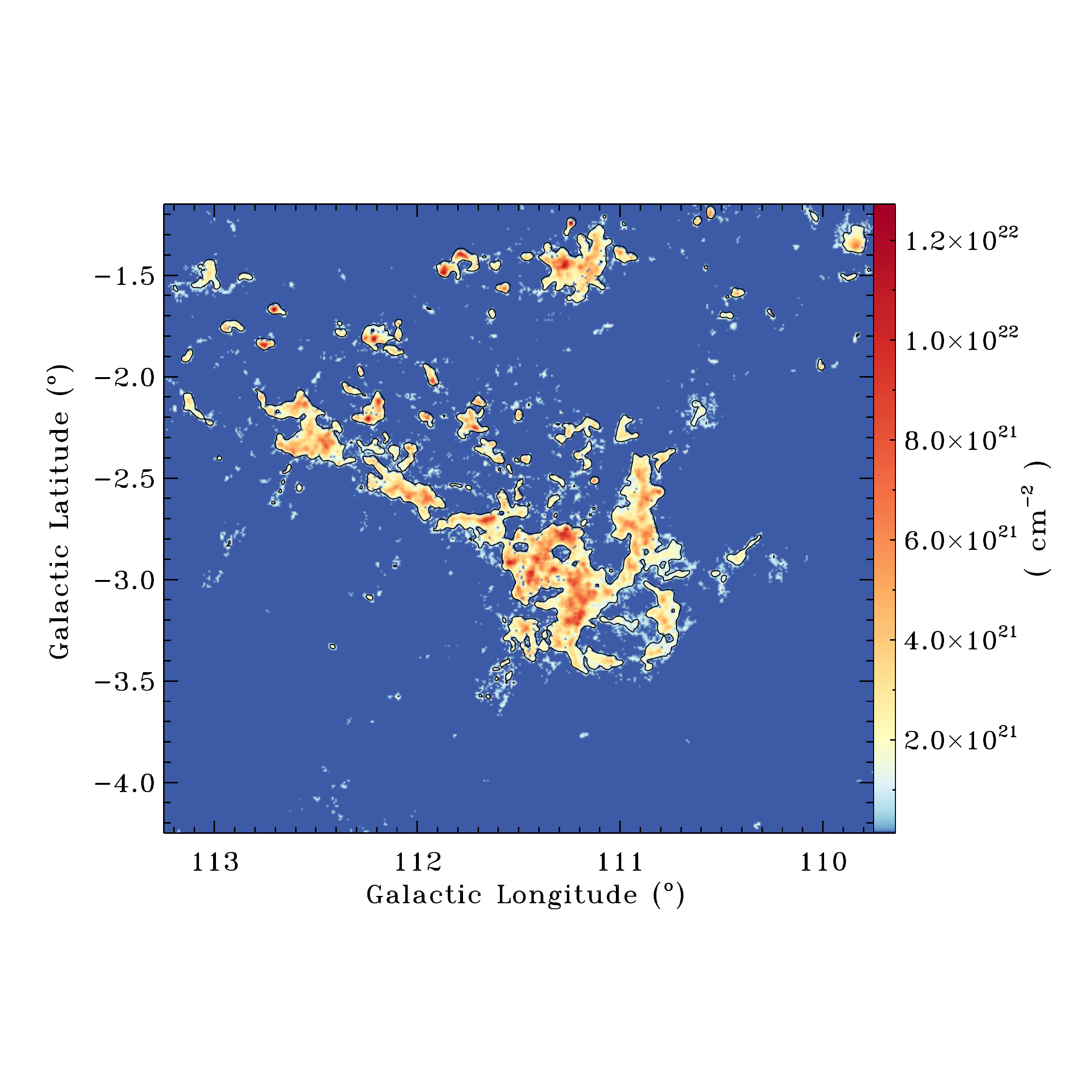}}
\subfigure[]{
\label{fig:Fig15b}
\includegraphics[trim=0cm 0cm 0cm 0cm, width = 0.49\linewidth , clip]{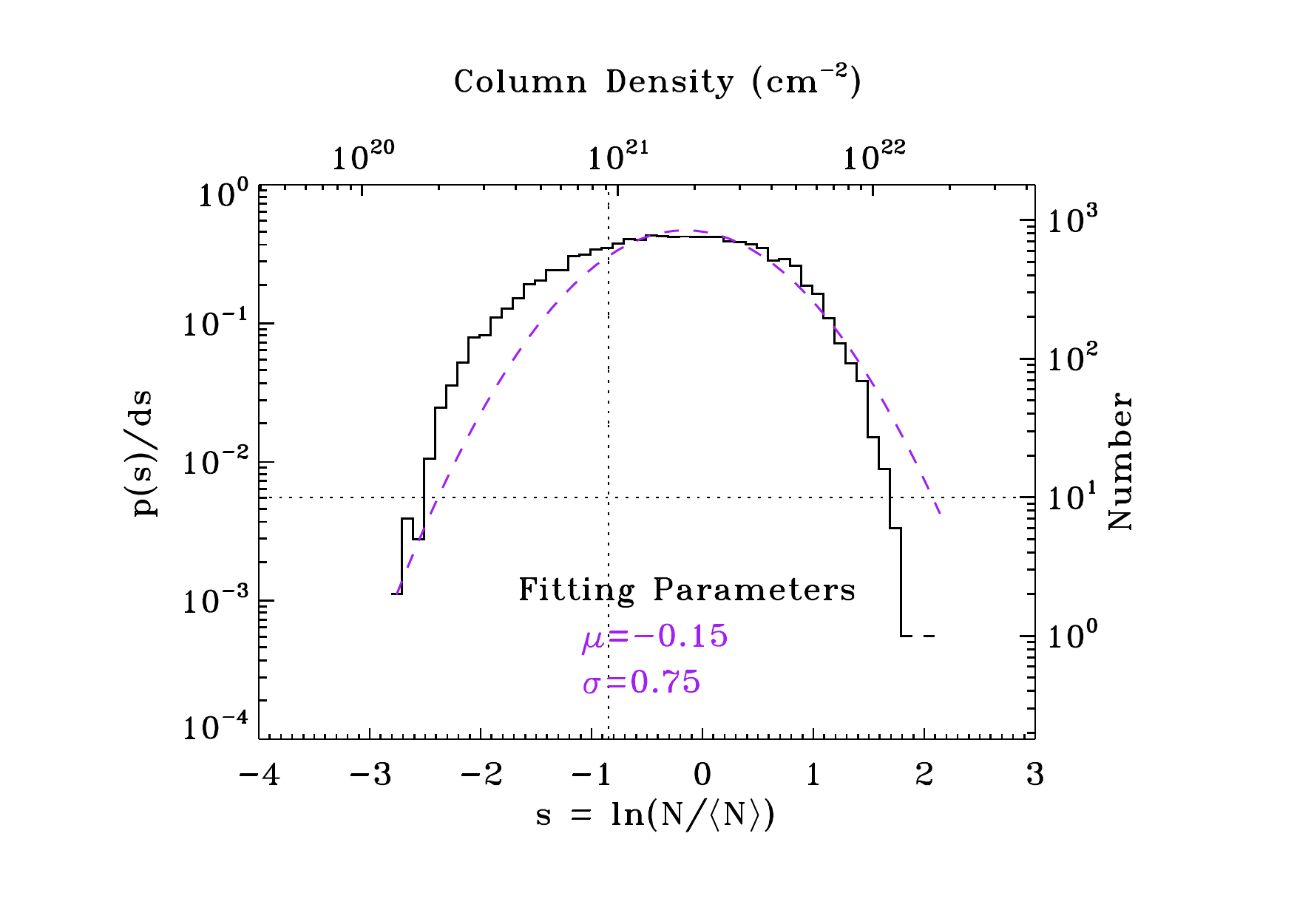}}
\caption{(a) Spatial distribution of the column density in Cas GMC derived under the assumption of LTE. (b) Probability density function of the $\rm{H_2}$ column density of Cas GMC. The vertical and horizontal dashed lines mark the detection limit and where the number of pixels within a bin is 10, respectively. $\mu$ and $\sigma$ are the mean and dispersion of the fitted Gaussian function of ln(N/$<$N$>$).}
\label{fig:Fig15}
\end{figure*}
The column density PDF (N-PDF) is a simple but very useful tool to describe the molecular cloud structure. Both theoretical work and numerical simulations predict that the N-PDFs of molecular clouds are characterized by log-normal functions when supersonic turbulence is the dominant dynamical process in molecular clouds and the N-PDFs tend to develop power-law tails at high densities once self-gravity becomes important \citep{Vazquez-Semadeni+etal+1994, Padoan+etal+1997, Klessen+etal+2000, Federrath+etal+2010}. This evolutionary trend has been supported by a broad range of observations with different kinds of tracers of column density \citep{Goodman+etal+2009, Kainulainen+etal+2009, Schneider+etal+2013}. However, \cite{Tassis+etal+2010} have shown through numerical simulations that log-normal N-PDFs are generic features of diverse model clouds that do not include supersonic turbulence, and they suggested that log-normal N-PDFs should not be interpreted as being a consequence of supersonic turbulence. \cite{Alves+etal+2017} argued that no log-normal PDFs are present if the completeness (the last closed contour of the sample) of the observation is taken into account. The wide-field survey toward the Cas GMC allows us to examine the N-PDF in this region.

Figure \ref{fig:Fig15a} displays the spatial distribution of the H$_2$ column density derived from the LTE method. The corresponding pixel-by-pixel N-PDF is shown in Figure \ref{fig:Fig15b}. According to \cite{Alves+etal+2017}, the last closed contour has a decisive role in determining the reliability of the log-normal peak. The completeness of our statistics is reflected in Figure \ref{fig:Fig15a} with the black contour, which corresponds to the detection limit. We note that none of the molecular cloud samples used in the study of \cite{Alves+etal+2017} were fully covered by observations. In this work, our observations fully cover the Cas GMC. We fit the PDF with a log-normal function from the detection limit N$\rm_{H2}$ = $\rm{9.2\times10^{20}cm^{-2}}$ to N$\rm_{H2}$ = $\rm{1.1\times10^{22}cm^{-2}}$ which is the value of the last bin that has the number of pixels greater than 10. The N-PDF can be well fitted in the range from $\rm{\sim10^{21}}$ cm$^{-2}$ to $\rm{\sim10^{22}}$ cm$^{-2}$ but drops drastically toward high densities when $\rm{N_{H2} >10^{22}}$ $\rm{cm^{-2}}$. The overall column density of the GMC is low, with an average column density of N$\rm_{H2}$ = $\rm{1.6\times10^{21}cm^{-2}}$, which is much below the threshold of star formation $6.3\times10^{21}$ cm$^{-2}$ \citep{Johnstone+etal+2004, Lada+etal+2010, Kainulainen+etal+2014}. The absence of a power tail at the high column density end in this region is consistent with the fact that this GMC is quiescent in star formation (see discussions in Section \ref{sec:Sec4.2}). These results all show that the Cas GMC is either in its early evolutionary stage or has passed its main star formation phase.

\subsection{Star Formation in the Cas GMC}\label{sec:Sec4.2}

To investigate the star formation activity in the Cas GMC, we used the archive data of WISE and 2MASS to identify young stellar objects (YSOs). The YSO identification scheme resembles the one described in \cite{Koenig+etal+2012}. The detailed criteria can be found in \cite{Koenig+etal+2014}. Here we just give a brief description of the identification process. Firstly, for the sources with detections in the WISE W1, W2, and W3 bands, the Class I and Class II YSOs are identified based on their infrared excess after the contamination of star-forming galaxies and AGNs is removed. Second, YSOs are identified among the remaining sources based on infrared excess in the H$-$K$_s$ versus W1$-$W2 color space. Thirdly, the WISE W4 (22\,$\mu$m) photometry is used to identify transition disks and to retrieve protostars from the AGN candidates. Finally, all YSOs identified above are re-examined to exclude the possible asymptotic giant branch (AGB) and classical Be stars (CBe) \citep{Rivinius+etal+2013}.

We totally identified 21 Class I protostars and 76 Class II YSOs in this region. The spatial distribution of the identified Class I and Class II YSOs is presented in Figure \ref{fig:Fig2}. Overall, most of the YSOs are associated with (i.e., no farther than 5 pc from) the Cas GMC if we assume the YSOs are at the same distance as the Cas GMC. Because no information on the distances of the identified YSOs is available at present, we also checked the possible association between the YSOs and the local gas. The result shows poor association between them. Compared with the Cas GMC, the local gas in this region is much less. Therefore, it is reasonable to assume that the YSOs in this region are born in the Cas GMC.

Overall, the distribution of YSOs in this region does not show obvious clustering, only somewhat concentrated near the \ion{H}{2} regions G113.009$-$01.393 and G111.236$-$01.236 and the western arc-shaped structure of Cloud 3. Compared with YSOs in the \ion{H}{2} regions, YSOs in the western arc-shaped structure of Cloud 3 are associated more closely with molecular gas, where the column density of the gas is close to or slightly larger than the threshold of star formation, N$\rm_{H2}{\sim6.3\times10^{21}}$ cm$^{-2}$. Half of the total LTE mass of the Cas GMC has column density below $3\times10^{21}$ cm$^{-2}$ and only 12$\%$ of the total mass has column density above the star formation threshold. The first value is comparable to that for the Taurus molecular cloud (TMC; $2.1\times10^{21}$ cm$^{-2}$) \citep{Goldsmith+etal+2008}, which is a nearby dark cloud inactive in star formation. Assuming an average YSO mass of 0.5 M$_{\sun}$, we calculated the star formation efficiency (SFE) of the Cas GMC to be 0.02$\%$ when we use the LTE mass of $2.1\times10^5\ M_{\sun}$. Considering the large distance of Cas GMC and the sensitivity of the WISE survey, the YSO sample we have obtained is not complete. The value of 0.02$\%$ is a lower limit of the actual star formation efficiency of the GMC. By comparison, this value is far below the SFE of the TMC (0.3$-$1.2$\%$) \citep{Goldsmith+etal+2008} and that of the Ophiuchus north cloud (0.3$\%$) \citep{Nozawa+etal+1991}. This may be a result of the lack of high-density gas in the Cas GMC.

\section{Summary}\label{sec:Sec5}
We have conducted a fully covered simultaneous survey of the $^{12}$CO, $^{13}$CO, and C$^{18}$O $J=1-0$ emission toward the Cas A SNR with a sky coverage of 3.5$\arcdeg\times3.1\arcdeg$. The GMC in this region is nearly free of C$^{18}$O emission, and its $^{12}$CO and $^{13}$CO emission comes from molecular gas with velocity in the range from $-55$ to $-25$ km s$^{-1}$. The spatial distribution and velocity structures of the GMC are revealed for the first time. The main results are presented as follows.
\begin{enumerate}
\item The total mass of the Cas GMC is 2.1$\times10^{5}\ M_{\sun}$ as derived from the $\rm{^{13}CO}$ emission and 9.5$\times10^5\ M_{\sun}$ from the $\rm{^{12}CO}$ emission. The GMC can be divided into three main clouds with masses on the order of 10$^4$$-$10$^5\ M_{\sun}$ based on the spatial and velocity distribution. Combined with the parallax distance information of two adjacent galactic masers, G111.23$-$01.23 and G111.25$-$00.76, we propose a distance of d = 3.4 kpc for the Cas GMC.
\item Two regions with broadened (6$-$7 km s$^{-1}$) or asymmetric $^{12}$CO line profiles are found in the vicinity of the Cas A SNR (within a 10$\arcmin\times$10$\arcmin$ region). The criterion of line broadening is $\Delta v > 6$ km s$^{-1}$. At distances of 2.5$-$13 pc from the Cas A SNR center, we have identified four clumps with masses in the range of 2000 to 3200 $M_{\sun}$ that encompass a less luminous cavity around the Cas A SNR. The southeastern clump, clump a, shows velocity dispersions as large as 5 km s$^{-1}$. All the four clumps possess velocity gradients toward the southeast corner of the SNR. The velocity gradients and the large velocity dispersions may be caused by the influence of the MS wind of the progenitor of the Cas A SNR.
\item Using the GAUSSCLUMPS algorithm, we have identified 547 $^{13}$CO clumps in the GMC, 54$\%$ of which are found to be gravitationally bound ($\alpha<2$). The CMF of the clumps follows a power-law distribution with an exponent of $-2.20$. The mass-radius relation $M\varpropto R_{eff}^{2.63}$ of the clumps is inconsistent with the Larson's third law of constant surface density and suggests non-uniform internal density distribution. The values of $\sigma_v^2/R_{eff}$ values of the clumps vary systematically with the surface densities.
\item The pixel-by-pixel distribution of the H$_2$ column density of the GMC can be well fitted with a log-normal PDF. The median value of the H$_2$ column density of the GMC is $1.6\times10^{21}$ cm$^{-2}$ and half the mass of the GMC is contained in regions with H$_2$ column density lower than $3\times10^{21}$ cm$^{-2}$, which is well below the column density threshold of star formation \citep{Johnstone+etal+2004}. The distribution of the YSO candidates in the region shows no agglomeration. The lack of high-density gas in the Cas GMC is consistent with its low SFE of 0.02$\%$, which is much lower than those of the two typical inactive star formation clouds, the Taurus and the Ophiuchus north regions.
\end{enumerate}

\acknowledgments
We thank the PMO-13.7 m telescope staff for their supports during the observation and the anonymous referee for constructive suggestions. This work is supported by the National Key R$\&$D Program of China (NO. 2017YFA0402701). We acknowledge the support by NSFC grants 11233007, 11127903, and 11503086. Y.M. acknowledges the support by NSFC grant 11503087 and by the Natural Science Foundation of Jiangsu Province of China (grant No. BK20141046). M.Z. acknowledges the support of funding from the European Union’s Horizon 2020 research and innovation program under grant agreement No. 639459 (PROMISE). We credit NASA/CXC/SAO for the X-ray FITS images of the Cas A SNR. This publication makes use of data products from the \emph{Wide-field Infrared Survey Explorer}, which is a joint project of the University of California, Los Angeles, and the Jet Propulsion Laboratory/California Institute of Technology, funded by the National Aeronautics and Space Administration. This publication also makes use of data products from the Two Micron All Sky Survey, which is a joint project of the University of Massachusetts and the Infrared Processing and Analysis Center/California Institute of Technology, funded by the National Aeronautics and Space Administration and the National Science Foundation. This work makes use of the SIMBAD database, operated at CDS, Strasbourg, France.

\software{GILDAS/CLASS \citep{Pety+etal+2005, GT+etal+2013}}
\section{Appendix}
\begin{figure*}[ht!]
\centering
\includegraphics[trim=0cm 0.5cm 0cm 1cm, width=\linewidth , clip]{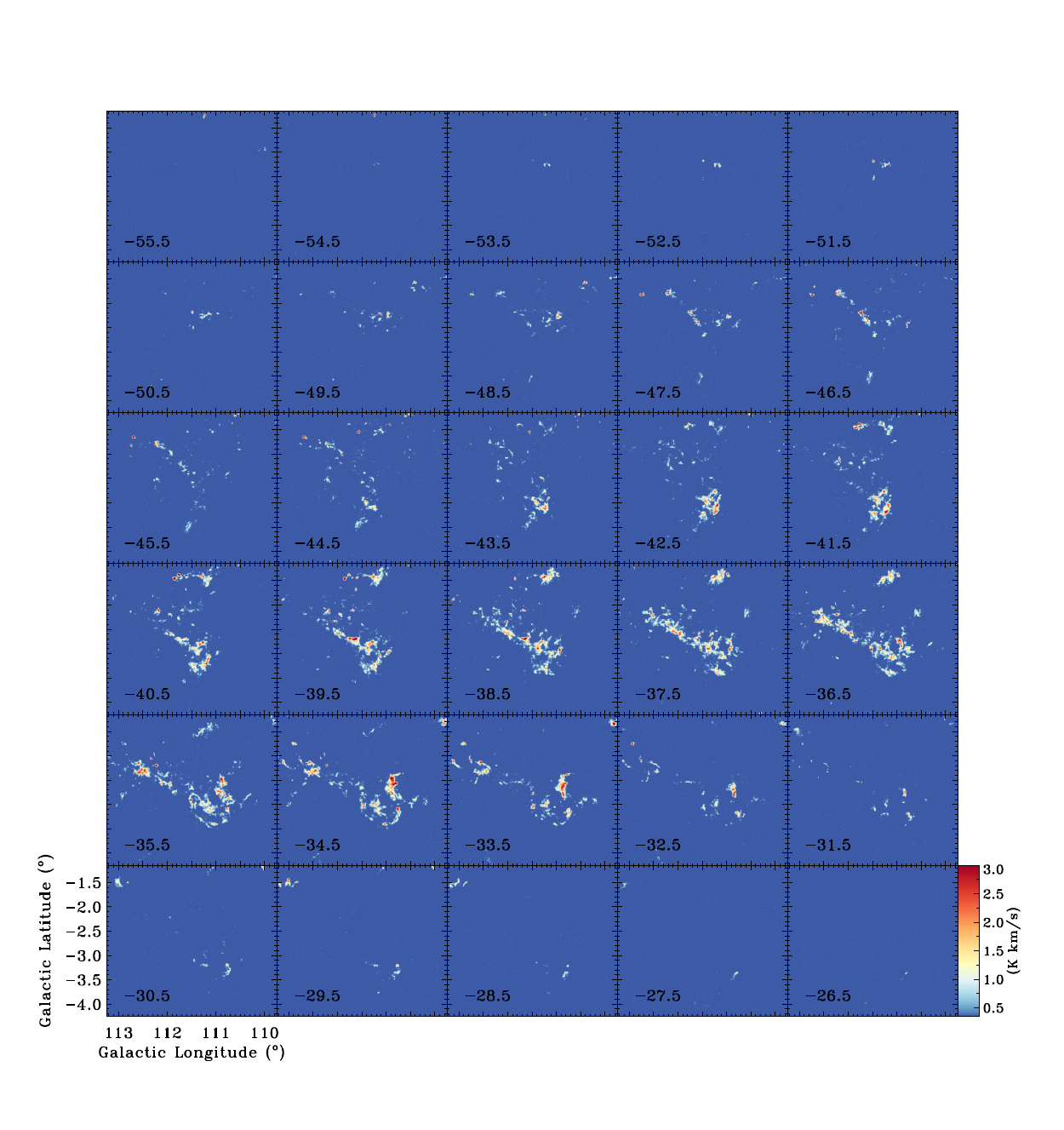}
\caption{$^{13}$CO $J=1-0$ velocity channel maps of Cas GMC integrated from $-$56 km s$^{-1}$ to $-$26 km s$^{-1}$ per 1 km s$^{-1}$. The central velocity of the integral interval is shown at the lower-left corner in each panel.\label{fig:Fig16}}
\end{figure*}

\begin{figure*}
\centering
\subfigure[]{
\includegraphics[trim=2cm 0cm 1cm 1cm, width=0.6\linewidth, clip]{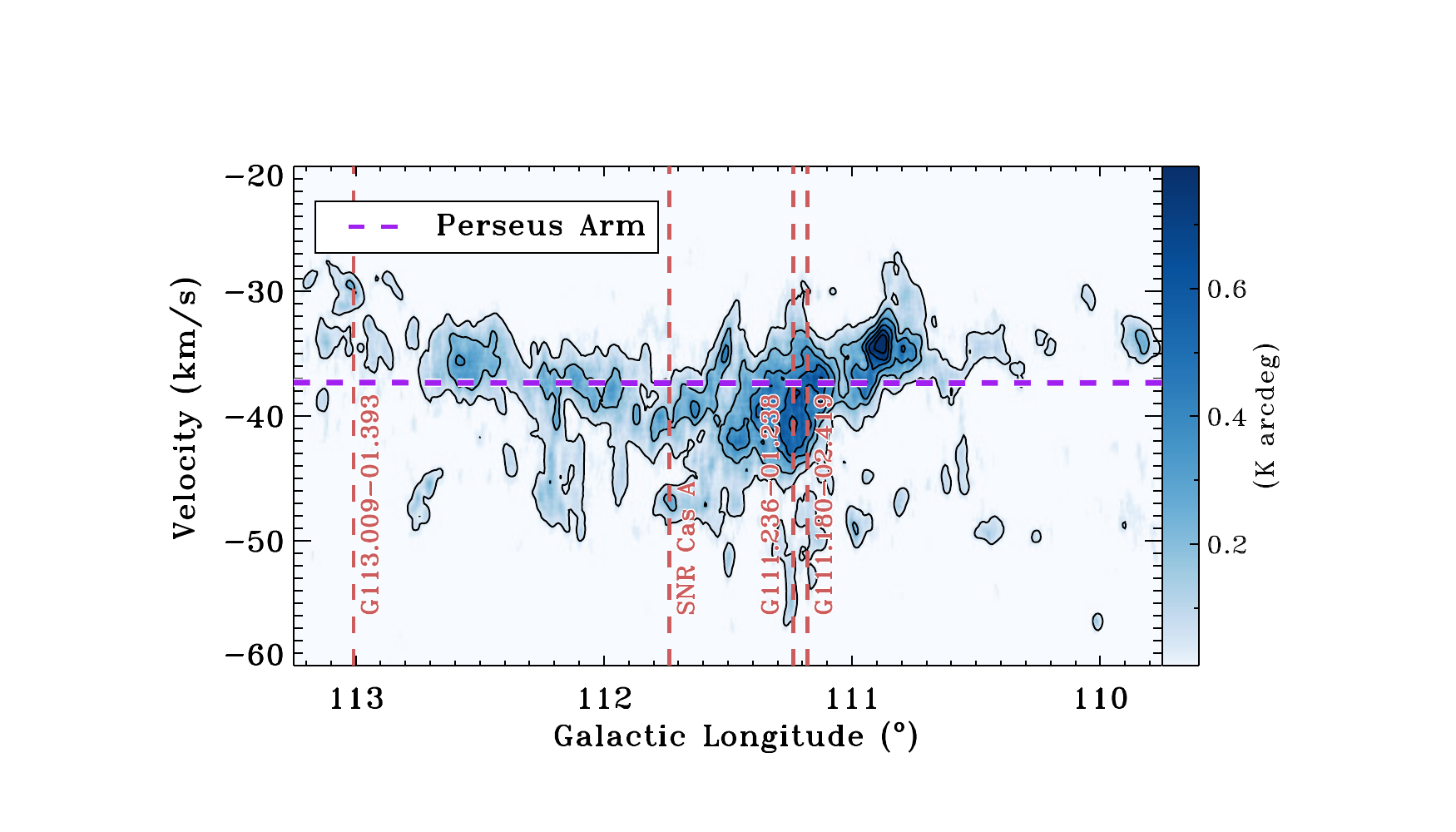}\label{Fig:lv_map_13CO}}
\subfigure[]{
\includegraphics[trim=0.5cm 3cm 1cm 5cm, width=0.38\linewidth, clip]{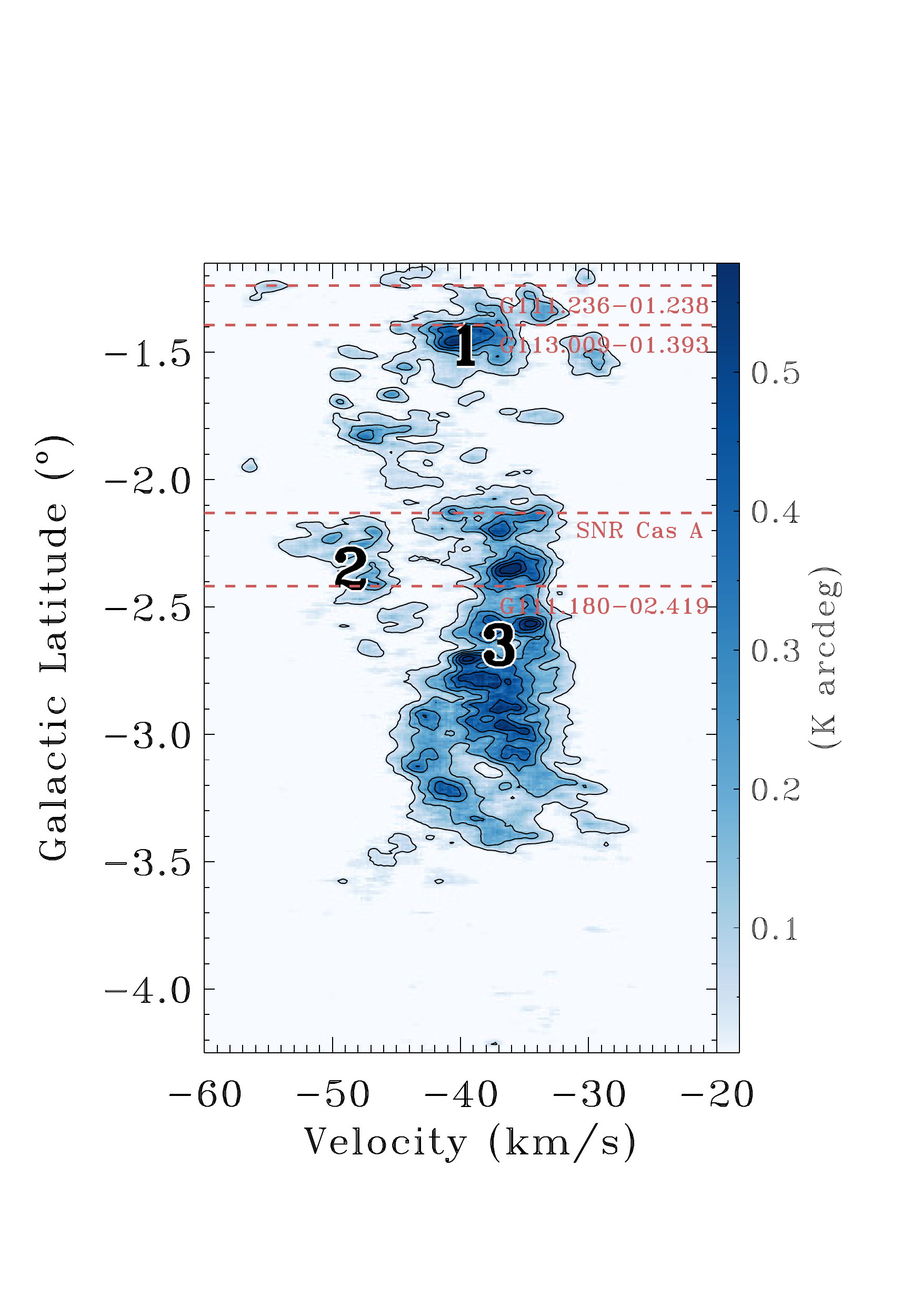}\label{Fig:bv_map_13CO}}
\caption{Same as Figure \ref{fig:Fig4} but with $^{13}$CO emission.}
\label{fig:Fig17}
\end{figure*}

\begin{figure*}
\subfigure[]{
\includegraphics[trim=1cm 3.5cm 2cm 3cm, width=0.5\linewidth, clip]{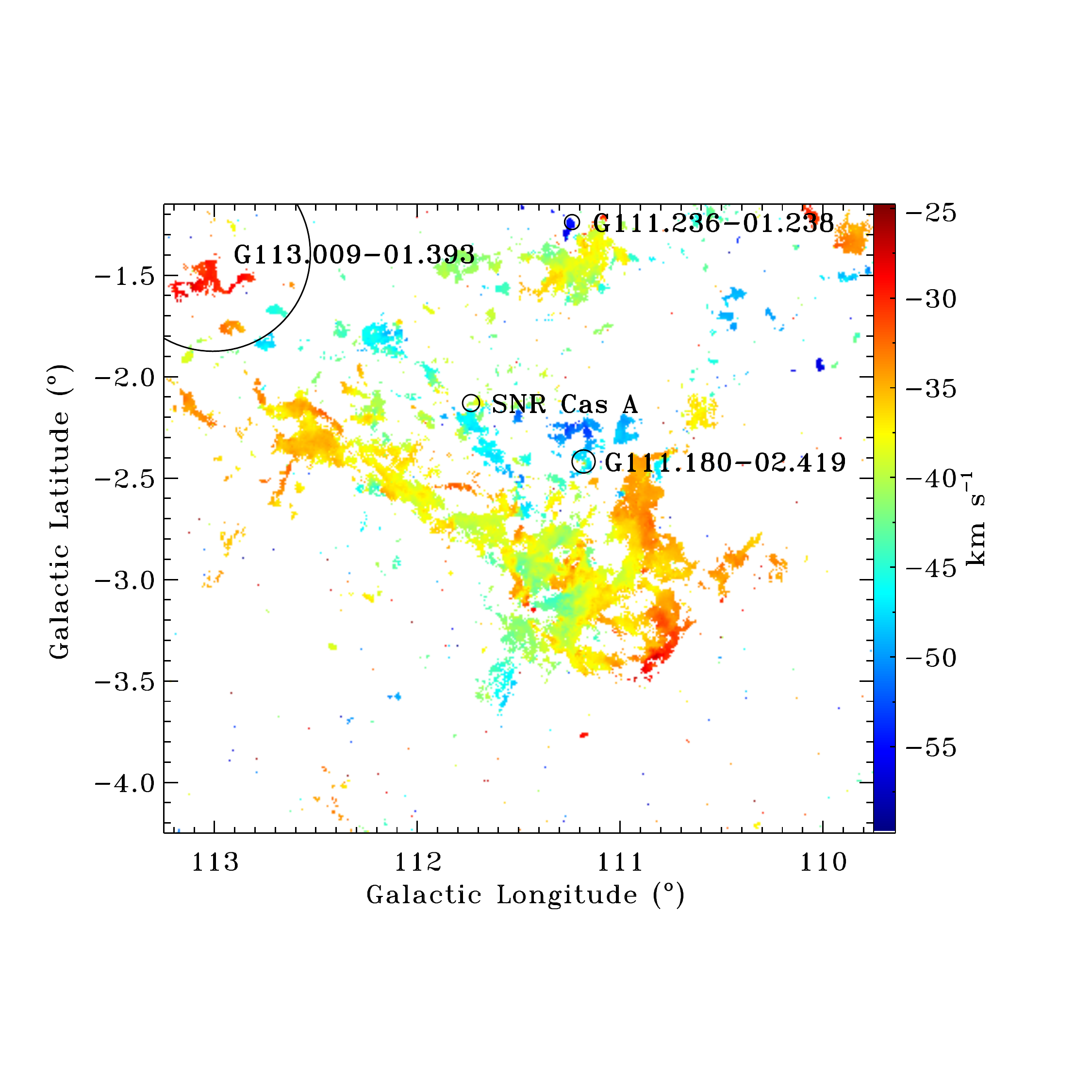}\label{Fig:m1_map_13CO}}
\subfigure[]{
\includegraphics[trim=1cm 3.5cm 2cm 3cm, width=0.5\linewidth, clip]{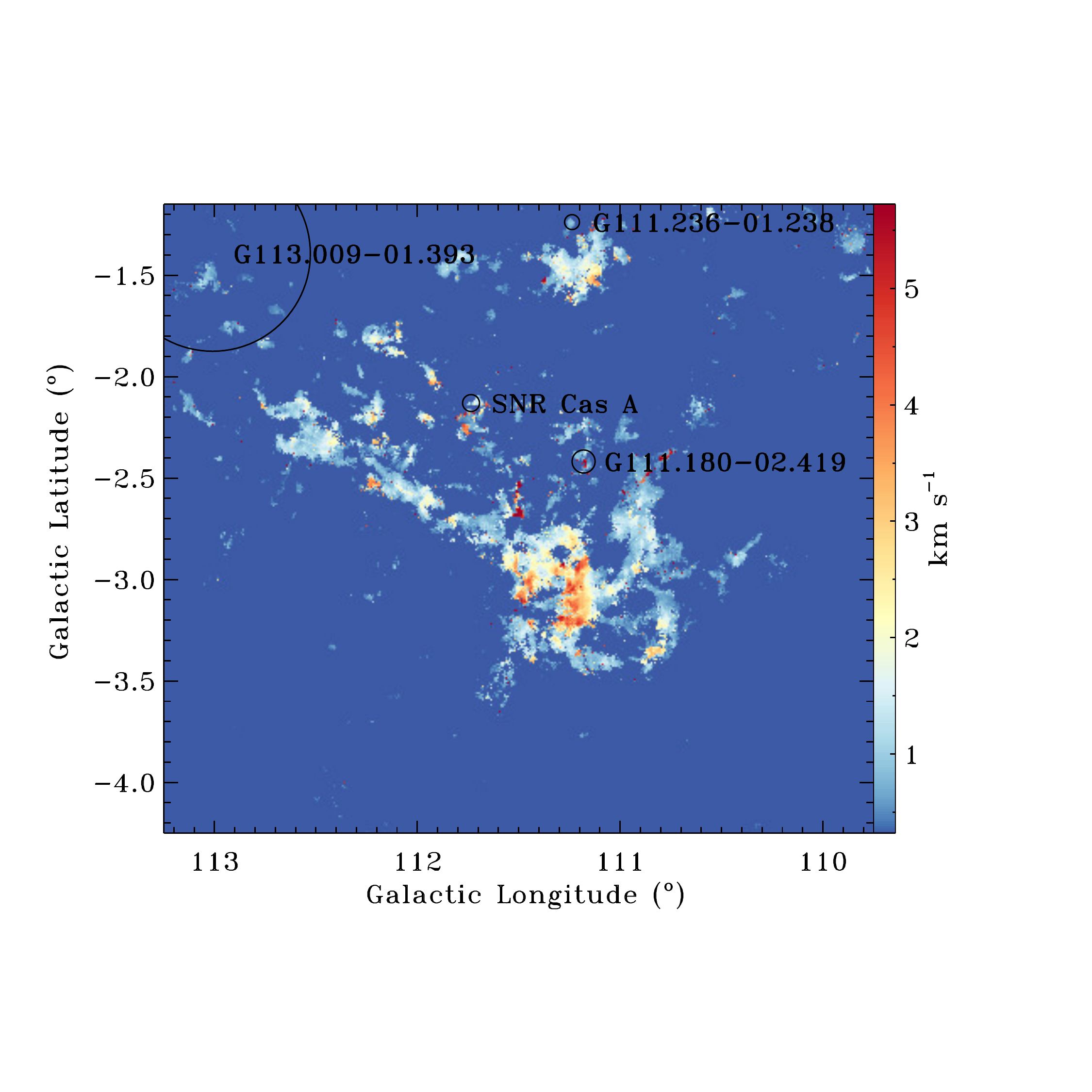}\label{Fig:m2_map_13CO}}
\caption{Same as Figure \ref{fig:Fig5} but with $^{13}$CO emission.}
\label{fig:Fig18}
\end{figure*}

We present the velocity channel maps, p-v diagrams, and moment 1 and 2 maps of the $^{13}$CO $J=1-0$ emission in Figures \ref{fig:Fig16}$-$\ref{fig:Fig18}, respectively.

\clearpage

\bibliographystyle{aasjournal}
\bibliography{references}

\end{document}